\documentclass[12pt]{article} 

\usepackage{amsmath,amsfonts,amsbsy,amssymb,array,accents,dsfont}
\usepackage{enumerate}
\usepackage{graphicx} 
\usepackage{caption} 
\usepackage{subcaption} 
\usepackage{amssymb}
\usepackage{upgreek}
\usepackage{bm}
\usepackage{slashed}
\usepackage[dvipsames]{xcolor}

\definecolor{myBlue}{RGB}{55, 100, 210}
\definecolor{myRed}{RGB}{200, 60, 40}

\PassOptionsToPackage{ 
     colorlinks=true,
     linkcolor=myBlue,
     urlcolor=blue,
     filecolor=black,
     citecolor=myRed,
}{hyperref}

\usepackage{mychet}
\let\includefigures=\iftrue
\let\useblackboard==\iftrue
\newfam\black

\def\PsiST{\Uppsi}

\def\epsilonb{{\boldsymbol\epsilon}}

\def\etab{{\boldsymbol\eta}}

\def\sl{\text{sl}}
\def\su{\text{su}}

\def\sltwo{\ensuremath{SL(2,\bR)}}
\def\sltwoc{\ensuremath{SL(2,\bC)}}

\def\sutwo{{SU(2)}}
\def\uone{U(1)}

\def\sqsphere{{\bS^3_\flat}}

\def\tight#1{\! #1 \!}  

\def\({\left(}
\def\){\right)}
\def\[{\left[}
\def\]{\right]}

\def\ie{{i.e.}}
\def\eg{{e.g.}}

\def\etc{{etc}}

\def\ST{{\sst\it\! ST}}
\def\WS{{\sst\it\! WS}}
\def\NS{{\it \!NS}}

\def\osc{{\rm osc}}
\def\tot{{\rm tot}}

\def\lstr{\ell_{\textit{s}}}

\def\gstr{g_{\textit s}^{\;}}

\def\pf{{\rm pf}}

\def\ntil{{n-2}}
\def\ktil{{k+2}}
\def\etatil{\alpha}

\def\phitil{{\phi}}

\def\sfX{{\mathsf X}}

\def\sfa{{\mathsf a}}
\def\sfb{{\mathsf b}}

\DeclareMathSymbol{\medhatsym}{\mathord}{largesymbols}{"62} 
\newcommand\lowermedhatsym{
  \text{\smash{\raisebox{-1.28ex}{%
    $\medhatsym$}}}}
\newcommand\medhat[1]{
  \mathchoice
    {\accentset{\displaystyle\lowermedhatsym}{#1}}
    {\accentset{\textstyle\lowermedhatsym}{#1}}
    {\accentset{\scriptstyle\lowermedhatsym}{#1}}
    {\accentset{\scriptscriptstyle\lowermedhatsym}{#1}}
}

\DeclareMathSymbol{\medtildesym}{\mathord}{largesymbols}{"65}
\newcommand\lowermedtildesym{
  \text{\smash{\raisebox{-1.2ex}{%
    $\medtildesym$}}}}
\newcommand\medtilde[1]{
  \mathchoice
    {\accentset{\displaystyle\lowermedtildesym}{#1}}
    {\accentset{\textstyle\lowermedtildesym}{#1}}
    {\accentset{\scriptstyle\lowermedtildesym}{#1}}
    {\accentset{\scriptscriptstyle\lowermedtildesym}{#1}}
}

\def\Phihat{\medhat\Phi}

\def\half{\frac12}

\def\hf{\coeff12}

\def\One{{\hbox{1\kern-1mm l}}}

\def\barray{\begin{array}}
\def\earray{\end{array}}
\def\be{\begin{equation}}
\def\ee{\end{equation}}
\def\bea{\begin{eqnarray}}
\def\eea{\end{eqnarray}}
\def\bal{\begin{align}}
\def\eal{\end{align}}
\def\nn{\nonumber}

\def\II{{\mathcal{I}}}
\def\LL{{\mathcal{L}}}
\def\MM{{\mathcal{M}}}
\def\NN{{\mathcal{N}}}
\def\OO{{\mathcal{O}}}

\newcommand{\bC}{{\mathbb C}}

\newcommand{\bH}{{\mathbb H}}

\newcommand{\bN}{{\mathbb N}}

\newcommand{\bR}{{\mathbb R}}
\newcommand{\bS}{{\mathbb S}}
\newcommand{\bT}{{\mathbb T}}

\newcommand{\bZ}{{\mathbb Z}}

\definecolor{cardinal}{rgb}{0.6,0,0}
\definecolor{darkgreen}{rgb}{0,0.4,0}
\definecolor{green}{rgb}{0,0.4,0}
\definecolor{golden}{rgb}{0.92, 0.7, 0}
\definecolor{midnight}{rgb}{0, 0, 0.5}
\definecolor{darkblue}{rgb}{0, 0, 0.7}

\numberwithin{equation}{section}

\mathchardef\mhyphen="2D

  \def\cC{\mathcal {C}}
\def\cD{\mathcal {D}} \def\cE{\mathcal {E}} 
\def\cG{\mathcal {G}}  \def\cI{\mathcal {I}}
\def\cJ{\mathcal {J}} \def\cK{\mathcal {K}} 
\def\cM{\mathcal {M}} \def\cN{\mathcal {N}} \def\cO{\mathcal {O}}
\def\cP{\mathcal {P}}  \def\cR{\mathcal {R}}
\def\cS{\mathcal {S}} \def\cT{\mathcal {T}} 
\def\cV{\mathcal {V}} \def\cW{\mathcal {W}} \def\cX{\mathcal {X}}
\def\cY{\mathcal {Y}} \def\cZ{\mathcal {Z}}

\def\one{{\hbox{\kern+.5mm 1\kern-.8mm l}}}
\def\zero{{\hbox{0\kern-1.5mm 0}}}

\def\id{\textrm{id}}

\def\id{{1 \kern-.28em {\rm l}}}

\def\journal#1&#2(#3){\unskip, \sl #1\ \bf #2 \rm(19#3) }
\def\andjournal#1&#2(#3){\sl #1~\bf #2 \rm (19#3) }

\def\ie{{\it i.e.}}
\def\eg{{\it e.g.}}

\def\etc{{\it etc}}

\def\sst{\scriptscriptstyle}

\def\half{\frac12}
\def\hf{{\textstyle\half}}

\def\One{{1\hskip -3pt {\rm l}}}

\catcode`\@=11
\def\slash#1{\mathord{\mathpalette\c@ncel{#1}}}
\def\II{{\cal I}}

\def\LL{{\cal L}}
\def\MM{{\cal M}}
\def\NN{{\cal N}}
\def\OO{{\cal O}}

\def\underrel#1\over#2{\mathrel{\mathop{\kern\z@#1}\limits_{#2}}}

\catcode`\@=12

\def\exp{{\rm exp}}

\def\ie{{\it i.e.}}
\def\eg{{\it e.g.}}

\def\mbar{{\bar m}}

\def\Qtil{{\widetilde Q}}
\def\LG{{\sst\bf LG}}
\def\jtil{{\tilde \jmath}}

\title{
Asymptotically Free {\texorpdfstring{${\rm AdS}_3/{\rm CFT}_2$}{}}
}

\author{Bruno Balthazar$^{\sfa}$, Amit Giveon$^{\sfb}$, David Kutasov$^{\sfa}$, and Emil J. Martinec$^{\sfa}$}

\affiliation{
\vskip 0.01cm
$^{\sfa}$Kadanoff Center for Theoretical Physics and Enrico Fermi Institute\\ University of Chicago, Chicago IL 60637\\ 
\vskip .5cm
$^{\sfb}$Racah Institute of Physics, The Hebrew University Jerusalem 91904, Israel
}

\abstract{%
We propose a new $AdS_3/CFT_2$ duality, in which the bulk string theory has a target spacetime $AdS_3$ times a squashed three-sphere $\bS^3_\flat$, and the dual ${\it CFT_{\rm 2}}$ is a symmetric product of sigma models on $\bR_\phi\times \bS^3_\flat$, deformed by a $\phi$-dependent $\bZ_2$ twist operator.
The duality maps the asymptotic region of $AdS_3$ to the region $\phi\to\infty$, where the twist interaction in the ${\it CFT_{\rm 2}}$ turns off.  
The $AdS_3$ backgrounds in question have $R_{AdS}<\lstr$, and so lie on the string side of the string/black hole correspondence transition.
As a consequence, the high energy density of states consists of a string gas in $AdS_3$ rather than an ensemble of BTZ black holes.
This property allows us to derive the dual ${\it CFT_{\rm 2}}$ by a systematic analysis of the worldsheet string theory on $AdS_3$.
}

\begin{document}
\hypersetup{pageanchor=false}
\begin{titlepage}
\maketitle
\thispagestyle{empty}
\end{titlepage}
\hypersetup{pageanchor=true}
\pagenumbering{arabic}

\toc
\thispagestyle{empty}

\vskip 1cm
\hrule


\section{Introduction and Summary}
\label{sec:introsummary}

\subsection{Introduction} 
\label{sec:intro}

The first example of the $AdS_3/CFT_2$ correspondence~\rcite{Maldacena:1997re} was obtained by studying type IIB string theory in the near-horizon geometry of a system of $n$ fivebranes wrapped around $\bS^1\times\cM_4$, with $\cM_4=\bT^4$ or $K3$, and $p$ strings wrapped around the $\bS^1$. Both types of branes were taken to be localized on the transverse $\bR^4$. The resulting geometry is 
\eqn[adsthree]{AdS_3\times \bS^3\times \cM_4}
with a non-zero field strength for a $B$ field. In the D1-D5 case, this $B$ field is the $(\!R,\!R)$ one, while for the NS5-F1 system it is the $(\NS,\NS)$ $B$-field. 

In the latter case, which we will focus on in this paper, the worldsheet theory \eqref{adsthree} is exactly solvable. The $AdS_3$ is described in this case by a supersymmetric sigma model with $\sltwo_L\times \sltwo_R$ affine Lie algebra symmetry, while the $\bS^3$ is described by the supersymmetric $SU(2)$ WZW model. The (total) level of the $SU(2)$ current algebra is equal to the number of fivebranes, $n$. The level of $\sltwo$, $k$, is determined by the consistency conditions of string theory, and is also equal to $n$. 

When the number of fivebranes is large, both the $AdS_3$ and $\bS^3$ are weakly curved. In particular, the radius of curvature of the $AdS_3$ is given by 
\eqn[kval]{R_{AdS}=\sqrt{k}\,\lstr ~,
}
\ie\ $R_{AdS}\gg \lstr$ for large $k(=n)$.  Thus, for many purposes, the supergravity approximation to string theory is a good one. On the other hand, when the number of fivebranes is of order one, the radius of curvature of $AdS_3$ (and that of $\bS^3$) is of order $\lstr$, and one has to use the full power of string theory to study the resulting background \eqref{adsthree}.

The construction of~\rcite{Maldacena:1997re} can be generalized to systems of curved NS5-branes, wrapping a cycle in a curved manifold. As we review in section \ref{sec:review}, a large class of such constructions leads to backgrounds of the form\footnote{The background \eqref{genads} involves in general an orbifold of $\bS^1\times \MM$, which we suppress here.}
\eqn[genads]{AdS_3\times \bS^1\times \MM~,}
where $\MM$ is a compact CFT. For backgrounds that are $(2,2)$ superconformal in spacetime, $\MM$ is typically an $N=2$ superconformal theory. The background \eqref{adsthree} is a special case of this construction, with $\MM=\frac{SU(2)}{U(1)}\times \cM_4$.

The level $k$ of the $\sltwo$ current algebra, and thus the radius of curvature of $AdS_3$~\eqref{kval}, is determined by the properties of the compact CFT $\cM$, in particular its central charge. Infinite sets of examples have been constructed, both with $k$ that is arbitrarily large, and with $k$ of order one; see \eg~\rcite{Elitzur:1998mm,Giveon:1999zm}. 
The level $k$ is in general non-integer, though it is generally rational in these constructions. 

It was pointed out in~\rcite{Giveon:2005mi} that models with $k<1$ (\ie\ $R_{AdS}<\lstr$, \kval) are qualitatively different from those with $k>1$ ($R_{AdS}>\lstr$). For $k>1$, the spacetime CFT has a normalizable $\sltwo$ invariant vacuum, and the high energy spectrum is dominated by BTZ black holes~\rcite{Strominger:1997eq}.%
\footnote{These two properties are related by modular invariance of the torus partiton sum and unitarity of the spacetime CFT~\rcite{Cardy:1986ie}.}
On the other hand, for $k<1$, neither the $\sltwo$ invariant vacuum nor the black holes are in the spectrum, and the high energy spectrum is that of perturbative strings in the background \eqref{genads}.

As discussed in~\rcite{Giveon:2005mi}, this phenomenon is related to the string/black hole correspondence of~\rcite{Horowitz:1996nw}, which connects the description of highly excited states in string theory as black holes to the description of highly excited states in perturbative string theory.  Many black holes have the property that the curvature at their horizon decreases with increasing mass, thus making a geometric picture more reliable. Conversely, as the black hole mass decreases, $\lstr$ corrections near the horizon grow, and the geometric picture becomes less reliable.  

When the string frame curvature at the horizon reaches the string scale, the effective description changes to that in terms of a perturbative gas of strings and/or D-branes. This transition happens in the vicinity of a particular mass (known as the correspondence point), and is believed to be smooth. In particular, at the correspondence point, the black hole and perturbative string entropies match, up to numerical coefficients independent of various parameters, such as the string coupling and charges carried by the states in question. 

For black holes in anti de Sitter spacetime the situation is different. The horizon curvature of large black holes is approximately (and for $AdS_3$ exactly) constant, set by the cosmological constant, and so the correspondence transition affects the asymptotic spectrum, in contrast to flat spacetime, where for sufficiently high energies the spectrum is always dominated by black holes. 

It was proposed in~\rcite{Giveon:2005mi} that the analog of the string/black hole correspondence of~\rcite{Horowitz:1996nw} in $AdS_3$ happens as a function of $k$, or the value of the cosmological constant in string units~\eqref{kval} (and for linear dilaton backgrounds, as a function of the slope $Q=\sqrt{2/k}$). It was argued that from the perspective of~\rcite{Horowitz:1996nw} it is natural to have a string/black hole transition at $k\sim 1$.

Despite the similarities, there are a number of important differences between the discussions of~\rcite{Horowitz:1996nw} and \rcite{Giveon:2005mi}. One is that changing $k$ corresponds to changing the theory, since $k$ is determined by the choice of compact CFT $\cM$ in \eqref{genads}.  A second difference is that while the transition of~\rcite{Horowitz:1996nw} is expected to be smooth (though see~\rcite{Chen:2021dsw} for a recent discussion), 
that of~\rcite{Giveon:2005mi} is sharp. Also, while some of the order one coefficients in the correspondence of~\rcite{Horowitz:1996nw} cannot be computed, since they involve non-perturbative $g_s$ effects, in $AdS_3$ one can calculate them since they have to do with classical string physics (\ie\ $\lstr$ effects).

In a separate development, the spectrum of perturbative string states in $AdS_3$ was studied in~\rcite{Maldacena:2000hw}.  It was shown there that it contains, in addition to normalizable states that are analogous to those familiar from higher dimensional $AdS$ spaces, a continuum of delta-function normalizable states that describe strings winding $w$ times about the spatial circle and carrying arbitrary radial momentum.  We will refer to these states as {\it long strings}.\footnote{The authors of~\rcite{Maldacena:2000hw} also constructed normalizable states corresponding to strings winding around the boundary circle. These states will play a role below, but from the perspective of our paper they do not correspond to long strings in our usage of the term, as we will explain.}  

The spectrum of spacetime scaling dimensions of long strings in the background \eqref{genads} is known to fall into a symmetric product structure~\rcite{Argurio:2000tb,Giveon:2005mi} (see~\rcite{Chakraborty:2019mdf} for a recent discussion and further references).  In particular, states with winding one are in one-to-one correspondence with delta-function normalizable states in the CFT  
\eqn[genadseff]{\bR_\phi\times \bS^1\times \cM~,}
where $\bR_\phi$ is a linear dilaton CFT with slope 
\eqn[lslope]{Q_\ell=(1-k)\sqrt{\frac{2}{k}}~,}
normalized such that the central charge is $c_\phi=1+3Q_\ell^2$.
The theory~\eqref{genadseff} was studied in~\rcite{Seiberg:1999xz}, where it was shown to be the effective theory on a long string. States with larger winding have scaling dimensions that coincide with those of twisted sectors of a symmetric product CFT, with building block \eqref{genadseff}.

The implications of these observations for the structure of the spacetime CFT are different for the two cases $k<1$ and $k>1$. For $k>1$ (\eg\ the original backgrounds~\eqref{adsthree} of~\rcite{Maldacena:1997re}), $Q_\ell$ in~\eqref{lslope} is negative. Thus, as $\phi\to\infty$, \ie\ as the long string approaches the boundary of $AdS_3$, the string coupling for the long strings $g_s\sim \exp(-\half Q_\ell\phi)$ diverges, and the description in terms of a symmetric product of \eqref{genadseff} fails. It also fails as $\phi\to-\infty$, since this is the region in the middle of $AdS_3$, where the string is not long. Thus, for $k>1$ one expects the description \eqref{genadseff} to only hold in a finite region of the radial direction of $AdS_3$.

For $k<1$, on the other hand, $Q_\ell$ \eqref{lslope} is positive, and the region near the boundary of $AdS_3$ is weakly coupled, both on the worldsheet and in spacetime. Thus, in that case the description of the spacetime CFT as a theory that asymptotes at large $\phi$ to the symmetric product of \eqref{genadseff} makes sense. In this paper we will propose that it indeed is such a theory. 

The notion of a CFT that asymptotes to a linear dilaton one may be unfamiliar to some readers, so we next illustrate it in a hopefully more familiar context. Suppose we are given a CFT which describes two scalar fields that live on $\bR_\phi\times \bS^1$ and two fermions, where $\bR_\phi$ has a linear dilaton of slope $Q$. If the theory is precisely that, all correlation functions on a given genus Riemann surface are singular, and a given correlation function becomes more and more singular as the genus increases.  The origin of these divergences is the region in $\bR_\phi$ in which $g_s\sim \exp(-\half Q\phi)\to\infty$. For positive dilaton slope $Q$, this is the region $\phi\to-\infty$, and vice versa. 

\begin{figure}[ht]
\centering
 \begin{subfigure}[b]{0.4\textwidth}
  \hskip 0cm
    \includegraphics[width=\textwidth]{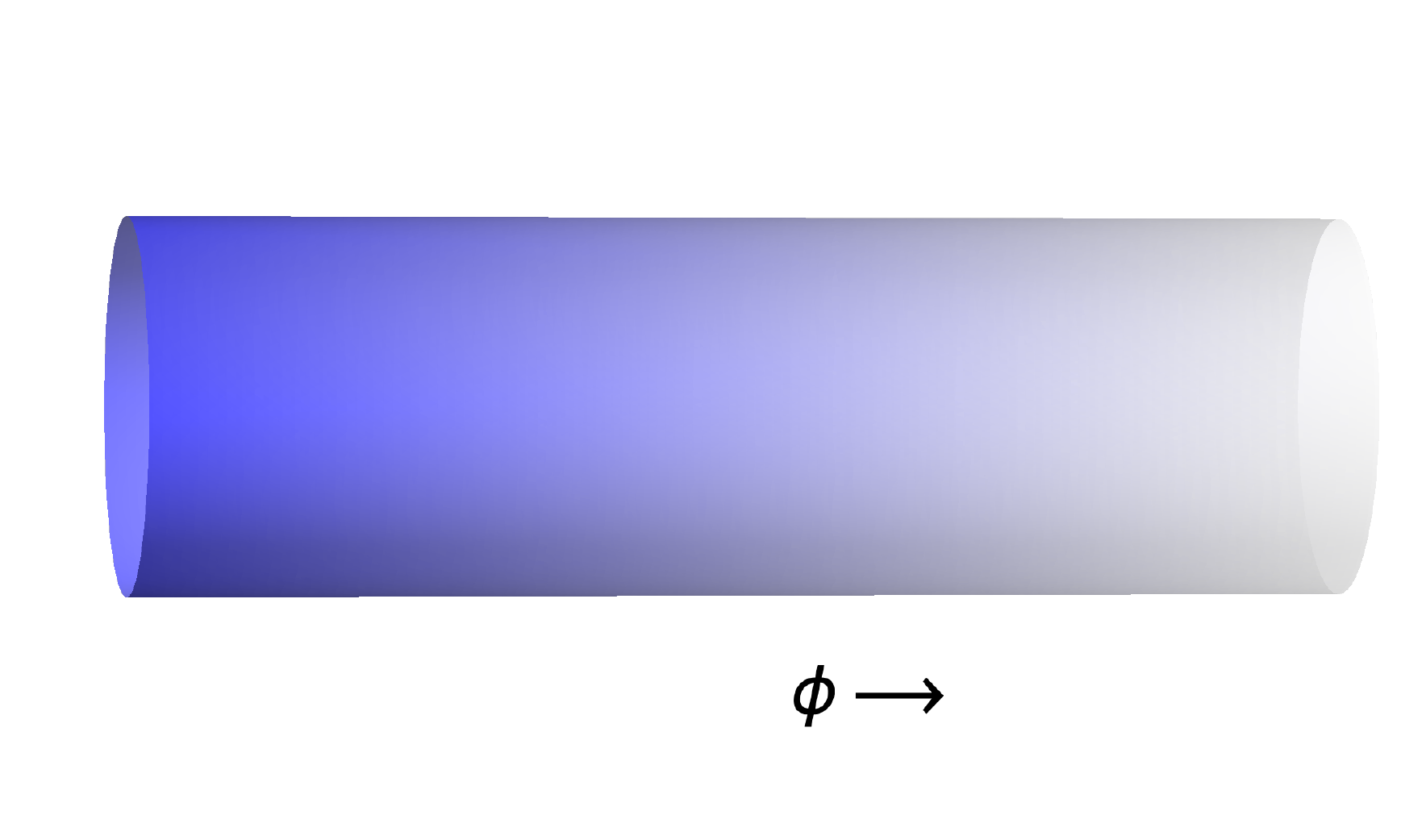}
    \caption{ }
    \label{fig:WindingTachyon}
  \end{subfigure}
\qquad\qquad
  \begin{subfigure}[b]{0.4\textwidth}
      \hskip 0cm
    \includegraphics[width=\textwidth]{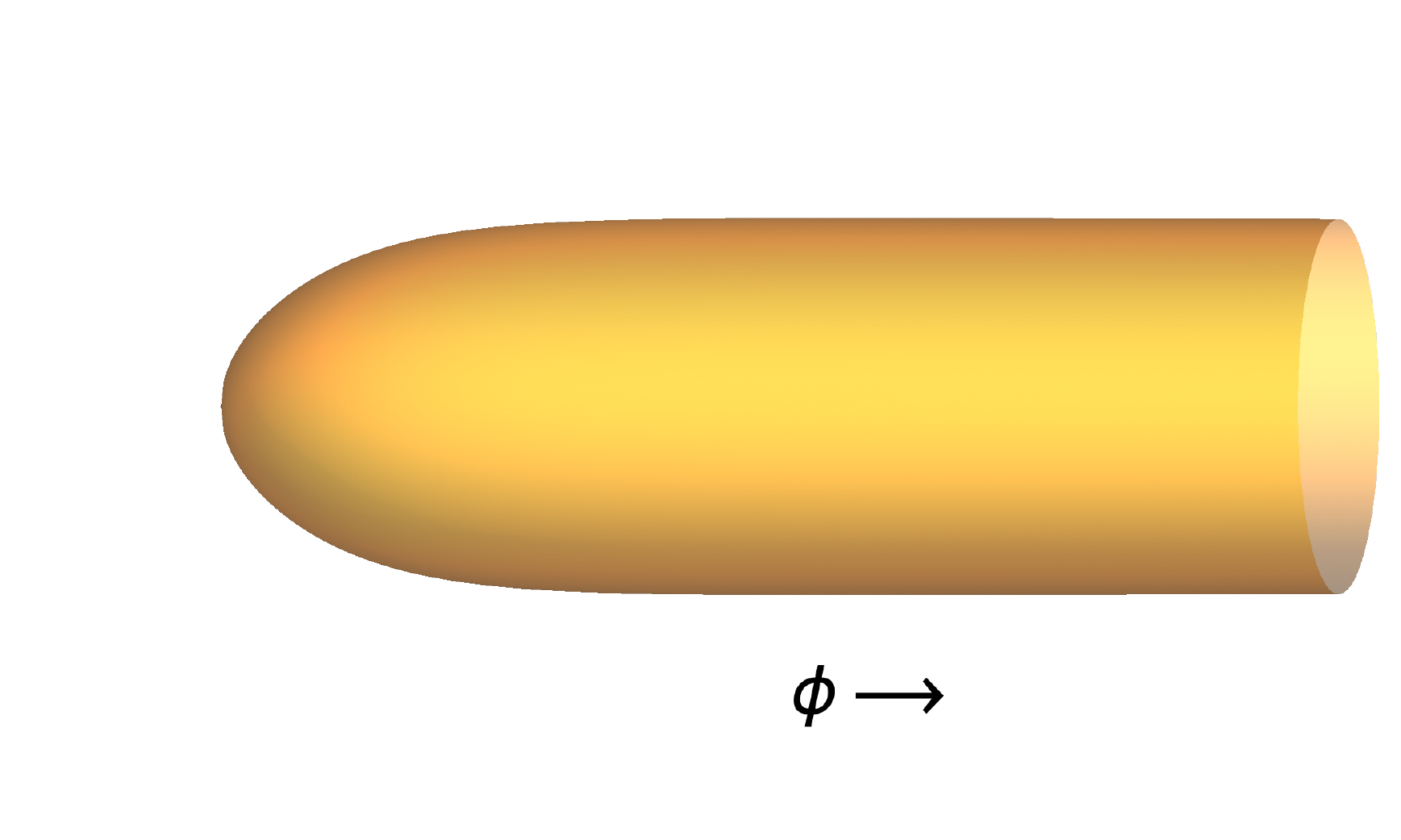}
    \caption{ }
    \label{fig:CigarPlot}
  \end{subfigure}
\caption{\it A sigma-model that approaches $\mathbb{R}_\phi\times \bS^1$ at large positive $\phi$ can be regularized (a) by adding a bosonic or $N=1$ supersymmetric Liouville potential that grows as $\phi\to-\infty$ (the blue shading indicates the height of the potential), or (b) by viewing $\mathbb{R}_\phi\times \bS^1$ as the large $\phi$ region of a bosonic or supersymmetric $\frac{\sltwo}{U(1)}$ (S)CFT. In both cases, the field $\phi$ is prevented from accessing the strongly coupled region at $\phi=-\infty$ by a ``wall'', and the resulting (S)CFT is non-singular.}
\end{figure}

Nevertheless, there are a number of well behaved CFT's that look (for positive $Q$) like the above in the weak coupling region $\phi\to\infty$. For example, one may replace the singular $\bR_\phi$ factor by Liouville theory, where a potential prevents the field $\phi$ from exploring the region $\phi\to-\infty$ (see \eg~\rcite{Seiberg:1992bj} for a discussion).  Or, one can combine the $\bR_\phi$ with one of the two fermions and form an $N=1$ Liouville background.  These two backgrounds are qualitatively depicted by the geometry in figure~\ref{fig:WindingTachyon}. 

Two other backgrounds with similar properties are obtained by combining $\bR_\phi$ with $\bS^1$ and forming the bosonic or superconformal $\frac{\sltwo}{U(1)}$ CFT. These can be viewed as sigma models on a cigar~\rcite{Elitzur:1990ubs,Mandal:1991tz,Witten:1991yr,Dijkgraaf:1991ba}, depicted in figure~\ref{fig:CigarPlot}, with a non-zero condensate of the Sine-Liouville or $N=2$ Liouville operator~\rcite{FZZref,Giveon:1999px,Giveon:1999tq,Kazakov:2000pm}.

\vspace{1mm}

All these theories look at large positive $\phi$ like a sigma-model on $\bR_\phi\times \bS^1$, with a ``wall'' at finite $\phi$, preventing this field from exploring the strong coupling region $\phi\to-\infty$. The wall is different in each case,\footnote{In particular, while in the backgrounds of figure 1a winding around the circle is conserved, in those of figure 1b it isn't.} but for some purposes its precise features are unimportant. In particular, all four models exhibit a continuum of delta-function normalizable states above a gap $\frac18 Q^2$ set by the asymptotic slope of the dilaton. 
The associated delta-function normalizable operators behave as $\phi\to\infty$ like 
\eqn[vvpp]{V_\beta\sim e^{\beta\phi}~,}
with $\beta=-\frac{Q}{2}+ip$.  The momentum $p$ can be positive or negative, describing outgoing and incoming waves. The presence of the wall implies that the two are not independent~-- if one sends in a wave with some particular $p$, a wave whose precise properties depend on the structure of the wall will come back. However, the fact that an operator with asymptotic behavior \eqref{vvpp} exists for all $p<0$, and has a dimension that can be read off at infinity, is insensitive to the precise structure of the wall but only to its presence, and in particular is the same for all four models described above. 

Another class of operators for which the above comments apply is non-normalizable operators, which behave like \eqref{vvpp} with $\beta$ real and  $\beta>-\frac Q2$. Such operators play an important role in Liouville theory~\rcite{Seiberg:1990eb,Ginsparg:1993is}, 
as well as backgrounds that involve the (supersymmetric) $\frac{\sltwo}{U(1)}$ CFT's~\rcite{Aharony:1998ub,Aharony:1999ks,Kutasov:2001uf}. 
One can think of them as obtained by analytic continuation in $\beta$ from the delta-function normalizable ones.

In contrast to the discussion above, the spectrum of {\it normalizable} states in asymptotically linear dilaton theories depends strongly on the structure of the wall. For example, if one regulates the strong coupling singularity above by a Liouville or $N=1$ Liouville wall, the resulting theory does not have any normalizable states (essentially because quantum mechanics in an exponential potential does not have any bound states). For the other two choices of wall, which give rise to the bosonic and supersymmetric $\frac{\sltwo}{U(1)}$ models, respectively, the CFT does contain normalizable states, corresponding to principal discrete series states in the underlying $\sltwo$ WZW model. The spectrum of these states is different in the two cases. 

The general conclusion from the above discussion is that non-singular CFT's that asymptote to linear dilaton models at large $\phi$ have a sector that can be analyzed without specifying the precise structure of the wall, namely the spectrum of delta-function normalizable and non-normalizable operators, that behave like $\exp(\beta\phi)$ with $\beta=-\frac Q2+ip$, and real $\beta>-\frac Q2$, respectively. Other observables, such as the spectrum of normalizable operators as well as correlation functions of non-normalizable and delta-function normalizable ones, do depend on the precise structure of the wall. 

Note that in the conformal field theories described above, the two asymptotic regions $\phi\to\pm\infty$ are treated asymmetrically. Operators are specified in the ``weak coupling'' asymptotic region of $\phi$-space, and the wall prevents the theory from exploring the other, ``strong coupling'' one. This is true despite the fact that we are dealing with CFT's rather than string theories, so the notion of coupling need not make sense, since there is no a priori reason for considering different genera, or summing over them.

\subsection{Summary} 
\label{sec:summary}

The comments in the previous subsection motivate us to study $AdS_3/CFT_2$ in the regime $k^2=R_{AdS}/\lstr<1$.  The absence of BTZ black holes in the spectrum suggests the possibility that the theory may be under more quantitative control in this case, since one does not have to resolve issues associated to black hole microstate structure. This expectation is further reinforced by the behavior of the linear dilaton $Q_\ell$ \lslope, which suggests that for $k<1$, fundamental strings are the right degrees of freedom for describing the dynamics at all energies~\rcite{Giveon:2005mi}.

In this paper we show that this expectation is realized. In particular, we show that type II string theory on 
\eqn[ourads]{AdS_3\times\sqsphere~,} 
where $\sqsphere$ is a squashed three-sphere, described by a certain exactly solvable worldsheet CFT, is dual to a spacetime CFT which has the structure 
\eqn[ourcft]{\left(\bR_\phi\times \sqsphere\right)^p/S_p}
at large positive $\phi$. Here $\phi$ is a free field with a linear dilaton $Q_\ell$ \lslope\ with $k=\frac{n}{n+1}$, $n=2,3,\dots$, and $\phi\to\infty$ is the weak coupling limit in the sense discussed in the previous subsection.  

The full spacetime CFT is described by the symmetric product \ourcft\ deformed by an operator in the $\bZ_2$ twisted sector of the symmetric product. The deforming operator, which we construct in section \ref{sec:wall}, decays exponentially at large positive $\phi$. Therefore, in that region its effects on the physics can be neglected, and the symmetric product structure \ourcft\ provides a good description. 

\begin{figure}[ht]
\centering
\includegraphics[width=.6\textwidth]{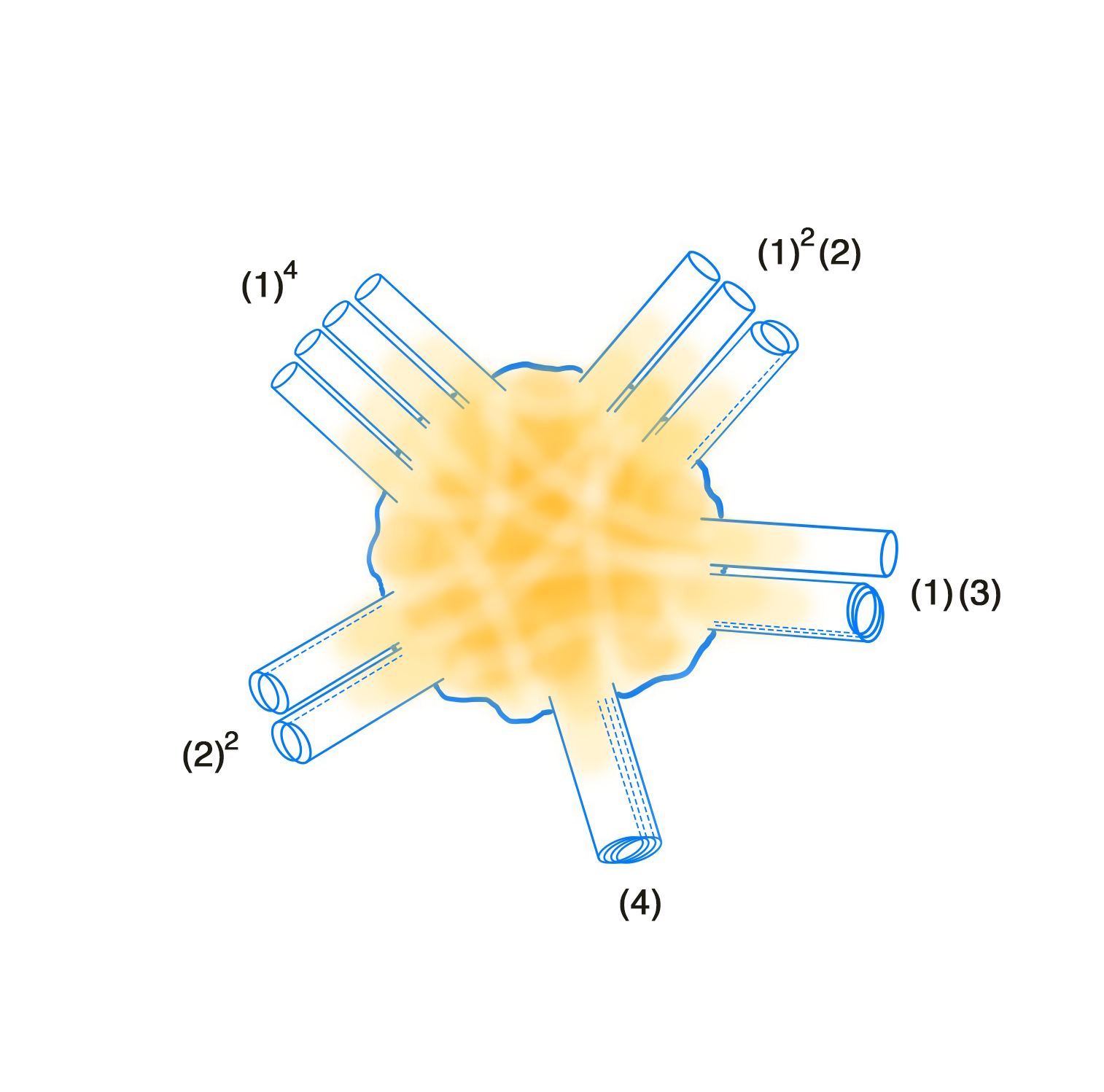}
\vspace{-0.5cm}
\caption{\it Cartoon of the deformed symmetric orbifold for $p=4$.  The coordinate $\phi$ increases radially outward from the center of the figure.  There are different asymptotic limits $\phi\to\infty$ corresponding to the different ways of partitioning long string winding, or equivalently according to the conjugacy classes of the symmetric group $S_4$.  At finite $\phi$, one loses the symmetric product structure, where the ``wall'' interconnects the different winding sectors.  More precisely, such effects occur at any $\phi$, but their effects are decaying exponentially as $\phi$ increases.}
\label{fig:bagpipes}
\end{figure}

Conversely, this operator grows exponentially as $\phi$ decreases. Hence, it has a large effect on the physics at finite $\phi$. In particular, the deformation introduces interactions among the constituent factors of the symmetric product \ourcft, and provides a wall that keeps the field $\phi$ away from the strong coupling region $\phi\to-\infty$.
A cartoon of the resulting geometry is sketched in figure~\ref{fig:bagpipes}. 

%
\vspace{1mm}
%

As in other examples of asymptotically linear dilaton backgrounds mentioned above, the coefficient multiplying the deforming operator in the spacetime Lagrangian is not a parameter in the theory -- it can be rescaled by a shift of $\phi$ (this is known as KPZ scaling in Liouville theory~\rcite{David:1988hj,Distler:1988jt}). The deformation can be thought of as introducing a scale $\phi^*$, such that physical observables that are dominated by the region $\phi\gg\phi^*$ are governed by the symmetric product \ourcft, while those that are centered in the region $\phi\lesssim\phi^*$ are sensitive to the deformation. As mentioned above, an example of the former is the scaling dimensions of delta-function normalizable and non-normalizable operators; examples of the latter are the spectrum of normalizable operators and correlation functions in the deformed symmetric product.  

The scale $\phi^*$ itself is arbitrary, due to the aforementioned shift symmetry $\phi\to\phi+{\rm const}$. This is analogous to the phenomenon of dimensional transmutation in asymptotically free quantum field theory, where a marginally relevant coupling is replaced in the quantum theory by an energy scale, above which the theory is free, and below which it is interacting. In our case, the analog of the energy scale is position in $\phi$, the analog of the high energy regime is $\phi\to\infty$, and the analog of the free theory at high energy is the symmetric product \ourcft, that describes the dynamics at large $\phi$. Since the symmetric product CFT is known to describe the kinematics of free strings~\rcite{Dijkgraaf:1997vv,Dijkgraaf:1997ku}, we may refer to the CFT \ourcft, and thus also to the dual AdS background \ourads, as {\it asymptotically free.}

We derive the duality between \ourads\ and the deformed version of \ourcft\ by a systematic analysis of the worldsheet theory.  In the bulk description \ourads, an important role is played by the sectors with non-zero winding, studied in~\rcite{Maldacena:2000hw}. Thus, after introducing type 0 and type II string theory on the background \ourads, in sections \ref{sec:type0short} and \ref{sec:typeIIshort}, respectively, we focus in sections \ref{sec:type0long}, \ref{sec:typeIIlong} on the spectrum of delta-function normalizable and non-normalizable operators in these sectors of the respective theories. These operators are insensitive to the deforming operator in the boundary theory, and thus can be mapped directly to the symmetric product \ourcft.  

We show that for winding one in $AdS_3$, the spectrum of delta-function normalizable and non-normalizable operators is in one-to-one correspondence with the spectrum of such operators in the building block of the symmetric product \ourcft. We match both the scaling dimensions of these operators in the spacetime CFT, and their wavefunctions in the radial $(\phi)$ direction.  

We also match single string operators with winding larger than one in $AdS_3$ with operators in the $\bZ_w$ twisted sectors of the symmetric product orbifold \ourcft. This involves in particular matching the radial coordinate of a string with winding $w$ in $AdS_3$, and the center of mass location of an operator in the $\bZ_w$ twisted sector of the symmetric orbifold.

One of the interesting aspects of our construction is the appearance, in the $AdS_3$ analysis, of a large number of dimension $(r,0)$ non-normalizable operators. In the symmetric product description \ourcft, they correspond to holomorphic operators in the building block of the symmetric product, such as $\partial\phi$ and its superpartner $\psi_\phi$, the $U(1)_L$ current in $\sqsphere$, the supercurrent and stress tensor of $\bS^3_\flat$, as well as various products of these operators. 

From the perspective of the spacetime theory, we expect such operators to be holomorphic in the region of $\phi$ space that is well described by the symmetric product \ourcft. We also expect their holomorphy to be violated in the region in $\phi$ where the deformation by the $\bZ_2$ twisted operator becomes significant. 

Interestingly, the $AdS_3$ analysis precisely reproduces these expectations.  Through that analysis, we find that the above operators are holomorphic at the level of non-normalizable operators, but their holomorphy is violated by a contribution from a normalizable operator in the background \ourads. Thus, the structure in the bulk theory is precisely what one would expect from the perspective of the deformed symmetric product structure of the spacetime CFT. In fact, by using this construction we can determine the deforming operator in the spacetime CFT from the $AdS$ analysis, which is how we arrive at its form as a particular normalizable operator in the $\bZ_2$ twisted sector of the orbifold \ourcft. 

The duality described in this paper has some uncommon features in the zoo of known holographic dualities. One is that, in the sense outlined above and described in detail in subsequent sections, the form of the spacetime CFT is {\it derived} from an analysis of the worldsheet one. The second is that in this case, both sides of the duality are under good control. 

In the bulk description this is due to the fact that string theory on $AdS_3$ with NS B-field can be studied precisely (at weak string coupling), even in regions where $R_{AdS}$ is of the order of the string scale, which is the case for the backgrounds \ourads. 
Furthermore, the absence of BTZ black holes suggests that the physics is visible in a weakly coupled regime amenable to perturbative analysis. 

In the boundary theory, it is due to the fact that while the full (deformed symmetric product \ourcft) theory is not weakly coupled, it approaches a weakly coupled theory in a region of field space (large positive $\phi$), and thus is expected to be under similar analytic control to other asymptotically linear dilaton theories, such as the ones mentioned in the previous subsection.


\section{Review of prior results}
\label{sec:review}

In this section we review some results from previous work that will play a role in our discussion below. We start by describing the class of backgrounds that we will study, and the way they appear in string theory. 


\subsection{Fivebranes, strings and Calabi-Yau singularities} 
\label{sec:GKP}

As discussed in section~\ref{sec:introsummary}, the $AdS_3$ vacua \eqref{adsthree} (with NS $B$-field) are obtained by adding to a system of $n\ge 2$ NS5-branes a large number, $p$, of fundamental strings. Before adding the strings, the decoupling limit $\gstr\to 0$ of the fivebranes has the geometry~\rcite{Callan:1991at}
\eqn[flatns]{\bR^{5,1}\times \bR_\phi\times \bS^3~,}
where $\bR^{5,1}$ is the $5+1$ dimensional submanifold that the fivebranes are stretched along, and $\bR_\phi\times \bS^3$ describes the space transverse to the fivebranes in spherical coordinates. The radial direction $\phi$ in that transverse space is described by a linear dilaton CFT, $\bR_\phi$, with slope
\eqn[Qcrit]{
Q=\sqrt{\frac{2}{n}} ~, 
}
and the $\bS^3$ is described by a level $n$ supersymmetric $SU(2)$ WZW model. As a check, the total worldsheet central charge is given by 
\eqn[totalcc]{c=6+\left(1+3Q^2\right)+\Big(3-\frac{6}{n}\Big)+10\cdot\half=15~,}
where the first three terms come from the bosonic fields on $\bR^{5,1}$, $\bR_\phi$ and $\bS^3$, respectively, while the last term is due to the ten free NSR fermions.
Note also that in \eqref{flatns} we took the fivebranes to wrap $\bR^5$, while in section~\ref{sec:intro} we took them to wrap a compact space, $\bS^1\times \cM_4$.

An alternative point of view on the background \eqref{flatns} starts with string propagation on $\bR^{5,1}\times {\it ALE}_n$, where ${\it ALE}_n$ is an ADE surface singularity.
For instance, the $A_{n-1}$ singularity\footnote{The discussion here and below can be generalized to $D$ and $E$ series singularities. In the geometric description, this involves replacing \eqref{aaannn} by the corresponding $D$ or $E$ series expression. In terms of \flatns\ it corresponds to different choices of modular invariant for the WZW model on $\bS^3$, which is known to have an ADE classification~\rcite{Cappelli:1987xt}.}  
has an algebraic description as the hypersurface
\eqn[aaannn]{z_1^n+z_2^2+z_3^2=0} 
in $\bC^3$, which describes a non-compact cone when $n\ge 2$.  The sigma-model on the cone \eqref{aaannn} has moduli corresponding to the flux of the NS $B$-field through the vanishing cycles of the cone. The description \eqref{flatns} corresponds to the case where these fluxes are set to zero
(see for example~\rcite{Aspinwall:1995zi,Ooguri:1995wj,Kutasov:1995te,Harvey:2001wm}).

The background \eqref{flatns} has the property that the string coupling goes to zero as $\phi\to\infty$. One can think of this region as the boundary in a holographic theory, known as Little String Theory (LST)~\rcite{Aharony:1998ub,Aharony:1999ks,Kutasov:2001uf}. As $\phi\to-\infty$ the string coupling diverges, which means that the description \eqref{flatns} is not useful there. 

There are a number of ways to understand and mitigate this singularity, one of which is to compactify $\bR^5$ to $\bS^1\times \bT^4$, and add to the geometry \eqref{flatns} $p$ fundamental strings wrapped around $\bS^1$ and localized inside the fivebranes at large negative $\phi$. The presence of the strings stops the string coupling from growing beyond a value of order $g_s^2\sim n V_{\bT^4}/p$. In the limit $p\to\infty$ with $n$, $V_{\bT^4}$ held fixed, the resulting string background \eqref{adsthree} is weakly coupled. From the perspective of the high energy theory, this geometry describes a state in LST that contains $p$ strings wrapped around a circle inside the fivebranes. From the low energy perspective it describes a two dimensional CFT.  

The crossover between the regime where the geometry is described by a linear dilaton spacetime $\bR_t\times \bS^1\times\bR_\phi\times \bS^3\times\cM_4$ and the one where it is described by \eqref{adsthree} is set by the ratio of the radius of the circle wrapped by the strings and the string length, $R_{\bS^1}/\lstr$. The $AdS_3$ limit sends this ratio to infinity. For finite values of this ratio, one can think of the resulting background as describing an irrelevant deformation of the corresponding (spacetime) CFT, that will be discussed in section  \ref{sec:deflin}.

The above construction can be generalized to describe singularities of Calabi-Yau (CY) manifolds and fivebranes wrapped around non-trivial cycles. A more detailed discussion of this construction appears in~\rcite{Giveon:1999zm}; here, we will give a few examples of classes of backgrounds that include the one we will focus on in this paper. 

A natural generalization replaces the ADE surface singularity~\eqref{aaannn} by a noncompact CY threefold singularity described locally by an equation in $\bC^4$, 
\eqn[bbbnnn]{z_1^n+z_2^2+z_3^2+z_4^2=0} 
with $n=2,3,\cdots$. As discussed in~\rcite{Giveon:1999zm}, in the decoupling limit of the conical singularity \bbbnnn, string theory on $\bR^{3,1}\times{\rm CY}$ is described by 
\eqn[ardoug]{\bR^{3,1}\times \bR_\phi\times \big( \bS^1\times LG_n \big)/\bZ_n~,}
where $LG_n$ is the $N=2$ minimal model (the supersymmetric coset model $\frac{\sutwo_n}{\uone}$), which can be described in terms of a chiral worldsheet superfield $Z_1$ with superpotential 
\eqn[lgnn]{W=Z_1^n~,}
and the $\bS^1$ is labeled by a coordinate $Y$.  We will discuss the radius of $Y$ in later sections. As mentioned in section 1 (see footnote 1), there is a $\bZ_n$ orbifold acting on $\bS^1\times LG_n$, which can be read off the GSO projection~\rcite{Giveon:1999zm}.

The slope of the linear dilaton $Q$ can be computed by generalizing \eqref{totalcc} to this case. It is convenient to parametrize it in terms of a ``level'' $k$, related to $Q$ by
\eqn[qqkk]{Q=\sqrt\frac{2}{k}~.}
One finds
\eqn[knad]{\frac{1}{k}-\frac{1}{n}=\half
~~\Longrightarrow~~
k=\frac{2n}{n+2} ~.}
For $2\le n<\infty$, $k$ takes values in the finite range $1\le k<2$.  

\bigskip
\noindent
Comments:
\begin{enumerate}[1)]
\item One can heuristically think of the background \ardoug\ as a description of the cone \eqref{bbbnnn} as follows. The polynomial \eqref{bbbnnn} is quasi-homogeneous, \ie\ it goes to a multiple of itself under the transformation $z_i\to\lambda^{\frac{1}{r_i}}z_i$, with $\lambda$ an arbitrary complex number, $r_1=n$ and $r_i=2$ for $i=2,3,4$. The infinite cylinder labeled by $(\phi,Y)$ in \ardoug\ can be thought of as describing $\ln\lambda$, while the Landau-Ginsburg model \lgnn\ describes the surface \bbbnnn, viewed as a surface in weighted projective space.  
\item Just as in the ALE case, the above construction can be alternatively thought of in terms of fivebranes, in this case wrapping the two dimensional surface 
\eqn[fivesurfad]{z_1^n+z_2^2=0~,} 
as well as the $\bR^3$ in \eqref{ardoug}.
\item The above discussion is a special case of a much more general construction, that involves string propagation in the vicinity of conical singularities of non-compact CY manifolds, described by generalizing \eqref{bbbnnn} to $F(z_1,z_2,z_3,z_4)=0$, with $F$ a quasi-homogeneous polynomial~\rcite{Gukov:1999ya,Giveon:1999zm}. All such models involve the analog of the $\lambda$ degree of freedom of point 1), and thus give rise to worldsheet theories that contain a factor $\bR_\phi\times \bS^1$. This factor is $N=2$ supersymmetric on the worldsheet, which is necessary for spacetime supersymmetry~\rcite{Banks:1987cy,Giveon:1999zm,Giveon:1999jg,Giveon:2003ku}.   
\end{enumerate}

\noindent
So far we have been discussing the background near a $CY_3$ singularity, or equivalently the near-horizon geometry of a curved fivebrane, which describes a $3+1$ dimensional vacuum of LST. We can now repeat the $5+1$ dimensional construction above, compactify the $\bR^3$ in \ardoug\ to $\bS^1\times \bT^2$, and add to the background $p$ strings wrapped around the $\bS^1$ at large negative $\phi$. This leads in the infrared to the background
\eqn[ardougads]{AdS_3\times \bT^2\times \big(\bS^1\times LG_n\big)/\bZ_n~,}
with the (total) level of the $\sltwo$ current algebra given precisely by \knad. 

The resulting $AdS_3$ backgrounds are quite stringy, since the radius of curvature of $AdS_3$, \kval, is of order $\lstr$, but if our goal is to study $AdS_3$ backgrounds with $k<1$, we need to go one step further. Thus, we replace the noncompact CY threefold singularity by a noncompact fourfold singularity
\eqn[cccnnn]{z_1^n+z_2^2+z_3^2+z_4^2+z_5^2=0
~~,~~~~
n=2,3,\dots } 
The surface wrapped by the fivebrane analogous to~\eqref{fivesurfad} is given in this case by 
\eqn[fivesurftwod]{z_1^n+z_2^2+z_3^2=0 ~.}
String theory in the decoupling limit is described by  
\eqn[twodlst]{R^{1,1}\times R_\phi\times \big(\bS^1\times LG_n\big)/\bZ_n ~.}
 The slope of the linear dilaton is given by \qqkk, but now 
\eqn[kntwod]{\frac{1}{k}-\frac{1}{n}=1 ~~\Longrightarrow~~ k=\frac{n}{n+1}<1 ~.}
Compactifying the spatial direction in $\bR^{1,1}$ in \twodlst\ on a circle, and adding the strings, leads to the background  
\eqn[klessone]{AdS_3\times \big(\bS^1\times LG_n\big)/\bZ_n ~,}
and now the level $k$ \kntwod\ is smaller than one. These are the models that we will study in this paper. Note that they are $(2,2)$ superconformal in spacetime. Indeed, type II string theory on a CY fourfold is $(2,2)$ supersymmetric. Adding the strings and going to the infrared enhances the symmetry to $(2,2)$ superconformal. 
Again, the addition of the $p$ strings stops the running of the string coupling at a value $g_s^2\sim 1/p$.

The orbifold $(\bS^1\times LG_n)/\bZ_n$ involved in~\eqref{ardoug}, \eqref{twodlst} describes a ``squashed'' $\bS^3$, that we will denote by $\sqsphere$.  When the radius of $\bS^1$ is $\sqrt{n}$ times the self-dual radius, one has the coset decomposition of the $\sutwo$ WZW model 
\eqn[sutwocoset]{
\sutwo_n = \left[\Big(\frac{\sutwo_n}{\uone}\Big)\times \uone\right]/\bZ_n
}
(so we could have described~\eqref{flatns} in the same manner as~\eqref{ardoug}, \eqref{twodlst}).  We can then dial the radius of the $\bS^1$ on the r.h.s. to any desired value, which on the l.h.s. corresponds to a $J^3\bar J^3$ marginal deformation which squashes the $\bS^3$ geometry~\rcite{Hassan:1992gi,Giveon:1993ph,Brennan:2020bju}.  We will make use of this equivalence of CFT's in what follows. In particular, we will see that in the non-critical models~\eqref{ardoug}, \eqref{twodlst}, the value of the $\bS^1$ radius that gives a spacetime supersymmetric theory differs from the one that figures in the decomposition of $\sutwo_n$ \eqref{sutwocoset}.


\subsection{\texorpdfstring{$SL(2,\bR)$}{} and \texorpdfstring{$SU(2)$}{} current algebra
\label{sec:currentalg}}

The structure of the models we will discuss is strongly constrained by their underlying worldsheet $\sltwo\times SU(2)$ supersymmetric affine symmetry. In this subsection we review some of their properties that will be used in later sections, and in the process establish some notations and conventions.

\subsubsection{\texorpdfstring{$SU(2)$}{}}

The $N=1$ superconformal $\sutwo_n$ WZW model consists of a bosonic $\sutwo_{n-2}$ WZW model, with currents $j^a_\su(z)$, and three free fermions $\psi^a_\su$, $a=3,+,-$ from which one can construct a $SU(2)_2$ current algebra. The total central charge is given by
\eqn[csu2]{c_{SU(2)}=c^{\rm bos}+c^{\rm ferm}=\frac{3(n-2)}{n}+\frac{3}{2}~.}
The bosonic currents and free fermions satisfy the OPE's
\eqn[bbhh]{
j^a_\su (z)\, j^b_\su(0) \sim \frac{n-2}{2 z^2}\,\delta^{ab}+i\epsilon^{ab}_{~\;c}\,\frac{j^c_\su(0)}{z}
~~,~~~~
\psi^a_\su(z)\psi^b_\su(0)\sim\frac{\delta^{ab}}{z}~.
}
The total $\sutwo$ currents are 
\eqn[bbii]{
J^a_\su=j^a_\su-\frac{i}{2}\epsilon^{a}_{~bc}\psi^b_\su\psi^c_\su ~.
}
This theory has $N=1$ superconformal symmetry with supercurrent
\eqna[Gsu]{G_\su&=\sqrt{\frac{2}{n}}\left(\psi^a_\su \, j^a_\su -i\psi^1_\su\psi^2_\su\psi^3_\su\right) \cr &=\sqrt{\frac{1}{n}}\left(\psi^+_\su j^-_\su+\psi^-_\su j^+_\su\right)+\sqrt{\frac{2}{n}}J^3_\su\psi^3_\su ~,}
where 
\eqna[jpsipm]{j^\pm_\su=j^1_\su\pm i j^2_\su,~~~~\psi^\pm_\su=\frac{\psi^1_\su\pm i\psi^2_\su}{\sqrt{2}}~.}
The fermions $\psi^a_\su$ and total currents $J^a_\su$ form $N=1$ supermultiplets of dimension $(\half,0)$ under~\eqref{Gsu}.

The current algebra primaries of the bosonic $\sutwo_{n-2}$ are given by the operators $v_{j';m',\mbar'}$, which have conformal weight
\eqn[hvjm]{h\left[v_{j';m',\mbar'}\right]=\frac{j'(j'+1)}{n} ~,}
and satisfy the following OPE's with the bosonic $SU(2)$ currents (in a particular normalization of the $v$'s),
\eqna[su2reps]{&j^3_\su(z) v_{j';m',\mbar'}(0)\sim \frac{m}{z} \, v_{j';m',\mbar'}(0)~,\cr
&j^\pm_\su (z)\, v_{j';m',\mbar'}(0)\sim \frac{\sqrt{j'(j'+1)-m'(m'\pm 1)}}{z} \, v_{j';m'\pm1}(0)~.}
Unitary representations lie in the range $j'=0,\half, 1,\cdots, \frac n2-1$; $m',\bar m'=-j', -j'\tight+1,\cdots, j'$.

The bosonic $\sutwo_{n-2}$ WZW theory admits a spectral flow transformation under which the modes of the currents transform as\footnote{In some of the equations below we will drop the subscript ``$\su$'' to avoid clutter.}
\eqna[sutwospecflow]{
j^\pm_n \to j^\pm_{n\pm w'}
~&,~~~~
j^3_n\to j^3_n + \frac{\ntil}2w' \, \delta_{n,0}~,
\\
\bar j^\pm_n \to \bar j^\pm_{n\pm \bar w'}
~&,~~~~
\bar j^3_n\to \bar j^3_n + \frac{\ntil}2\bar w' \, \delta_{n,0}~,
}
with $w',\bar w'\in\bZ$ obeying the restriction $w'-\bar w'\in 2\bZ$. Under spectral flow,
the operator $v_{j';m',\mbar'}$ flows to an operator 
$v^{(w',\bar w')}_{j';m',\mbar'}$ 
which has conformal weight and $j^3$ charge
\eqn[sutwoqnos]{h\big[v^{(w',\bar w')}_{j';m',\mbar'}\big]= \frac{j'(j'+1)}{n} + m' w' + \frac{\ntil}{4}(w')^2
~~,~~~~
j_0^3\big[v^{(w',\bar w')}_{j';m',\mbar'}\big]=m'+\frac{\ntil}{2}w'~, }
and similarly for the right-moving spectral flow.  The flow is an automorphism of the affine Lie algebra, that takes the highest weight state to current algebra descendants.%
\footnote{The operator $v^{(w',\bar w')}_{j';m',\mbar'}$ is a descendant in the representation with spin $j'$ for $w'\in2\bZ$, and in the representation with spin $n/2-1-j'$ for $j'\in 2\bZ+1$.}
One can understand it via the decomposition of $v_{j';m',\mbar'}^{(w',\bar w')}$ into its $\uone$ and $\frac\sutwo\uone$ components 
\eqna[bospf]{
v_{j';m',\mbar'}^{(w',\bar w')} = \lambda_{j';m',\mbar'} \,
\exp\left[i\sqrt{\frac{2}{n-2}}\left(\Big(m'+\frac{n-2}{2}\,w'\Big)\,y_\su + 
\Big(\mbar'+\frac{n-2}{2}\,\bar w'\Big)\,\bar y_\su\right) \right]~,
}
where $y_\su,\bar y_\su$ bosonize the $\uone$ currents
\eqn[jbos]{
j^3_\su=i\sqrt{\frac{n-2}{2}}\, \partial y_\su
~~,~~~~
\bar j^3_\su=i\sqrt{\frac{n-2}{2}}\, \bar\partial \bar y_\su ~,
}
and $\lambda_{j';m',\mbar'}$ is an operator in the $\frac\sutwo\uone$ coset model (and thus neutral under the $\uone$ currents $j^3_\su,\bar j^3_\su$).  This {\it parafermion decomposition}%
\footnote{So called because the decomposition of the currents $j^\pm_\su$ results in an exponential of $y$ times a $\bZ_n$ parafermion operator~\rcite{Fateev:1985mm}.}
shows that the spectral flow quanta $w',\bar w'$ act as ``winding numbers'' conjugate to the left and right zero mode momenta $m',\mbar'$ (though of course there is no conserved winding number in $\sutwo=\bS^3$).
One can see from \eqref{bospf} that the operator $v^{(w',\bar w')}_{j';m',\mbar'}$ is a Virasoro primary for any $w',\bar w'\in \bZ$, but in general it is not a current algebra primary.

In the supersymmetric theory, one can also perform spectral flow with respect to the total $SU(2)$ algebra, $J^a_n$. This combines the bosonic spectral flow described above with a spectral flow for the fermions.  To describe it explicitly, it is convenient to bosonize the fermions $\psi^\pm_\su$: 
\eqn[sutwobos]{
\psi^+_\su\psi^-_\su = i\partial H_\su 
~~,~~~~
\psi^\pm_\su = e^{\pm i H_\su} ~,
}
and similarly for $\bar\psi^\pm_\su$. Here $H_\su$ is normalized in the standard CFT way, $H_\su(z)H_\su(w)\sim -\ln(z-w)$. We make the same choice for other scalar fields that appear in our construction, such as $y_\su$ \jbos.

The supersymmetric spectral flow takes the operator $v_{j';m',\mbar'}$ to
\eqn[aarr]{V^{(w',\bar w')}_{j';m',\mbar'}=e^{i w' H_\su+i\bar w'\bar H_\su} \; v^{(w',\bar w')}_{j';m',\mbar'}~.} 
Using the superconformal current of $\sutwo_n$ given by~\eqref{Gsu}, one can check that $V^{(w',\bar w')}_{j';m',\bar{m}'}$ is a superconformal primary (but, again, not a current algebra primary). 

The supersymmetric $SU(2)_n$ WZW model admits a super-parafermionic decomposition~\eqref{sutwocoset} in terms of an $N=2$ supersymmetric $\frac{SU(2)_n}{U(1)}$ coset CFT and a free superfield $(\psi^3_\su, J^3_\su)$. The spectral flow \eqref{aarr} does not act on the $\frac{SU(2)_n}{U(1)}$ coset, but only on the $U(1)$ part. 

The various $U(1)$ currents can be bosonized as~\eqref{jbos}, \eqref{sutwobos}, as well as
\eqna[su2bos]{
&J^3_\su=i\sqrt{\frac{n}{2}}\,\partial Y
~~,~~~~
J^R_\su=i\sqrt{\frac{n-2}{n}}\,\partial Z \equiv ia\,\partial Z~,}
where $J^R_\su$ is the $U(1)$ R-symmetry current of the $N=2$ supersymmetric $\frac{SU(2)_n}{U(1)}$ coset CFT,
\eqn[JRsu2]{J^R_\su=\psi^+_\su\psi^-_\su+\frac{2}{n}J^3_\su ~.}
Note that it is orthogonal to $J^3_\su$, as is implied by the decomposition of the SCFT $SU(2)_n$ described above.

The scalar fields defined in~\eqref{jbos}, \eqref{sutwobos}, \eqref{su2bos} are related by a field space rotation
\eqna[YHrot]{
Y &= \sqrt{\frac{n-2}{n}} \, y_\su + \sqrt{\frac2n}\, H_\su~,
\\
Z &= -\sqrt{\frac2n}\,y_\su + \sqrt{\frac{n-2}{n}}\, H_\su ~.
}
In this rotated basis, the $SU(2)_n$ operator with general fermion charges $\eta_\su,\bar\eta_\su$ and bosonic spectral flow $w',\bar w'$
\eqna[Vparaf]{V^{(\eta_\su,\bar\eta_\su,w',\bar{w'})}_{j';m',\bar{m}'}&\equiv e^{i\eta_\su H_\su+i\bar{\eta}_\su\bar{H}_\su}v^{(w',\bar{w'})}_{j';m',\bar{m}'}}
can be decomposed in terms of an exponential operator carrying the total $J^3_\su$ charge, together with a charge-neutral super-parafermion operator
\eqna[pfdecomp]{
V_{j';m',\mbar'}^{(\eta_\su,\bar\eta_\su,w',\bar w')} &=  
 \Lambda_{j';m',\mbar'}^{(\alpha,{\bar\alpha})} \,\exp\Bigl[ i\sqrt{\frac{2}{n}} \Bigl(m'+\alpha+\frac{n}{2} w'\Bigr)Y
+ i\sqrt{\frac{2}{n}} \Bigl(\mbar'+{\bar\alpha}+\frac{n}{2} \bar w'\Bigr)\bar Y \Bigr]~, }
where $\etatil=\eta_\su-w'$, $\bar\alpha=\bar\eta_\su-\bar w'$, and $\Lambda_{j;m',\mbar'}^{(\alpha,{\bar\alpha})} $ is the super-parafermion operator of the supersymmetric $\frac{SU(2)_n}{U(1)}$ coset CFT, whose left and right scaling dimensions are
\eqna[pfspec]{
h\big[\Lambda_{j';m',\mbar'}^{(\alpha,\bar\alpha)}\big] &= \frac{j'(j'+1)}{n}-\frac{(m'+\alpha)^2}{n}+\frac{ \alpha^2}{2} ~,
\\
\bar h\big[\Lambda_{j';m',\mbar'}^{(\alpha,\bar\alpha)}\big]  &= \frac{j'(j'+1)}{n}-\frac{(\mbar'+\bar{\alpha})^2}{n}+\frac{ \bar{\alpha}^2}{2}  ~.
}
The super-parafermion operators can be further decomposed in terms of an exponential carrying the R-charge and the bosonic parafermion introduced in~\eqref{bospf}
\eqna[bospf2]{
\Lambda_{j';m',\mbar'}^{(\alpha,{\bar\alpha})} &= 
\lambda_{j';m',\mbar'}^{~} \;\exp\left[ i\frac{2}{\sqrt{n(\ntil)}} \left(
\Bigr(-m'+\frac{\ntil}{2}\alpha\Bigr) Z 
+  \,\Bigr(-\mbar'+\frac{\ntil}{2}{\bar\alpha}\Bigr) \bar Z\right)\right] ~.
}
We will use interchangeably the descriptors supersymmetric $\frac\sutwo\uone$ coset model, $N=2$ minimal model, $N=2$ Landau-Ginsburg model, and parafermion theory to refer to the same CFT.

Another aspect of the superconformal field theory $SU(2)_n$ that will play a role below is the moduli space of theories obtained by deforming the action by the operator $\int\! d^2z J^3_{su} \bar{J}^3_{su}$.  This deformation clearly does not act on the $\frac\sutwo\uone$ part of the operators, \eg\ on the operators $\Lambda_{j';m',\mbar'}^{(\alpha,\bar\alpha)}$ in \eqref{pfdecomp}. The action of the deformation on the $U(1)$ factor corresponds to changing the radius of $Y$.  The parafermion decomposition~\eqref{pfdecomp} is then modified to
\eqn[pfdecompsq]{
V_{j';m',\mbar'}^{(\eta_\su,\bar\eta_\su,w',\bar w')} =  
\Lambda_{j';m',\mbar'}^{(\alpha,{\bar\alpha})} \,
\exp\Bigl[ i \Bigl(p_Y Y + \bar p_Y\bar Y\Bigr) \Bigr]~. }
The spectrum of the deformed exponential in $Y,\bar Y$ can be written as~\rcite{Yang:1988bi,Giveon:1993ph,Brennan:2020bju} 
\eqna[squonespec]{
(h,\bar h) &= \half(p_Y^2,\bar p_Y^2)~,
\\
\left(p_{Y}^{~},\bar p_Y^{~}\right) &= \frac{1}{\sqrt{2n}}\Big(\frac{p+nP}{R}\pm R(\ell+nL)\Big)~,
}
with the quantum numbers related to those of~\eqref{pfdecomp} via
\eqna[paramrel]{
2(m'+\alpha) = p+\ell ~~&,~~~~  w' = P+L
\quad,\qquad P,L\in\bZ~,
\\
2(\bar m'+{\bar\alpha}) = p-\ell ~~&,~~~~\bar w' = P-L 
\quad,\qquad p,\ell \in\{0,...,n\tight-1\} ~.
}
Here, $\alpha,{\bar\alpha}$ are spectral flow parameters under the $N=2$ superconformal symmetry of the $\frac\sutwo\uone$ coset model, and $w',\bar w'$ are spectral flow parameters in the $\uone$ CFT.  Essentially, the circular orbits of the vector $\uone$ isometry of $\sutwo$ (generated by $J^3_\su+\bar J^3_\su$) expand in size by a factor $R$, while the circular orbits of the corresponding axial $\uone$ isometry (generated by $J^3_\su-\bar J^3_\su$) shrink by the same factor (for a discussion in the present context, see~\rcite{Brennan:2020bju}).  As mentioned in section~\ref{sec:intro}, we will refer to this deformed geometry as a squashed $\bS^3$, and will denote it by $\sqsphere$. It will play an important role in our construction of spacetime supersymmetric theories of the sort discussed in the previous subsection.

\vskip 1cm

\subsubsection{\texorpdfstring{$SL(2,\bR)$}{}}

The above discussion can be repeated, with a few interesting twists, for the case of $\sltwo$. The supersymmetric $\sltwo_k$ WZW model consists of a bosonic $\sltwo_{k+2}$ WZW model, with bosonic currents $j^a_\sl(z)$, and three free fermions $\psi_\sl^a$, which give a $\sltwo_{-2}$ affine Lie algebra. The total central charge is given by
\eqn[csl2]{c_{SL(2)}=c^{\rm bos}+c^{\rm ferm}=\frac{3(k+2)}{k}+\frac{3}{2} ~.}

The bosonic currents and free fermions satisfy the OPE's
\eqn[aazz]{j^a_\sl(z)j^b_\sl(0)\sim\frac{k+2}{2 z^2}\eta^{ab}+i\epsilon^{ab}_{~\;c}\frac{j^c_\sl(0)}{z}~,~~~\psi^a_\sl(z)\psi^b_\sl(0)\sim\frac{\eta^{ab}}{z} ~,}
where $\eta^{ab}=(+,+,-)$, and $\epsilon^{123}=1$. The total currents
\eqn[bbaa]{J^a_\sl=j^a_\sl-\frac{i}{2}\epsilon^a_{~bc}\psi^{b}_\sl\psi^c_\sl}
have level $(k+2)+(-2)=k$.
The theory has $N=1$ superconformal symmetry with supercurrent
\eqna[Gsl]{G_\sl&=\sqrt{\frac{2}{k}}(\eta_{ab}\,\psi^a_\sl \,j^b_\sl+i\psi^1_\sl\psi^2_\sl\psi^3_\sl)\cr &=\sqrt{\frac{1}{k}}\left(\psi_\sl^+ j^-_\sl+\psi_\sl^- j^+_\sl\right)-\sqrt{\frac{2}{k}}J^3_\sl\psi_\sl^3 ~,}
where 
\eqn[jpsisl]{j^\pm_\sl=j^1_\sl\pm i j^2_\sl,~~~\psi^\pm_\sl=\frac{\psi^1_\sl\pm i \psi^2_\sl}{\sqrt{2}}~.}
Primary operators under the bosonic $\sltwo_{k+2}$ current algebra are operators $\Phi_{j;m,\mbar}^{(0)}$, which have conformal weight
\eqn[bbbbgggg]{h\big[\Phi_{j;m,\mbar}^{(0)}\big]=-\frac{j(j-1)}{k} ~,}
where the meaning of the superscript $(0)$ will be clarified below.  These operators satisfy the OPE's with respect to the bosonic $\sltwo$ currents
\eqna[sl2reps]{&j^3_\sl(z) \Phi_{j;m,\mbar}^{(0)}(0)\sim \frac{m}{z}\Phi_{j;m,\mbar}^{(0)}(0)~,\cr
&j^\pm_\sl(z) \Phi_{j;m,\mbar}^{(0)}(0)\sim \frac{m\mp (j-1)}{z}\Phi_{j;m\pm1,\mbar}^{(0)}(0) ~.}
Some of the operators $\Phi^{(0)}_{j;m,\mbar}$ correspond to normalizable or delta-function normalizable states.  Normalizable states belong to unitary discrete representations $\cD_j^\pm$ of the bosonic theory, known as the principal discrete series, for which $j$ lies in the range 
\eqn[unitaryrangeSL]{
\half<j<\half\big( k+1 \big) ~,
}
with $m-j\in\bN_0$ for $\cD^+$, and $-j-m\in\bN_0$ for $\cD^-$ (where $\bN_0$ are the non-negative integers).  Delta-function normalizable states belong to the principal continuous series 
$\cC_{j,\alpha}$, $j\in \half+i\bR$ and $\alpha\in[0,1)$ ($\alpha$ is the fractional part of $m$).  
In string theory, the principal discrete series representations $\cD_j^-$ describe in-states for normalizable states of strings in $AdS_3$, while $\cD_j^+$ describes the corresponding out-states.  One can think of the two types of representations as bound states and scattering states, respectively.  Note that the $AdS_3$ vacuum corresponds to $j=1$ and is thus non-normalizable for $k<1$, according to \unitaryrangeSL.

In CFT and string theory on $AdS_3$ one also needs to consider non-normalizable operators, that do not satisfy the bound \eqref{unitaryrangeSL}. These operators give rise to local operators in the spacetime CFT~\rcite{Kutasov:1999xu}.  A convenient semi-classical description of such operators in Euclidean $AdS_3=\frac{\sltwoc}{\sutwo}\equiv \bH_3^+$ parametrizes the target space via the matrix~\rcite{Teschner:1997fv}
\eqn[hmat]{
h = 
\left(\begin{matrix} 1 & 0 \\ \gamma & 1 \end{matrix}\right)
\left(\begin{matrix}
e^\phi & 0 \\
0 & e^{-\phi}  
\end{matrix}\right)
\left(\begin{matrix} 1 & \bar\gamma \\ 0 & 1 \end{matrix}\right)
=
\left(\begin{matrix}
e^\phi & e^\phi \bar\gamma \\
e^\phi \gamma & e^{-\phi} + e^{\phi} \gamma\bar\gamma
\end{matrix}\right)~,
}
on which $g\in\sltwoc$ acts via $h\to g^{-1}h(g^{-1})^\dagger$.  The functions
\eqn[scalingfns]{
\Phi_j(x,\bar x) = \frac{2j-1}{\pi}\biggl(\bigl(x,1)\cdot h\cdot \biggl(\begin{matrix} \bar x \\ 1\end{matrix}\biggr)\biggl)^{-2j}
= \frac{2j-1}{\pi} \Big( |\gamma-x|^2e^{\phi} + e^{-\phi} \Big)^{-2j}
}
are eigenfunctions of the Laplacian on $\bH_3^+$. The complex parameter $x$ labels points on the boundary. $\Phi_j(x,\bar x)$ transforms as a tensor of weight $(j,j)$ under $\sltwoc$.

In CFT on $AdS_3$, the functions \eqref{scalingfns} are promoted to operators $\Phi_j(x,\bar x;z,\bar z)$, as their arguments $\phi,\gamma,\bar\gamma$ are now two dimensional fields. One can think of these operators as analogs of the operators $v_{j;m,\bar m}$ in the CFT on $\bS^3$ \eqref{hvjm}, except here they are written in a position basis on the boundary. An analogous formalism for CFT on $\bS^3$ was developed in~\rcite{Zamolodchikov:1986bd}.

The $\sltwo$ currents can also be written in the position basis on the boundary, as 
\eqna[Jofx]{
j(x;z) &= -j^+_\sl(z)+2x j^3_\sl(z)-x^2 j^-_\sl(z) ~, 
\\[.1cm]
\psi(x;z) &= -\psi^+_\sl(z)+2x \psi^3_\sl(z)-x^2 \psi^-_\sl(z)  ~,
\\
J(x;z) &= j(x;z) + \half\psi(x;z)\partial_x\psi(x;z) ~.
}
The current algebra and the transformation properties of the local operators $\Phi_j(x,\bar x;z,\bar z)$ are given in this basis by 
\eqna[JPhialgebra]{
J(x;z)J(y;w) &\sim k\frac{(y-x)^2}{(z-w)^2} + \frac{1}{z-w}\big[(y-x)^2\partial_y-2(y-x)\big] J(y;w)~,
\\
J(x;z)\Phi_h(y,\bar y;w,\bar w) &\sim \frac{1}{z-w}\big[(y-x)^2\partial_y + 2h(y-x)\big]\Phi_h(y,\bar y; w,\bar w)~.
}
At large $\phi$ (\ie\ near the boundary of $AdS_3$), the operators $\Phi_j$ behave as 
\eqn[Phiasymp]{
\Phi_j(x,\bar x) \sim
e^{(j-1)Q\phi}\, \delta^2(\gamma-x)+\cO\big(e^{(j-2)Q\phi)}\big) 
+ \frac{2j-1}{\pi} \frac{e^{-jQ\phi}}{|\gamma-x|^{4j}} + \cO\big(e^{-(j+1)Q\phi}\big) 
+\cdots~,
}
where $Q=\sqrt{2/k}$, $\phi$ has been rescaled by $Q/2$ relative to~\eqref{hmat}, \eqref{scalingfns}, and the meaning of the ellipses will be explained below. For $j>1/2$, the operator~\eqref{Phiasymp} is non-normalizable due to a divergence of the corresponding wavefunction as $\phi\to\infty$, in which case the leading term in~\eqref{Phiasymp} shows that $\Phi_j$ reduces to a local operator on the conformal boundary.%
\footnote{
The norm, given by 
$$
\int\! d\phi d\gamma d\bar\gamma \,e^{Q\phi}\, \big|\Phi_j \big|^2 \sim \int\! d\phi \, e^{Q(2j-1)\phi} \,(\cdots) ~,
$$
shows that the operators $\Phi_j$ in~\eqref{Phiasymp} are non-normalizable for $j>\half$.
}

The semiclassical operators $\Phi_j$ in equation~\eqref{scalingfns}, obey a reflection symmetry~\rcite{Teschner:1997fv,Giveon:2001up}
\eqn[reflection]{
\Phi_j(x,\bar x) = \frac{2j-1}{\pi} \int\! d^2 x' \, |x-x'|^{-4j}\, \Phi_{1-j}(x',\bar x')~,
}
which for real $j$ allows us to restrict our attention to $j>\hf$, and for $j=\hf+is$ says that the operators with $s>0$ and $s<0$ are not independent, as one would expect. 

For some purposes it is convenient to transform the local operators $\Phi_j(x,\bar x;z,\bar z)$ from position space $(x,\bar x)$ to momentum space $(m,\bar m)$, as in~\eqref{bbbbgggg}, \eqref{sl2reps}.  These representations are related by
\eqna[Phiexpn]{
\Phi_{j;m,\mbar}^{(0)} &= \int\! d^2x\, x^{j+m-1}\bar x^{j+\mbar-1} \, \Phi_{j}(x,\bar x) ~.
}
Plugging the asymptotic expansion \eqref{Phiasymp} into \eqref{Phiexpn}, we find that the operators $\Phi_{j;m,\mbar}^{(0)}$ behave at large $\phi$ like
\eqna[Phimodeasymp]{
\Phi_{j;m,\mbar}^{(0)} &\sim
e^{(j-1)Q\phi} \,\gamma^{j+m-1}\bar\gamma^{j+\mbar-1}+\cO(e^{(j-2)Q\phi})  \\
&\hskip .5cm
+(2j\tight-1) \frac{\Gamma(j\tight+m)\Gamma(j\tight-\mbar)\Gamma(1\tight-2j)}{\Gamma(m\tight-j\tight+1)\Gamma(-\mbar\tight-j\tight+1)\Gamma(2j)}
\,e^{-jQ\phi} \, \gamma^{m-j}\bar\gamma^{\mbar-j} + \cO(e^{-(j+1)Q\phi}) +\cdots ~.
}
The momentum variables $(m,\bar m)$ must satisfy the constraint $m-\mbar\in\bZ$, necessary for the single-valuedness of the integral \eqref{Phiexpn} (but are otherwise unrestricted and in particular unrelated to $j$); $m-\bar m$ is the spatial momentum on the boundary, which is quantized since the spatial coordinate lives on a circle.

From the expansion \eqref{Phimodeasymp} we see that for general $j$, when the momentum variables $(m,\bar m)$ take some specific values, the coefficient of the leading decaying (normalizable) term diverges. An example is $-(m+j)\in \bN_0$.
These divergences correspond to values of $j,m$ at which the local, non-normalizable operator $\Phi_j$ can create a normalizable state from the vacuum. It is an analog of the LSZ reduction in standard QFT, and is discussed in detail in a closely related context (the $\frac{\sltwo}{U(1)}$ coset) in~\rcite{Aharony:2004xn}. That paper also discusses other singularities of \eqref{Phimodeasymp} that do not have this interpretation.

To implement the above procedure in our case we proceed as follows. Consider the case $m=\bar m$, $m+j\to 0$, as an example. The operator \eqref{Phimodeasymp} diverges in this limit; therefore, we need to take the limit more carefully, to ensure that the operator remains finite. To do that, we define
\eqn[modphiop]{\widetilde{\Phi}_{j;-j,-j}^{(0)}\equiv\lim_{m,\bar m\to -j}(m+j)\Phi_{j;m,\mbar}^{(0)}.}
Looking back at the expansion \eqref{Phimodeasymp}, we see that in this limit the leading, non-normalizable, contribution to $\Phi_{j;m,\mbar}^{(0)}$ disappears, and we are left with a finite operator that behaves at large~$\phi$ like $e^{-Qj\phi}$. For $j>\half$ this operator is normalizable, and thus it describes a normalizable state, at least semi-classically. 

In the exact theory the situation is more interesting. The FZZ correspondence~\rcite{FZZref,Giveon:1999px,Maldacena:2000hw,Kazakov:2000pm,Giveon:2016dxe,Martinec:2020gkv} asserts that the normalizable operator \eqref{modphiop} has additional contributions from sectors with non-zero winding associated to the ellipses in equations~\eqref{Phiasymp}, \eqref{Phimodeasymp}. These sectors and contributions will play an important role in our discussion below, and we will postpone a detailed discussion of them to a later point in the paper. Here we simply note that the question whether the operator \eqref{modphiop} is indeed normalizable or not depends on the nature of these contributions. In the semi-classical limit $k\to\infty$ with $j$ fixed, they are known to be rapidly decaying at large $\phi$, and thus the operator \modphiop\ is normalizable. If $k$ is large but $j$ scales like $k$, they actually can be dominant at large $\phi$, and can even make this operator non-normalizable. The condition that the operator remains normalizable is in fact the origin of the upper bound on $j$ in equation \eqref{unitaryrangeSL}, which is valid for arbitrary $k$.

Another correction to the semi-classical picture above in the full quantum theory, which is related to the one we just discussed is the fact that the reflection symmetry \eqref{reflection} is modified~-- the r.h.s. of \eqref{reflection} is multiplied by a factor $\cR(j)$, with
\eqn[Rj]{\cR(j)=\frac{\Gamma\left(1-\frac{2j-1}{k}\right)}{\Gamma\left(1+\frac{2j-1}{k}\right)} }
(see \eg~\rcite{Giveon:2001up} for a more detailed discussion).
This factor goes to one in the classical limit $k\to\infty$, and for the delta-function normalizable case is a pure phase. This modification means that the subleading terms in \eqref{Phiasymp}, \eqref{Phimodeasymp} are multiplied by $\cR(j)$ as well.  

Plugging the expansion \eqref{Phiexpn} into \eqref{JPhialgebra}, we can compute the OPE's of the operators $\Phi_{j;m,\mbar}^{(0)}$ with the $\sltwo$ currents. It is straightforward to check that this gives the OPE's~\eqref{sl2reps}.

The bosonic $\sltwo$ CFT again admits a spectral flow transformation under which the currents transform as
\eqna[sltwospecflow]{
j^\pm_n \to j^\pm_{n\pm w}
~&,~~~~
j^3_n\to j^3_n + \frac{\ktil}2w \, \delta_{n,0}~,
\\
\bar j^\pm_n \to \bar j^\pm_{n\pm w}
~&,~~~~
\bar j^3_n\to \bar j^3_n + \frac{\ktil}2 w \, \delta_{n,0}~,
}
where $w$ is an integer, and we have dropped the subscript ``$\sl$'' to avoid clutter. The operator $\Phi_{j;m,\mbar}^{(0)}$ flows to an operator $\Phi^{(w)}_{j;m,\mbar}$ which has conformal weight and $j^3$ charge
\eqn[aaqq]{h[\Phi^{(w)}_{j;m,\mbar}]=-\frac{j(j-1)}{k}-m w-\frac{\ktil}{4}w^2,~~~j_0^3[\Phi^{(w)}_{j;m,\mbar}]=m+\frac{\ktil}{2}w ~.}
The operator $\Phi^{(w)}_{j;m,\mbar}$ is again a Virasoro primary for any $w\in \bZ$, but it is not a current algebra primary. The OPE's~\eqref{sl2reps} generalize to
\eqna[sl2windingreps]{&j^3_\sl(z) \Phi^{(w)}_{j;m,\mbar}(0)\sim \frac{m+\frac{k+2}{2}w}{z}\Phi^{(w)}_{j;m,\mbar}(0)~,\cr
&j^\pm_\sl(z) \Phi^{(w)}_{j;m,\mbar}(0)\sim \frac{m\mp (j-1)}{z^{\pm w+1}}\Phi^{(w)}_{j;m\pm1,\mbar}(0) ~.}
Note that the left and right spectral flows in \eqref{sltwospecflow} are identical because we are working on the universal cover of $\sltwo$. Thus, the timelike direction in the group manifold is non-compact and has no winding.  In contrast to the $\sutwo$ case, this spectral flow is not an automorphism of representations, rather it generates new representations of the current algebra~\rcite{Maldacena:2000hw}.

As mentioned above, string winding number is not conserved, and so operators described in a given sector have contributions from other sectors.  Related to this is the phenomenon of FZZ duality~\rcite{FZZref,Giveon:1999px,Maldacena:2000hw,Kazakov:2000pm,Giveon:2016dxe,Martinec:2020gkv} which relates $\cD^-$ unitary representations in winding sector $w$ to $\cD^+$ unitary representations in winding sector $w-1$; in particular highest weight states are identified via
\eqn[FZZduality]{
\Phi_{j;-j,-j}^{(w)} \equiv \Phi_{\frac k2+1-j;\frac k2+1-j,\frac k2+1-j}^{(w-1)} ~.
}
Note from equation~\eqref{aaqq} that the conformal dimension and $j^3$ eigenvalues match. The rest of the map between the representations follows from the spectral flow of the generators~\eqref{sltwospecflow}.  
Thus for instance when we are counting states we should not include both sets of representations $\{\cD^+_{(w),j}\}$ and $\{\cD^-_{(w),j}\}$, but only one or the other.  Our conventions are such that $\cD^-_{(w),j}$ representations with $w\le -1$ describe in-states bound to $AdS_3$ that wind $|w|$ times around the azimuthal direction.  The remaining $\cD^-_{(w),j}$ representations with $w\ge0$ map via FZZ duality to the set of $\cD^+_{(w),j}$ representations with $w\ge1$, which are charge conjugates of the $\cD^-_{(w),j}$ representations with $w\le-1$, and which thus describe out-states.

As in the $SU(2)$ case, in the supersymmetric case we can also consider spectral flow with respect to the total $\sltwo_{k}$ algebra. To do this, we combine the operators $\Phi^{(w)}_{j;m,\mbar}$ with a contribution from the fermions
\eqn[sltwobos]
{
\psi^+_\sl\psi^-_\sl = i\partial H_\sl 
~~,~~~~
\psi^\pm_\sl = e^{\pm i H_\sl} ~,
}
and consider the operators 
\eqn[superPhi]{
\Phihat^{(w)}_{j;m,\mbar} = e^{-i w (H_\sl + \bar H_\sl)} \, \Phi^{(w)}_{j;m,\mbar} ~,
} 
whose conformal weight and $J^3$ charges are 
\eqn[aass]{h\big[\Phihat^{(w)}_{j;m,\mbar}\big] = -\frac{j(j-1)}{k}-m w - \frac{k}{4}w^2
~~,~~~~
J_0^3\big[\Phihat^{(w)}_{j;m,\mbar}\big] = m+\frac{k}{2}w ~.}
The operators $\Phihat^{(w)}_{j;m,\mbar}$ are again superconformal primaries, but not current algebra primaries.

\vskip 1cm

\subsection{Symmetries of string theory on {\texorpdfstring{$SL(2,\bR)_k\times SU(2)_n$}{}}}
\label{sec:STsymm}

The $\sltwo\times\sutwo$ current algebra symmetry of the worldsheet theory lifts to a corresponding symmetry in spacetime~\rcite{Giveon:1998ns,Kutasov:1999xu}. Given the total $SU(2)_n$ currents $J^a_\su(z)$ on the worldsheet, which are the top components of superfields whose bottom components are the $\psi^a_\su(z)$, one can construct worldsheet vertex operators for the $SU(2)$ currents in spacetime. These are given by\footnote{As discussed in~\rcite{Kutasov:1999xu}, \eqref{Kst} is a special case of a general construction that associates a dimension $(1,0)$ current in the spacetime CFT to every dimension $(1,0)$ worldsheet current that is the top component of a dimension $(\half,0)$ worldsheet superfield.}
\eqn[Kst]{
\cK^a(x) \simeq 
\int\!d^2z\,e^{-\varphi-\bar{\varphi}} \psi_\su^a(z)\bar \psi(\bar x;\bar z)\,\Phi_1(x,\bar x;z,\bar z)~,
}
where $\simeq$ means that we have omitted an overall numerical constant, which can be fixed using the techniques of~\rcite{Kutasov:1999xu}, $\bar\psi(\bar x;\bar z)$ was written in~\eqref{Jofx}, and $\Phi_1(x,\bar x)$ is the local operator~\eqref{scalingfns} for $j=1$. 

The operators \Kst\ are holomorphic in $x$, in that their $\bar x$ derivative is BRST exact (and so formally decouples from string theory correlation functions).  They satisfy a spacetime current algebra
\eqn[Kcurrentalg]{
\cK^a(x)\cK^b(y) \sim \frac{\half n\,\delta^{ab}\, \cI}{(x-y)^2} + \frac{\epsilon^{ab}_{~\;c} \,\cK^c(y)}{x-y}
}
with level $n\,\cI$, where the operator
\eqn[Ioperator]{
\cI(x,\bar x) \simeq 
\int \! d^2z\, e^{-\varphi-\bar{\varphi}}\psi(x,z)\bar \psi(\bar x,\bar z) \Phi_1(x,\bar x;z,\bar z)
}
plays the role of the identity operator in that it commutes with the currents $\cK^a(x)$ (again up to BRST exact quantities).

In order to determine the transformation properties of operators in the theory under the global part of the $SU(2)$ current algebra in the spacetime CFT, it is useful to recall that a corollary of the construction of~\rcite{Kutasov:1999xu} is that 
\eqn[ointwssp]{\oint\! dx\, \cK^a(x)=\oint\! dz\, J^a_{\rm su} ~.}
In other words, the action of the global $SU(2)$ generators in the spacetime theory can be read off from that of the worldsheet ones.

The $\sltwo$ current algebra that underlies the worldsheet theory, \eqref{aazz}, \bbaa, lifts to a spacetime Virasoro algebra with central charge $c=6k\cI$ generated by
\eqna[Tst]{
\cT(x) & \simeq 
\int\!d^2z\,e^{-\varphi-\bar\varphi}\big[\partial_x \psi(x;z)\partial_x\Phi_1+2\left(\partial^2_x\psi(x;z)\right)\Phi_1\big]\,\bar \psi(\bar x;\bar z)~,
\\[.1cm]
\cT(x)\cT(y) &\sim \frac{3k\,\cI}{(x-y)^4} + \frac{2\cT(y)}{(x-y)^2} + \frac{\partial_y \cT}{x-y} ~.
}
In analogy to \ointwssp, the global part of the spacetime Virasoro algebra is generated by the global part of the worldsheet $\sltwo$ current algebra, 
\eqn[globalsltwo]{L_0=-J^3_{\rm sl,0}\;;\;\;L_{\pm 1}=-J^\pm_{\rm sl,0} ~.}

\noindent
As discussed in \rcite{Giveon:2001up}, the operator $\cI$, which appears in the expressions for the levels of current algebras and the central charge of the spacetime CFT, behaves in string theory on $AdS_3$ as a non-trivial dimension zero operator. In particular, its expectation values in different correlation functions are different. It was pointed out in~\rcite{Porrati:2015eha} that the reason for this is that the standard formalism for computing correlation functions in string theory on $AdS_3$ can be viewed as taking place at a fixed chemical potential for the number of fundamental strings, rather than at a fixed number of strings $p$. The two are related as usual by a Legendre transform. Performing this transform leads to standard spacetime conformal and current algebras. 



\section{Type 0 string theory on \texorpdfstring{$AdS_3\times \bS^3$}{}}
\label{sec:type0short}


In this section and the next, we will put together the elements described in the previous section, and construct type 0 string theory on $AdS_3\times \bS^3$. Our ultimate goal is to study the $AdS_3$ backgrounds corresponding to the systems described in subsection \ref{sec:GKP}, which are described by type II string theory on this background, with a particular squashing of the $\bS^3$. The type 0 theory provides a good warm-up exercise, and is also useful since one can view the type II theory as a chiral orbifold of the type 0 one. 

Type 0 string theory on $AdS_3\times \bS^3$ has only spacetime bosons, which belong to the $(\NS,\NS)$ and $(\!R,\!R)$ sectors of the worldsheet theory. To define the theory we need to choose a GSO projection, which in this case is non-chiral (\ie\ it acts on both the left and right-movers). In the $(\NS,\NS)$ sector one can think of it as the projection 
\eqn[nonchiralGSO]{
(-1)^F=(-1)^{\bar F} ~,
}
where $F$ and $\bar F$ are the left and right-moving fermion numbers, respectively. As usual, the superconformal $(\beta,\gamma)$ ghosts also carry odd fermion number.


\subsection{Tachyon and graviton operators}
\label{sec:tachyon}

The lowest lying state in the $(\NS,\NS)$ sector is the type 0 tachyon, described (in the $(-1,-1)$ picture) by the vertex operator
\eqn[VNS]{e^{-\varphi-\bar\varphi}\,v_{j';m'\mbar'}\Phi_j(x,\bar x;z,\bar z) ~. }
The mass-shell condition sets
\eqn[Lzerotach]{
-\frac{j(j-1)}{k} + \frac{j'(j'+1)}{n} = \half ~.
}
Using \eqref{kntwod} this can be written as
\eqn[massmod]{\frac1k\left(j-\half\right)^2=\frac1n\left(j'+\half\right)^2-\frac14 ~.}
For $SU(2)$ spin $j'$ up to $(2j'+1)^2=n$, the type 0 tachyon lies below the 
Breitenlohner-Freedman (BF) bound~-- the corresponding $j$ \eqref{massmod} belongs to the principal continuous series. These are tachyonic modes in $AdS_3$; their presence indicates an instability of the type 0 vacuum we are studying. They do not appear in the supersymmetric theories we will discuss later.   

At the next level, we have the gravity sector of the model, consisting of the metric, NS $B$-field and dilaton. In the $(-1,-1)$ picture, the corresponding vertex operators include
\eqna[WXxspace]{
\cW_{j'}(x,\bar x) &= e^{-\varphi-\bar\varphi}\,\psi(x)\bar\psi(\bar x)\, \Phi_j(x,\bar x) \, v_{j'}~,
\\[.1cm]
\cX_{j'}(x,\bar x) &= e^{-\varphi-\bar\varphi}\, \Phi_j(x,\bar x)\,\big(\psi_\su\bar\psi_\su\,  v_{j'}\big)_{j'+1} ~.
}
Some comments about these operators:
\begin{enumerate}[1)]
\item Recall that $j'=0,\half,1,\cdots \half(n-2)$.
\item We have suppressed the dependence of the various operators on the worldsheet coordinates $(z,\bar z)$. Of course, as usual in string theory, the full BRST invariant operators, which are obtained from \eqref{WXxspace} by integrating over $z$, or by multiplying by the ghosts $c\bar c$, do not have any $z$ dependence.
\item We have also suppressed the values of $j^3_\su$ and its right-moving analog, $(m',\bar m')$. These values are exhibited, for example, in \eqref{VNS}.
\item The notation $\big(\psi_\su\bar\psi_\su\,  v_{j'}\big)_{j'+1}$ means that we are coupling a spin one representation of the total $SU(2)$ coming from the fermions with a spin $j'$ representation coming from the bosons into a representation with spin $j'+1$. This needs to be done for the left-movers and right-movers separately. See \eg~\rcite{Kutasov:1998zh} for further discussion.
\end{enumerate}

\noindent
The mass-shell conditions of string theory applied to \eqref{WXxspace} set
\eqn[Lzerograv]{
-\frac{j(j-1)}{k} + \frac{j'(j'+1)}{n} = 0 ~,
}
whose solution is
\eqn[WXspins]{
j = \half\Big( 1 + \sqrt{\frac{(2j'+1)^2+n}{n+1}} \Big) ~.
}
Note that in \eqref{WXspins} we have chosen a sign in solving the quadratic equation for $j$ \eqref{Lzerograv}. This was done so that $j>\half$, which as we discussed in the previous section means that the operators \eqref{WXxspace} give rise to local operators in the spacetime CFT. The (left and right-moving) spacetime scaling dimensions of these operators are given by~\rcite{Kutasov:1999xu}
\eqna[spacescale]{
h_{\ST}\left(\cW_{j'}\right) &= j-1~,
\\[.1cm]
h_{\ST}\left(\cX_{j'}\right) &= j~,
}
with $j$ related to $j'$ via \eqref{WXspins}. The expression for $j$ implies that it is outside the range~\eqref{unitaryrangeSL}, in which one finds normalizable states that belong to the principal discrete series. Therefore, the local operators in the spacetime CFT that correspond to the vertex operators \eqref{WXxspace} do not have corresponding normalizable states, a situation familiar from theories such as Liouville and $\frac{\sltwo}{U(1)}$. 

For $j'=0$, the operator $\cW_{j'=0}$ is proportional to the operator $\cI$ \eqref{Ioperator}, which as we discussed is proportional to the identity operator. The fact that it does not have a corresponding normalizable state is essentially the statement that the $\sltwo$ invariant vacuum is non-normalizable for $k<1$, a fact mentioned above. 

The operators $\cW_{j'}$ and $\cX_{j'}$ can be thought of as gravitons with polarization in $AdS_3$ and $\bS^3$, respectively. We can also construct gravitons whose left-moving polarization is in $\bS^3$ and right-moving polarization in $AdS_3$, or the other way around. An example is the vertex operators
\eqn[hybridvertex]{\cK_{j'}=e^{-\varphi-\bar\varphi}\big(\psi_\su v_{j'}\big)_{j'+1}\bar\psi(\bar x)\, \Phi_j(x,\bar x)~,
}
where $j$ is again related to $j'$ via \eqref{WXspins}.
This set of operators is a mixture of the two sets of operators in equation~\eqref{WXxspace}~-- their left-moving part looks like that on the second line, while the right-moving part looks like the first line. Accordingly, their spacetime dimensions are given by 
\eqn[dimkk]{h_{\ST}\left(\cK_{j'}\right) = j~~,~\;\;\bar h_{\ST}\left(\cK_{j'}\right) = j-1  ~.}
Thus, these are spin one operators. For $j'=0$, they become holomorphic, and are nothing but the $SU(2)$ currents \eqref{Kst}. 

So far we discussed the first two levels in the $(\NS,\NS)$ sector of the type 0 string on $AdS_3\times S^3$, corresponding to the type 0 tachyon and gravity sector of this non-critical string theory. We can in principle continue to higher levels, and construct the infinite towers of operators familiar from critical string theory. All these operators are non-normalizable, and correspond to local operators in the spacetime CFT. None of them are in the region~\eqref{unitaryrangeSL}, and thus they do not correspond to normalizable states in the spacetime theory.  

In our discussion above we have constructed local operators in the spacetime CFT as vertex operators in the underlying string theory. Below, we will find it convenient to work in a basis of eigenstates of $J^3_{\rm sl,0}$, which as we mentioned corresponds to $L_0$ in the spacetime CFT, see \globalsltwo. For this we would like to perform the integral transform \eqref{Phiexpn}. As discussed in section \ref{sec:review}, this is a subtle process, since this integral diverges for values of $m$ and $j$ that correspond to normalizable (principal discrete series) states.   

To see what's going on, consider for example the operator on the second line of \eqref{WXxspace}. Applying the integral transform to it, we find a momentum space operator $\cX_{j'}(m,\bar m)$, given by the vertex operator
\eqn[chimm]{\cX_{j'}(m,\bar m) = e^{-\varphi-\bar\varphi}\, \Phi^{(0)}_{j;m,\bar m}\,\big(\psi_\su\bar\psi_\su\,  v_{j'}\big)_{j'+1} ~.
}
The operator \eqref{chimm} is an eigenoperator of $J^3_{\rm sl,0}$ with eigenvalue $m$, and similarly for the right-movers. Since the spacetime $L_0$ operator is equal to $-J^3_{\rm sl,0}$, it is natural to set $m$ to the negative of the scaling dimension of $\cX_{j'}$, which via \eqref{spacescale} is $m=\bar m=-j$. This means that the integral \eqref{Phiexpn} diverges, as reflected in the pole of the coefficient of the second term in \eqref{Phimodeasymp}. 

As we discussed in section \ref{sec:review}, the physical vertex operator for this value of $(m,\bar m)$ is the residue of this pole, \eqref{modphiop}. It is superficially normalizable, since it does not contain the leading, non-normalizable, term in \eqref{Phimodeasymp}. However, the full quantum vertex operator receives contributions from other winding numbers~\rcite{FZZref,Giveon:1999px,Maldacena:2000hw,Kazakov:2000pm,Giveon:2016dxe,Martinec:2020gkv}, which may or may not be normalizable. For the case of the operators \eqref{WXxspace} these contributions are non-normalizable, and thus so is the full vertex operator \eqref{chimm}. 

Note also that in the discussion of section \ref{sec:review}, we denoted the residue of the pole by $\widetilde{\Phi}_{j;-j,-j}^{(0)}$. In equation~\eqref{chimm} we omitted the tilde on $\Phi^{(0)}$ to simplify the notation, but it should be understood to be there. The same is true in many equations below.

The conclusion of the above discussion is that we can associate to the local operator \eqref{WXxspace} a momentum space version that looks like \eqref{chimm} with $m=\bar m=-j$. Because the spin $j$ of~\eqref{WXspins} lies outside the normalizable range of the unitary discrete series, these operators are all non-normalizable.

Repeating the discussion above for the operators $\cW_{j'}(x,\bar x)$ on the first line of \eqref{WXxspace}, we arrive at the following momentum space version of these operators: 
\eqna[WX]{
\cW_{j'} &= e^{-\varphi-\bar\varphi}\, \left(\psi_\sl\bar\psi_\sl\, \Phi^{(0)}_{j}\right)_{j-1;m=\mbar=-j+1} \, v_{j'}~,
\\[.1cm]
\cX_{j'} &= e^{-\varphi-\bar\varphi}\, \Phi^{(0)}_{j;m=\mbar=-j}\,\big(\psi_\su\bar\psi_\su\,  v_{j'}\big)_{j'+1}~,
}
with obvious generalizations for the other operators discussed above. Here $\left(\psi_\sl\bar\psi_\sl\, \Phi^{(0)}_{j}\right)_{j-1}$ denotes the $\sltwo$ representation with total spin $j-1$ obtained by coupling a spin one $\sltwo$ representation coming from the fermions and a spin $j$ representation from $\Phi_{j}$. In the first line of \eqref{WX} we also specified the eigenvalues of $J^3_{\rm sl;0}$ and $\bar J^3_{\rm sl;0}$, $m$ and $\bar m$, respectively. Below we will mostly work with momentum space operators, such as \eqref{WX}, mainly because they are more natural in sectors with non-zero winding.


\subsection{Ramond sector operators}
\label{sec:Ramond}

So far we discussed the $(\NS,\NS)$ sector of type 0 string theory on $AdS_3\times \bS^3$. We next turn to the $(\!R,\!R)$ sector. 
At the lowest level, we find the $(\!R,\!R)$ gauge fields in this background. In the $\left(-\half,-\half\right)$ picture, they are described by the vertex operators
\eqn[VRR]{e^{-\half\varphi-\half\bar\varphi} \,
S \,v^{~}_{j'}
\Phi_{j}^{(0)}~, }
where $S$ is a spin field for the six free fermions $\psi_{\rm su}$, $\psi_{\rm sl}$. In \eqref{VRR} we have suppressed the spinor indices on $S$ as well as the values of $(m,\bar m)$ and $(m',\bar m')$, since they are all coupled by the requirement that \eqref{VRR} transforms covariantly under $\sltwo\times SU(2)$. 

The mass shell condition for \eqref{VRR} reads
\eqn[LzeroRR]{
j-\half = \frac{1}{\sqrt{n+1}}\Big(j'+\half\Big)  ~,
}
so the lower end of this range 
($j'<\frac{n}{2\sqrt{n+1}}-\half$)
lies within the unitary range~\eqref{unitaryrangeSL} for the $\sltwo$ spin $j$.

To construct the spectrum of $(R,R)$ gauge fields on $AdS_3\times \bS^3$, we proceed in two steps. First, we search for BRST invariant operators, which can be done separately for left and right-movers on the worldsheet, and then impose the GSO projection to put the left and right-movers together. 

Since the total $SU(2)$ charge commutes with the BRST operator, we can restrict the analysis to a given $SU(2)$ representation. To further simplify the analysis, we can start by constructing the BRST invariant operator corresponding to the highest weight state in a given total $SU(2)$ representation, and then fill in the rest of the representation by acting with the generators of the total $SU(2)$.  

For the above highest weight operators, the holomorphic analysis yields the following BRST invariant vertex operators: 
\begin{align}\label{RRcohomology}
{\cO}^{(1)}&=e^{-\frac{\varphi}{2}}e^{\frac{i}{2}H_\sl+\frac{i}{2}H_3+\frac{i}{2}H_\su}\Phi_{j;m=-j}^{(0)}v^{w=0}_{j';j'} ~,
\nn\\
{\cO}^{(2)}&=e^{-\frac{\varphi}{2}}e^{\frac{i}{2}H_\su}v^{w=0}_{j';j'}\left(e^{-\frac{i}{2}H_\sl+\frac{i}{2}H_3}\Phi_{j;m=-j}^{(0)}-\frac{1}{2j}e^{\frac{i}{2}H_\sl-\frac{i}{2}H_3}\Phi_{j;m=-j-1}^{(0)}\right) ~,
\nn\\
{\cO}^{(3)}&=e^{-\frac{\varphi}{2}}e^{\frac{i}{2}H_\sl}\Phi_{j;m=-j}^{(0)}\left(e^{\frac{i}{2}H_3-\frac{i}{2}H_\su}v^{w=0}_{j';j'}-\frac{1}{\sqrt{2j'}}e^{-\frac{i}{2}H_3+\frac{i}{2}H_\su}v^{w=0}_{j';j'-1}\right) ~,
\\
{\cO}^{(4)}&=e^{-\frac{\varphi}{2}}\left(e^{-\frac{i}{2}H_\sl+\frac{i}{2}H_3-\frac{i}{2}H_\su}\Phi_{j;m=-j}^{(0)}v^{w=0}_{j';j'}\right.
+\frac{1}{\sqrt{2j'}}e^{-\frac{i}{2}H_\sl-\frac{i}{2}H_3+\frac{i}{2}H_\su}\Phi_{j;m=-j}^{(0)}v^{w=0}_{j';j'-1}
\nn\\
&\hskip .5cm
-\frac{1}{2j}e^{\frac{i}{2}H_\sl-\frac{i}{2}H_3-\frac{i}{2}H_\su}\Phi_{j;m=-j-1}^{(0)}v^{w=0}_{j';j'}
\left.+\frac{1}{\sqrt{2j'}2j}e^{\frac{i}{2}H_\sl+\frac{i}{2}H_3+\frac{i}{2}H_\su}\Phi_{j;m=-j-1}^{(0)}v^{w=0}_{j';j'-1}\right) \;,
\nn
\end{align}
where the spin field $S$ in \VRR\ has been written in the bosonized representation in terms of $H_\su$ and $H_\sl$ defined in \eqref{sutwobos} and \eqref{sltwobos}, respectively, as well as $H_3$ defined by
\eqn[Hdefs]{
i\partial H_3 = \psi^3_\sl \psi^3_\su
~~,~~~~
\frac{\psi^3_\su\pm \psi^3_\sl}{\sqrt{2}} = e^{\pm i H_3}~,
}
and similarly for the right-movers. The signs in \eqref{RRcohomology} are determined using the conventions for cocycle factors described in Appendix \ref{app:cocycles}. 

The combinations of spin field polarizations and bosonic spins result in projections onto total $\sltwo\times SU(2)$ spins $(j_\tot^{~},j'_\tot)$
\eqn[dimspin]{\cO^{(1)}: (j-\hf, j'+\hf)~;\quad \cO^{(2)}: (j+\hf, j'+\hf)~;\quad \cO^{(3)}: (j-\hf, j'-\hf)~;\quad \cO^{(4)}: (j+\hf, j'-\hf) ~.
} 
By acting on~\eqref{RRcohomology} with the total $J^-_{\su;0}$, we can obtain the rest of the operators in the $SU(2)$ representations indicated in \eqref{dimspin}.

To perform the GSO projection in the $(\!R,\!R)$ sector of the type 0 theories, we define the worldsheet fermion number $F$ of operators via their eigenvalue $\pm1$ under the simultaneous shifts
\eqn[Fnum]{
H_a \to H_a + \epsilon_a \pi
~~,~~~~
\varphi\to \varphi + i\pi~,
}
for some fixed vector $\epsilonb$ with entries $\epsilon_a=\pm1$, and similarly for right-movers, with  $\epsilonb\to\bar\epsilonb$.
In the NS sector, these shifts act on the bosonized representation as the usual fermion parity (including the superconformal ghost contribution).
In the Ramond sector, the GSO projection is a chirality projection on spacetime spinors.

As in the critical string, the non-chiral GSO projection~\eqref{nonchiralGSO} of string theory on $AdS_3\times \bS^3$ comes in two varieties, type 0A and type 0B, which both restrict to~\eqref{nonchiralGSO} according to the definition of fermion parity via the shift~\eqref{Fnum}.  The difference between type 0A and type 0B amounts to how $(-1)^F$ and $(-1)^{\bar F}$ act in the $(\!R,\!R)$ sector. Choosing the half-shifts $\epsilonb=\bar\epsilonb=\epsilonb_0\equiv(1,1,1)$ to be the same on left and right%
\footnote{We denote this joint vector by $\epsilonb_0$ for reasons that will become clear in section \ref{sec:typeIIshort}.} leads to a projection where the $(\!R,\!R)$ bispinor fields have the same left and right chiralities in $AdS_3\times \bS^3$; the allowed sectors are conventionally denoted $(R+,R+)$ and $(R-,R-)$.  This defines the type 0B theory.  Alternatively, one can flip the sign of an odd number of entries in $\bar\epsilonb$ relative to $\epsilonb$, in which case the $(\!R,\!R)$ operators that survive the projection are $(R+,R-)$ and $(R-,R+)$; this is the type 0A theory.  We will choose $\epsilonb=\epsilonb_2\equiv (-1,1,1)$ for this alternate definition of fermion number in the Ramond sector. 

With the definition \Fnum\ of the operator $(-1)^F$ in the Ramond sector, one has the following action of this operator on the holomorphic vertex operators \eqref{RRcohomology}. For $\epsilonb=\epsilonb_0$, one has   
\eqn[actionepsilon]{(-1)^F\cO^{(i)}=-\cO^{(i)}\;\;{\rm for}\;\; i=1,4;\;\;\; (-1)^F\cO^{(i)}=\cO^{(i)}\;\;{\rm for}\;\; i=2,3;}
while for $\epsilonb=\epsilonb_2$ all the signs are reversed.

Combining the two worldsheet chiralities, we find the GSO invariant combinations
\eqna[GSOeven]{&0A:~~{\cal O}^{(1)} \bar{{\cal O}}^{(2)},{\cal O}^{(1)} \bar{{\cal O}}^{(3)},{\cal O}^{(2)} \bar{{\cal O}}^{(1)},{\cal O}^{(2)} \bar{{\cal O}}^{(4)},{\cal O}^{(3)} \bar{{\cal O}}^{(1)},{\cal O}^{(3)} \bar{{\cal O}}^{(4)},{\cal O}^{(4)} \bar{{\cal O}}^{(2)},{\cal O}^{(4)} \bar{{\cal O}}^{(3)}~,\cr
&0B:~~{\cal O}^{(1)} \bar{{\cal O}}^{(1)},{\cal O}^{(2)} \bar{{\cal O}}^{(2)},{\cal O}^{(3)} \bar{{\cal O}}^{(3)},{\cal O}^{(4)} \bar{{\cal O}}^{(4)},{\cal O}^{(1)} \bar{{\cal O}}^{(4)},{\cal O}^{(2)} \bar{{\cal O}}^{(3)},{\cal O}^{(3)} \bar{{\cal O}}^{(2)},{\cal O}^{(4)} \bar{{\cal O}}^{(1)}  ~.}
Note that in the $0A$ $(0B)$ theory, all the states have odd (even) $(j_\tot-\bar j_\tot)-(j'_\tot-\bar j'_\tot)$.


\subsection{The general case}
\label{sec:generalops}

We can also consider vertex operators associated to excited strings and general winding sectors.  In the $(\NS,\NS)$ sector one has
\eqn[Vgen]{
e^{-\varphi-\bar\varphi} \, \cP(\partial^r\!\psi,\partial^s\! j) \,
\bar\cP(\bar\partial^{\bar r}\!\bar\psi,\bar\partial^{\bar s\,} \!\bar j) \,
\Phi^{(w)}_{j;m,\mbar}\, v_{j';m',\mbar'} ~.
}
Here $\cP,\bar\cP$ are general polynomials in currents, fermions, and derivatives, subject to the BRST constraints and GSO projection. The $L_0$ constraint reads
\eqn[L0excited]{
-\frac{j(j-1)}{k} - w m - \frac k4\,w^2 + \frac{j'(j'+1)}{n} + N_\osc = \delta_0~, }
where $N_\osc\in\hf\bN_0$ is the oscillator excitation level (the conformal dimension of $\cP$), and $\delta_0=\hf$. 
In the $(\!R,\!R)$ sector, the mass-shell condition takes again the form \eqref{L0excited}, with $N_\osc\in\bN_0$ and $\delta_0=\frac14$.  

In the $w=0$ sector, the $j$ values of all of the excited operators (those with $N_\osc >0$ in~\eqref{L0excited}) lie outside the unitary range~\eqref{unitaryrangeSL}; there are no associated normalizable string states.  The states of the type 0 string in the zero winding sector are thus field-theoretic in nature, consisting of the closed string tachyon together with a small number of $(\!R,\!R)$ fields~\eqref{GSOeven}.  Only a subset of $\sutwo$ angular modes can be dressed by unitary discrete series representations of $\sltwo$, corresponding to bound states in $AdS_3$ (a subset of angular modes of the tachyon dressed by continuous series representations of $\sltwo$ represent unstable modes of the background).  

More highly excited string bound states can be found in discrete series winding sector representations $\cD_\pm^{(w)}$ with $w\ne 0$.  The solution to the mass shell constraint~\eqref{L0excited} takes on a different character in nonzero winding sectors in that we can adjust $m$ to keep $j$ within the unitary range of discrete series representations.  Thus one can have normalizable string states with arbitrarily high oscillator excitation so long as $w\ne 0$. 

While we have focused on the spectrum of type 0 string theory on $AdS_3\times \bS^3$, we noted in section~\ref{sec:currentalg} that the theory has a moduli space where the $\bS^3$ is squashed by a $J^3_\su\bar J^3_\su$ deformation.  This generalization will play an important role in our discussion of the type II theory below.


\section{Long strings in type 0 on \texorpdfstring{$AdS_3\times \bS^3$}{} }
\label{sec:type0long}

In the previous section we described type 0 string theory on $AdS_3\times \bS^3$, focusing on the sector with winding zero on $AdS_3$. As we will see, the sectors with non-zero winding are particularly instructive in studying the theory. Thus, we now turn to them.  

Winding around the spatial circle on the boundary of $AdS_3$ is not a good quantum number, except near the boundary, since that circle is contractible. Indeed, a string wrapped around the circle on the boundary can shrink to zero size and disappear, by moving into the bulk of $AdS_3$. However, below we will be studying operators that are non-normalizable, or delta-function normalizable, and those can be labeled by their winding around the circle, measured at the boundary. We will loosely refer to such operators as {\it long string} operators.

There are two constructions in the literature of strings winding around the boundary circle in $AdS_3$. Seiberg and Witten (SW)~\rcite{Seiberg:1999xz} showed that in string theory on $AdS_3\times \NN$ a single long string (\ie\ a string singly wound around the boundary circle) is described as a state of radial momentum $p$ in a superconformal field theory whose asymptotic geometry is
\eqn[mmll]{\MM=\bR_\phi\times\NN ~,}
where $\bR_\phi$ is a linear dilaton CFT with slope $Q_\ell$ \eqref{lslope},
describing the radial direction of $AdS_3$.  Here $\phi\to+\infty$ corresponds to the region near the boundary, and the SW description is only valid there, by construction.  The radial dependence of a long string state is described in the theory $\cM$ \mmll\ by an operator $\exp(\beta\phi)$, with $\beta=-Q_\ell/2+ip$, where $p$ is the radial momentum.

The behavior of the dilaton on $\bR_\phi$ is qualitatively different for $k>1$ and for $k<1$. In the former case, the strong coupling region is near the boundary, and therefore the SW description breaks down there (in the sense described in section \ref{sec:introsummary}). Thus, it is only valid in a finite range of $\phi$, where it is sufficiently large that we can talk about a long string, but not so large that the description is invalidated by the strong coupling.

On the other hand, for $k<1$, the case we study here, the region near the boundary corresponds to weak coupling in~\eqref{mmll}, and the situation is similar to that in Liouville theory~-- we can label operators by their behavior as $\phi\to\infty$, although correlation functions typically require an understanding of the finite $\phi$ region.

Recall that in our case, the compact CFT $\NN$ in \eqref{mmll} is 
\eqn[sutwon]{\NN=SU(2)_n ~,}
the superconformal $SU(2)$ WZW model at total level $n$, or a squashed version thereof obtained by a marginal $J^3_{\rm su}\bar J^3_{\rm su}$ deformation. The corresponding level of $\sltwo$ is \eqref{kntwod}
\eqn[valuekk]{k=\frac{n}{n+1}~,}
and the slope of the linear dilaton in the SW theory, \eqref{lslope}, can be written as
\eqn[longss]{Q_\ell=\sqrt\frac{2}{k_\ell}\;\;{\rm with}\;\; k_\ell=n(n+1)  ~.}

The second description of long strings is due to Maldacena and Ooguri (MO)~\rcite{Maldacena:2000hw}, who constructed vertex operators in worldsheet string theory corresponding to long strings with arbitrary winding. For strings with winding one, the SW and MO constructions are closely related~-- the MO vertex operators are in one-to-one correspondence with operators in the SW theory. Note that this is only true for non-normalizable and delta-function normalizable operators, since their existence is not sensitive to the detailed properties of the finite $\phi$ region. 

As mentioned in section \ref{sec:introsummary}, MO also constructed normalizable (discrete series) operators in sectors with non-zero winding, but those do not need to match anything in the SW theory, since the normalizable spectrum depends on properties of the finite $\phi$ region, which is not part of the SW description  \eqref{mmll}. 

As we will discuss later in this section (around equation \eqref{windingdim}), the long strings with winding larger than one, whose vertex operators were constructed in MO, extend the SW description~\eqref{mmll} of one singly wound long string to the Fock space of multiply-wound long strings, which has the structure of a symmetric product $\MM^p/S_p$, where $p$ is the number of strings creating the $AdS_3$ vacuum. 
More precisely, the MO spectrum of single long strings with a particular $w$ corresponds to the spectrum of $\bZ_w$ twisted operators in the symmetric orbifold. Of course, for perturbative string states one keeps $w$ fixed and sends $p\to\infty$; for windings $w\sim p$, the perturbative string description of~\rcite{Maldacena:2000hw} is not expected to be valid.%

From the point of view of the symmetric product orbifold, operators describing singly wound long strings belong to the untwisted sector. Thus, each operator $\OO$ in the SW theory corresponds to an operator of the form 
\eqn[singletr]{\sum_{i=1}^p\OO_i~,}
where $\OO_i$ is the operator $\OO$ in the $i$'th copy of $\MM$. We will discuss such operators extensively below, and will mostly omit the sum \eqref{singletr}. This sum is implicit in many of our formulae.

In the rest of this section, we will start by discussing some examples of the above correspondence between vertex operators in type 0 on $AdS_3\times \bS^3$ and the corresponding SW theory, $\bR_\phi\times \bS^3$. We will then describe the general picture suggested by these examples and other considerations.

A word of caution on notation: some of the operators in the SW theory are similar to ones that appear in the worldsheet theory on $AdS_3$. For example, $v_{j;m,\bar m}$ denotes a worldsheet operator in $SU(2)_{n-2}$ WZW, and also the corresponding operator in the SW theory. We will try to make it clear which of the two theories we are discussing at every stage; hopefully, this will not lead to too much confusion.


\subsection{Operators associated to \texorpdfstring{$\bR_\phi$}{} in the block theory}
\label{sec:phiops}

The first operator we will consider is\footnote{It is important to emphasize that \eqref{ebphi} is the form of the operator at large positive $\phi$. An operator with this asymptotic behavior is guaranteed to exist on general grounds, but its form at finite $\phi$ depends on the modification of the spacetime sigma model there, which we will not be discussing in this section.} 
\eqn[ebphi]{e^{\beta\phi}}
in the SW theory \eqref{mmll}. Here $\beta$ is given by 
\eqn[defbeta]{\beta=-\frac{Q_\ell}{2}+ip ~,}
where the radial momentum $p$ is real for delta-function normalizable operators. Non-normalizable operators can be obtained by continuation to real positive $ip$. The dimension of the operator \eqref{ebphi} in the spacetime CFT is 
\eqn[dimexp]{h_{\ST}=\bar h_{\ST} = -\half \beta(\beta+Q_\ell)= \frac{p^2}{2}+\frac{Q_\ell^2}{8} ~.}
The worldsheet vertex operator corresponding to \eqref{ebphi} is given by 
\eqn[ebphiws]{e^{\beta\phi}~~\longleftrightarrow~~e^{-\varphi-\bar\varphi} \, e^{i(H_\sl+\bar H_\sl)} \, \Phi_{j;m,\bar m}^{(-1)}~,}
where $\Phi_{j;m,\bar m}^{(-1)}$ is a special case of the construction reviewed in section~\ref{sec:review}; see \eqref{aaqq}, \eqref{sl2windingreps}. 
To show that the r.h.s. of \ebphiws\ is BRST invariant, one can use equations~\eqref{aazz}, \eqref{Gsl}, \eqref{sl2windingreps}, \eqref{sltwobos} to obtain the OPE between $G_\sl(z)$ and~\eqref{ebphiws}. We find that it is BRST closed, provided the mass-shell condition,
\eqn[massshell]{m=\bar m=\frac{j(j-1)}{k}+\frac{k+2}{4}}
is satisfied.

The spacetime scaling dimension of the operator \eqref{ebphiws}, \eqref{massshell} is given by 
\eqn[stscale]{h_{\ST}=\bar h_{\ST}=\frac{k}{2}-m=-\frac{j(j-1)}{k}+\frac{k-2}{4}~,}
where we used \eqref{sl2windingreps} and \eqref{globalsltwo}.  Delta-function normalizable operators correspond to 
\eqn[longs]{j=\half+is~,}
which yields
\eqn[hhlong]{h_{\ST}=\bar h_{\ST}=\frac{s^2}{k}+\frac{Q_\ell^2}{8}~,}
where we used \eqref{lslope}. Comparing \eqref{dimexp} with \eqref{hhlong}, we see that they agree if we take 
\eqn[matchwave]{p=s\sqrt\frac{2}{k}=sQ~.}
This relation has a natural interpretation. In the SW theory, the radial wavefunction corresponding to the operator on the l.h.s. of \eqref{ebphiws} is
\eqn[spacewave]{\Psi(\phi)\simeq e^{(\beta+\frac{Q_\ell}{2})\phi}=e^{ip\phi} ~.}
On the other hand, the wavefunction corresponding to the operator on the r.h.s. of \eqref{ebphiws} in string theory on $AdS_3$ is 
\eqn[wswave]{\Psi(\phi)\simeq e^{Q(j-\half)\phi}=e^{isQ\phi}~.}
The condition \eqref{matchwave} is the statement that the worldsheet and spacetime radial wavefunctions coincide. As we will see below, this is a general feature of the holographic operator map. 

\bigskip
\noindent
A few comments about the preceding discussion are useful at this point:
\begin{enumerate}[1)]
\item
We mentioned before that the correspondence \ebphiws can be extended from delta-function normalizable operators \eqref{defbeta}, \eqref{longs} to non-normalizable ones, for which $\beta$ and $j$ are real. This is done by continuing to imaginary $p$, $s$, while maintaining the relation \eqref{matchwave}. 
\item The worldsheet vertex operator on the right hand side of \eqref{ebphiws} corresponds to a particular mode of the operator that is dual to the one on the left hand side. To construct other modes, we can repeatedly apply to it the spacetime $L_{-1}$ and $\bar L_{-1}$ operators, represented in the worldsheet theory by $-J_{\rm sl,0}^-$ and  $-\bar J_{\rm sl,0}^-$, respectively, see \globalsltwo. Equivalently, we can conjugate it by these operators:
\eqn[localop]{
\cO(x,\bar x) = e^{-x J_{\rm sl,0}^- - \bar{x} \bar{J}_{\rm sl,0}^-} \cO(0) \, e^{x J_{\sl,0}^-+\bar{x} \bar{J}_{\rm sl,0}^-}
}
with $\cO(0)$ given by the r.h.s. of \eqref{ebphiws}.
\item 
An interesting special case of the above construction is the operator obtained by sending $\beta\to 0$ in \eqref{ebphi}, the identity operator in the block of the symmetric product, \eqref{mmll}. Note that this operator is non-normalizable, since its wavefunction \spacewave\ goes like $\exp\left(Q_\ell\phi/2\right)$ as $\phi\to\infty$. Looking back at equation~\eqref{defbeta}, we see that this operator corresponds to $ip=Q_\ell/2$. Using the relations \eqref{massshell}, \eqref{longs}, \eqref{matchwave}, we find that it has 
\eqn[jmid]{j=1-\frac{k}{2}~~,~~~~ m=\bar m=\frac{k}{2} ~.}
Thus, we conclude that the worldsheet vertex operator
\eqn[wsidd]{\II^{(-1)}=e^{-\varphi-\bar\varphi}e^{i(H_\sl+\bar H_\sl)}\,\Phi_{1-\frac{k}{2};\frac{k}{2},\frac{k}{2}}^{(-1)}}
corresponds to the identity operator in the block of the symmetric product $\MM^p/S_p$.
\end{enumerate}

\noindent
To see that the operator \eqref{wsidd} is proportional to the identity operator, we proceed as follows. First, we conjugate it as in \eqref{localop}, to construct the operator  
\eqn[idminusone]{
\II^{(-1)}(x,\bar x) = e^{-x J_0^- - \bar{x} \bar{J}_0^-} \II^{(-1)} \, e^{x J_0^-+\bar{x} \bar{J}_0^-} ~.
}
The operator $\II^{(-1)}(x,\bar x)$ has spacetime scaling dimension $(0,0)$. To show that it is proportional to the identity, we want to show that it satisfies 
\eqn[derzero]{\partial_x\II^{(-1)}(x,\bar x)=\partial_{\bar x}\II^{(-1)}(x,\bar x)=0 ~.}
A direct calculation leads to 
\eqn[commJ]{\partial_x\II^{(-1)} = [J_0^-,\II^{(-1)}]=e^{-\varphi-\bar{\varphi}}e^{i \bar{H}_{\sl}}\left(\sqrt{2} \, \psi^3_{\sl} \,
\Phi^{(-1)}_{1-\frac{k}{2};\frac{k}{2},\frac{k}{2}}
+e^{iH_{\sl}} \, j^-_{\sl;0} \, \Phi^{(-1)}_{1-\frac{k}{2};\frac{k}{2},\frac{k}{2}}\right)  ~.}
It turns out that this operator is BRST exact. To see this, consider the BRST exact operator\footnote{Here $j$, $m$ are arbitrary.} 
\eqna[Lmbda]{\big\{Q_{\it BRST},e^{-2\varphi-\bar\varphi}\partial\xi e^{i\bar{H}_{\sl}}\Phi^{(-1)}_{j;m,\bar{m}}\big\}
&= e^{-\varphi-\bar\varphi} \,
e^{i \bar{H}_{\sl}}\left[\sqrt{2}{\left(\frac{k+2}{2}-m\right)} \, \psi^3_{\sl} \,
\Phi^{(-1)}_{j;m,\bar{m}}\right.\cr
\hskip 1.5cm 
+e^{iH_{\sl}} \, j^-_{\sl;0} \,
&\Phi^{(-1)}_{j;m,\bar{m}}+\left.(m\tight+j\tight-1)
\left(\partial e^{i H_{\sl}}+e^{iH_{\sl}}\partial\varphi \right) 
\Phi^{(-1)}_{j;m-1,\bar{m}}\right] ~.}
For $j=1-\frac{k}{2},\, m=\bar{m}=\frac{k}{2}$, \eqref{Lmbda} reduces to \eqref{commJ}. We conclude that $\partial_x\II^{(-1)}(x,\bar x)$ is BRST exact and thus vanishes in correlation functions of BRST invariant operators.

A similar analysis, with left and right-movers exchanged on the worldsheet and in spacetime, shows that $\partial_{\bar x}\II^{(-1)}(x,\bar x)$ vanishes. Thus, the operator $\II^{(-1)}(x,\bar x)$ is a dimension zero operator, whose correlation functions are independent of position. It is natural to interpret it as the identity operator in the block of the symmetric product (summed over all blocks, as in \eqref{singletr}).

In fact, this operator can be thought of as the winding $-1$ representation of the operator $\II$ discussed in section~\ref{sec:review}, equation~\eqref{Ioperator}. 
In the $m,\mbar$ basis, that operator is 
$e^{-\varphi-\bar\varphi}e^{i(H_\sl+\bar H_\sl)}\,\Phi^{(0)}_{1;-1,-1}$.  
We mentioned above that operators with winding zero have in general non-trivial contributions with non-zero winding, and \eqref{wsidd} represents the same operator $\II$ but emphasizes a different winding sector contribution to it. 

One way to see this relation is to note that the operator \eqref{wsidd} has non-singular OPE's with $\psi^3_{\rm sl}$ and $J^3_{\rm sl}$. Thus, it belongs to the coset $\frac{\sltwo}{U(1)}$. Its behavior in the radial direction can be read off by plugging $j=1-\frac k2$ into \eqref{Phimodeasymp}, which gives $\Phi_{1-\frac{k}{2};\frac{k}{2},\frac{k}{2}}^{(-1)}\sim \exp(-\phi/Q)$. This is precisely the behavior of the $N=2$ Liouville superpotential, which is known to be related to the operator $\cI$ \eqref{WXxspace}, \eqref{WX}, by FZZ duality~\rcite{Giveon:1999px,Giveon:1999tq,Giveon:2001up}. In particular, it is normalizable for $k>1$, and non-normalizable for our case, $k<1$.

Note also that the relation between operators in different winding sectors only applies to special operators. As an example, in \eqref{ebphiws} we constructed a worldsheet vertex operator which corresponds to the operator \eqref{ebphi} in the block of the symmetric product discussed above. For $\beta=0$, this vertex operator exists in the winding zero sector as well, but for $\beta\not=0$ it does not. Thus, we see that most operators that exist at non-zero winding do not exist at winding zero. This is closely related to the fact that for $k>1$ the FZZ duality is only valid for normalizable states. For $k<1$, some of those become non-normalizable.

The discussion around equations \derzero-\Lmbda\ has an interesting generalization to non-zero $\beta$ in \eqref{ebphi}. On the field theory side, one has
\eqna[derebphi]{\partial_xe^{\beta\phi}=\beta\partial_x\phi \,  e^{\beta\phi} ~.}
In the worldsheet theory, we compute 
\eqn[Dxebphi]{\left[J_0^-,e^{-\varphi-\bar\varphi} \, e^{i(H_\sl+\bar H_\sl)} \, \Phi_{j;m,\bar m}^{(-1)}\right]=e^{-\varphi-\bar\varphi} \, e^{i\bar H_\sl}\left(\sqrt{2} \, \psi^3_\sl \, \Phi_{j;m,\bar m}^{(-1)}+e^{i H_\sl} \, j^-_{\sl;0} \, \Phi_{j;m,\bar m}^{(-1)}\right) ~,}
where $m,\bar{m}$ are given by~\eqref{massshell}.
Comparing to \eqref{Lmbda}, we see that up to a BRST commutator, we have the duality 
\eqna[dualbb]{\beta\partial_x\phi \, e^{\beta\phi} ~~\longleftrightarrow~~ 
(m+j-1)(\partial\varphi+i\partial& H_\sl)e^{-\varphi-\bar\varphi} \, e^{i(H_\sl+\bar H_\sl)} \, \Phi_{j;m-1,\bar m}^{(-1)}\cr
\hskip 1.5cm 
&{-\sqrt{2}\left(m-\frac{k}{2}\right)e^{-\varphi-\bar\varphi}\psi^3_\sl e^{i\bar{H}_\sl}\Phi_{j;m,\bar m}^{(-1)} ~.}
}
In the limit $\beta\to 0$, in which the l.h.s. goes to zero, the r.h.s. does as well, since $m+j-1$ and $m-k/2$ vanish for the values \eqref{jmid}. Of course, in that limit, the calculation reduces to the one we did before. 
However, since for small $\beta$ both the l.h.s. and the r.h.s. of \eqref{dualbb} go like $\beta$, we can divide them by $\beta$ and then take the limit $\beta\to 0$. In this limit we get the following correspondence:
\eqna[dualbbzero]{\partial_x\phi ~\longleftrightarrow~ e^{-\varphi-\bar\varphi}\left[
\frac{1}{\sqrt{2}(1-k)}(\partial\varphi+i\partial H_\sl) \, e^{i(H_\sl+\bar H_\sl)} \, 
\Phi^{(-1)}_{1-\frac{k}{2};\frac{k}{2}-1,\frac{k}{2}}{-\,\psi^3_\sl\, e^{i\bar{H}_\sl}\,\Phi_{1-\frac{k}{2};\frac{k}{2},\frac{k}{2}}^{(-1)}}\right] ~.
}
Thus, we find that the worldsheet theory contains an operator with spacetime scaling dimension $(1,0)$ (the r.h.s. of \eqref{dualbbzero}) that corresponds in the SW theory (the block of the symmetric product CFT) to the operator $\partial_x\phi$ (the l.h.s.). 

\bigskip
\noindent
We pause again for comments about this result. 
\begin{enumerate}[1)]
\item
In taking the limit $\beta\to 0$ above, we implicitly made an assumption about the holographic map \eqref{ebphiws} described above. That map in general involves a $\beta$-dependent multiplicative constant that we have not determined (and cannot determine without access to information about correlation functions which we currently do not have). The precise value of that constant is not important for our considerations, but we have assumed that it does not vanish as $\beta\to 0$. Of course, we made that assumption already in the discussion of the operator $\II^{(-1)}$ above.
\item 
It is interesting to ask whether the operator \dualbbzero\ is holomorphic. In the boundary theory, this is the question whether $\bar\partial\partial\phi=0$. In the large $\phi$ region we are working in, the answer is expected to be yes. Indeed, in the SW long string effective field theory, $\phi$ is a free scalar field with background charge. However, this holomorphy is expected to be broken in the full theory, since the SW theory must be modified at finite $\phi$, where the symmetric product picture has to smoothly connect to the physics of short strings in $AdS_3$.  
For example, it might be that the $\phi$ field cannot explore that region of field space due to the kind of wall familiar from Liouville theory, the $\frac\sltwo\uone$ coset model (the cigar CFT), \etc. Such walls in general make the $\phi$ field interacting and break the holomorphy of $\partial\phi$. It is thus interesting, that in the worldsheet theory, naively it appears that the operator on the r.h.s. of \dualbbzero\ is holomorphic. 
The reason is that to compute the action of $\partial_{\bar x}$ on it, we need to do exactly the calculation that we did for $\II^{(-1)}$ above. That calculation appears to only involve (non-trivially) the right-movers on the worldsheet, and for those, the operator under consideration is exactly the same as there. Thus, it appears  that \dualbbzero\ is holomorphic. We will return to this issue in section~\ref{sec:wall}, and will see how this tension is resolved.
\item
In equation \dualbbzero\ we constructed a holomorphic operator in the spacetime CFT by studying worldsheet operators in the sector with $AdS_3$ winding $-1$. There is another worldsheet construction of holomorphic operators in the spacetime CFT, in the sector with winding zero~\rcite{Kutasov:1999xu}, reviewed briefly in section~\ref{sec:STsymm}. It is natural to ask what is the relation between the two constructions. We will see below that some holomorphic operators appear in both sectors, and some only appear in the sector with non-zero winding. The operator \dualbbzero\ does not have a counterpart in the winding zero sector and thus belongs to the second class.  We will see below that the holomorphic operators of the first class are conserved, while those of the second class are not.
\end{enumerate}

\vskip .5cm


\subsection{Operators in the block theory that involve the \texorpdfstring{$\bS^3$}{}}
\label{sec:sphereops}

We next move on to some generalizations of the holographic correspondence above. One simple generalization is to add non-trivial wavefunctions on the three-sphere. In the boundary theory this corresponds to considering operators of the form 
\eqn[ebphivjm]{v_{j';m',\bar m'} \, e^{\beta\phi}~,}
where $v_{j';m',\bar m'}$ is a primary of the $SU(2)_L\times SU(2)_R$ affine Lie algebra describing a non-trivial spherical harmonic on the sphere, as before. The dimension of this operator in the spacetime CFT is 
\eqn[dimvjm]{h_{\ST}=\bar h_{\ST}=-\half\beta(\beta+Q_\ell)+\frac{j'(j'+1)}{n} ~.}
The dual operator in the worldsheet string theory is    
\eqn[ebphivjmws]{v_{j';m',\bar m'} \, e^{\beta\phi}~~\longleftrightarrow~~
e^{-\varphi-\bar\varphi} \, e^{i(H_\sl+\bar H_\sl)} \, v_{j';m',\bar m'} \, \Phi_{j;m,\bar m}^{(-1)} ~.}
Imposing the worldsheet consistency conditions as before, we get 
\eqn[massshellvjm]{m=\bar m=\frac{j(j-1)}{k}+\frac{k+2}{4}-\frac{j'(j'+1)}{n} ~.}
It is straightforward to check that this operator has the correct spacetime scaling dimension, \dimvjm if $j$ is related to $\beta$ as before, via \eqref{defbeta}, \eqref{longs}, \eqref{matchwave}. For $j'=0$ \ebphivjm\ reduces to \ebphi, and \ebphivjmws\ to \ebphiws. 

Another interesting generalization concerns holomorphic operators. In \dualbbzero\ we have constructed an operator with spacetime scaling dimension $(1,0)$. It is natural to ask whether there are other operators with that scaling dimension in the sector with winding $-1$. In particular, the theory contains an $SU(2)$ current algebra discussed in section~\ref{sec:review}, and it is interesting to ask whether this sector contains additional spacetime currents that transform in the adjoint representation of that $SU(2)$. We turn next to this question. 

One vertex operator that transforms in the adjoint of the $SU(2)$ of~\rcite{Kutasov:1999xu} is  
\eqn[suws1]{\cK^{a\,(-1)} = e^{-\varphi-\bar\varphi} \, \psi_\su^a  \, e^{i\bar H_\sl} \, \Phi_{j;m,\bar m}^{(-1)}~,}
where $a=1,2,3$ is the $SU(2)$ adjoint label, and $(m,\bar m)$ are given by \massshell.
The operator~\eqref{suws1} corresponds in the spacetime CFT to one with $h_{\ST}-\bar h_{\ST}=1$. We believe that it corresponds to the following operator in the block of the symmetric product,
\eqn[totsu2]{\cK^a(x) e^{\beta\phi} ~.}
Here, $\cK^a(x)$ is the generator of the affine Lie algebra of level $n$ in the block, and $\beta$ is related to $j$ by equations \defbeta, \longs, \matchwave, as before. In particular, in the limit $\beta\to 0$ we get the correspondence  
\eqn[dualcurr]{\cK^{a} ~~\longleftrightarrow~~ e^{-\varphi-\bar\varphi} \, \psi_\su^a \,  e^{i\bar H_\sl} \, \Phi^{(-1)}_{1-\frac{k}{2};\frac{k}{2},\frac{k}{2}}  ~. }
In other words, the situation is like that for the operator $\cI$ above. The currents on the r.h.s. of \dualcurr are the winding $-1$ representations of the $SU(2)$ currents~\eqref{Kst} (or rather, the operator in the $m$ basis that results from the LSZ procedure described in the discussion around~\eqref{modphiop}) resulting from FZZ duality. They are non-normalizable, in agreement with the fact that the $\sltwo$ invariant vacuum is not a normalizable state for $k<1$.  We will see below that they are conserved, as expected.

We note in passing that while for $\beta=0$ the vertex operator~\eqref{suws1} describes an operator that also exists in the sector with winding zero, given by equation~\eqref{Kst}, for non-zero $\beta$ the operators~\eqref{suws1}, \eqref{totsu2} do not have an analog in the $w=0$ sector. This is another example of the phenomenon mentioned above for the  operators~\ebphiws, \massshell, which do not correspond to any operators in the $w=0$ sector, except for $\beta=0$, where they give the $w=-1$ representation of the operator $\II$, equation~\eqref{Ioperator}. We will see below that this pattern is much more general. 

Coming back to the discussion of spin one operators in the spacetime CFT, we have constructed before the operators $\partial\phi$ and $\cK^a$ in the winding one sector, equations~\eqref{dualbbzero} and~\eqref{dualcurr}. It is natural to ask whether there are additional operators with this general structure, in particular ones that transform in the adjoint representation of $SU(2)$.  A natural ansatz is 
\eqn[moresu2]{e^{-\varphi-\bar\varphi} \, K^a \, e^{i(H_\sl+\bar H_\sl)} \,
\Phi^{(-1)}_{1-\frac{k}{2};\frac{k}{2}-1,\frac{k}{2}}~,}
where $K^a$ is a worldsheet current (\ie\ a dimension $(1,0)$ holomorphic operator). 
One possible choice for this current is 
\eqn[fermsu2]{
K^a 
= \alpha \, j^a_\su -\frac i2\epsilon^{abc}\psi^b_\su\psi^c_\su
\equiv \alpha \, k^a_{\rm bos} + k^a_{\rm ferm}~,
}
where the two terms on the r.h.s.
are the bosonic and fermionic worldsheet $\sutwo$ currents (see section~\ref{sec:currentalg}), and $\alpha$ is a constant to be determined. 

In order for the operator \eqref{moresu2} to be BRST invariant, the current \eqref{fermsu2} must be the bottom component of a worldsheet superfield. One can use~\eqref{bbhh} to check that its OPE with the superconformal generator~\eqref{Gsu} is
\eqn[opeGJ]{G_\su(z) K^a(0)\sim \sqrt{\frac{2}{n}}\left(\alpha\frac{n-2}{2}+1\right)\frac{\psi^a_\su}{z^2}+\cdots~.}
Thus, for $\alpha=-\frac{2}{n-2}$ the $1/z^2$ term in \eqref{opeGJ} is absent, and the operator \eqref{moresu2}, which in this case takes the form
\eqn[onemoresu2]{e^{-\varphi-\bar\varphi}\left(k^a_{\rm ferm}-\frac{2}{n-2} k^a_{\rm bos}\right) e^{i(H_\sl+\bar H_\sl)} \,
\Phi^{(-1)}_{1-\frac{k}{2};\frac{k}{2}-1,\frac{k}{2}}~,}
is BRST invariant. It corresponds to a second $SU(2)$ current (in addition to \eqref{dualcurr}), which according to the  discussion of this section is holomorphic, up to BRST commutators. 

A third $SU(2)$ current is obtained by taking 
\eqn[thirdsu2]{K^a=\psi^a_\su\psi^3_\sl ~.}
Plugging \eqref{thirdsu2} into \eqref{moresu2}, and requiring that the resulting operator is BRST invariant, we find the following. The $SU(2)$ part of the operator is an $N=1$ superconformal primary, but the $\sltwo$ part is not (the OPE of the superconformal generator with $\psi^3_\sl e^{iH_\sl}\Phi^{(-1)}_{1-\frac{k}{2};\frac{k}{2}-1,\frac{k}{2}} $ has a $1/z^2$ term). This problem can be fixed by replacing 
\eqn[replace]{\psi^3_\sl \, e^{iH_\sl} \,
\Phi^{(-1)}_{1-\frac{k}{2};\frac{k}{2}-1,\frac{k}{2}}
~~\longrightarrow~~ 
\psi^3_\sl \, e^{iH_\sl} \, \Phi^{(-1)}_{1-\frac{k}{2};\frac{k}{2}-1,\frac{k}{2}}+\frac{\sqrt{2}}{k-2}e^{2 i H_\sl}\Phi^{(-1)}_{1-\frac{k}{2};\frac{k}{2}-2,\frac{k}{2}}~,}
where the relative constant has been chosen such that the $1/z^2$ term cancels between the two contributions. Plugging \replace\ into \eqref{moresu2} we find a third dimension $(1,0)$ current that transforms in the adjoint of $SU(2)$ and takes the form 
\eqn[finalthird]{e^{-\varphi-\bar\varphi} \,\psi^a_\su \,e^{i\bar H_\sl}\left(
\psi^3_\sl \, e^{iH_\sl} \, \Phi^{(-1)}_{1-\frac{k}{2};\frac{k}{2}-1,\frac{k}{2}}+\frac{\sqrt{2}}{k-2}e^{2 i H_\sl}\Phi^{(-1)}_{1-\frac{k}{2};\frac{k}{2}-2,\frac{k}{2}}\right) ~.
}
Thus, we see that string theory on $AdS_3\times \bS^3$ contains, in the winding $w=-1$ sector, three sets of dimension $(1,0)$ operators that transform in the adjoint representation of the $SU(2)_L$ current algebra \Kst, and are holomorphic modulo BRST commutators. A natural interpretation of these operators in the SW theory ~\eqref{mmll}, \eqref{sutwon}, is the following. One, given by \eqref{dualcurr}, appears to be the $w=-1$ version of the spacetime $\sutwo$ current~\eqref{Kst}. The other two combinations\footnote{The precise linear combinations remain to be determined.} of \dualcurr, \eqref{onemoresu2}, \eqref{finalthird},  form an $SO(4)_1=\sutwo_1\times\sutwo_1$ current algebra acting on four fermions.  These fermions transform as a vector under $SO(4)$, mirroring the worldsheet structure. We can denote them by $\chi^i$, with $i=1,2,3,4$ running over the four dimensions of the target space $\bR_\phi\times\bS^3$. In terms of these fermions, the $SO(4)_1$ currents take the usual form 
\eqn[sofourvec]{K^{ij}=\chi^i\chi^j~.}
In the $\sutwo\times\sutwo$ language, the fermions $\chi$ are bispinors $\chi^{\alpha\dot\alpha}$, $\alpha,\dot\alpha=\pm$, and the currents \sofourvec\ take the form 
\eqn[sofour]{
k^{a}_{(1)} = 
\chi^{\alpha\dot\alpha}\chi^{\beta\dot\beta}\sigma^a_{\alpha\beta}\epsilon_{\dot\alpha\dot\beta}
~~,~~~~
k^{a}_{(2)} = \chi^{\alpha\dot\alpha}\chi^{\beta\dot\beta}\epsilon_{\alpha\beta}^{~}\sigma^a_{\dot\alpha\dot\beta} ~,
}
where $\sigma^a$ are the Pauli matrices and $\epsilon$ the two-dimensional Levi-Civita tensor.  There is a similar structure on the right-moving side of the spacetime CFT, obtained by exchanging the left and right-movers on the worldsheet. 

It is worth noting that in~\eqref{onemoresu2}, \eqref{finalthird} we constructed the holomorphic operators, but it is easy to modify them to include a factor of $e^{\beta\phi}$, as in~\eqref{totsu2}, by changing $j$ and adjusting some constants. It is also possible to add a non-trivial spherical harmonic on the sphere, as in \eqref{ebphivjm}, \eqref{ebphivjmws}. We will not describe the details here. 

Continuing in our program of mapping out the low-lying operators in $AdS_3\times \bS^3$ and their analogues in the dual SW theory, we next consider the operators
\eqn[kkbar]{e^{-\varphi-\bar\varphi}\psi^a_\su\bar\psi^b_\su \Phi_{j;m,\bar m}^{(-1)}}
with $m$, $\bar m$ given again by \eqref{massshell}. 
This operator looks the same as~\eqref{dualcurr} as far as the left-movers are concerned, but with the right-movers treated the same as the left-movers, which is not the case in~\eqref{dualcurr}. It is natural to conjecture that the dual operator in the SW theory is a left-right symmetric analog of \eqref{totsu2}, 
\eqn[kkbarSW]{\cK^a(x)\bar \cK^b(\bar x)e^{\beta\phi}}
with $\beta$ related to $j$ as before \eqref{defbeta}, \eqref{longs}, \eqref{matchwave}. 

Another interesting set of operators is 
\eqn[kkbarhh]{e^{-\varphi-\bar\varphi}\psi^a_\su\bar\psi^b_\su e^{i(H_\sl+\bar H_\sl)}\Phi_{j;m,\bar m}^{(-1)} ~.}
In this case, BRST invariance imposes the constraint 
\eqn[massshellnew]{m=\bar m=\frac{j(j-1)}{k}+\frac{k}{4} ~.}
The spacetime dimension of these operators is 
\eqn[dimspace]{h_{\ST}=\bar h_{\ST}=-\half\beta(\beta+Q_\ell) +\half}
with $\beta$ related to $j$ as before. 

The operators \eqref{kkbarhh} transform in the $(3,3)$ representation of the total $SU(2)_L\times SU(2)_R$. To understand them better, it is useful to note that there are additional operators with the same dimensions that are obtained by replacing $\bar\psi^b_\su$ by $\bar\psi^3_\sl$. For the resulting operators to be BRST invariant, one has to make a modification similar to that described around equation~\eqref{replace}. 
This gives a triplet of operators in the $(3,1)$ of $SU(2)_L\times SU(2)_R$. A similar replacement for the left-movers gives a $(1,3)$, and a $(1,1)$. Altogether, this construction gives rise to sixteen operators that transform as $2\otimes 2=3\oplus1$ under $SU(2)_L$ and similarly under $SU(2)_R$.

To understand the operators \kkbarhh\ and their generalizations described above in the spacetime theory, it is natural to use the description in terms of the fermions $\chi$ around equation~\sofourvec. They correspond to the operators 
\eqn[fermbil]{\chi^i\bar\chi^j e^{\beta\phi}}
with $i,j=1,2,3,4$, as before. In terms of the bispinor description they involve the bilinears
$\chi^{\alpha\dot\alpha}\bar\chi^{\beta\dot\beta}$. To understand the $3\oplus1$ decomposition under the total $SU(2)_L$ found in the worldsheet construction, note that the part of it that acts on the fermions is the diagonal $SU(2)_2$ in the product $SU(2)_1\times SU(2)_1$. Under this $SU(2)$, the fermions $\chi^i$ transform as a $3\oplus1$, where the $3$ is the first three components of the vector $\chi^i$ (say), and the $1$ the remaining one.

So far we discussed the $(\NS,\NS)$ sector of the SW CFT, which we saw comes from the $(\NS,\NS)$ sector of the worldsheet theory. It is interesting to examine the $(\!R,\!R)$ sector of the spacetime theory, which comes from the $(\!R,\!R)$ sector of the worldsheet one. This motivates us to consider vertex operators of the form\footnote{Here we set the $\phi$ exponent to zero. It is straightforward to consider operators with a nontrivial $\phi$ profile as well.}
\eqna[dimquarter]{
\sum_{\etab,\bar\etab} 
\Big[e^{-\half\varphi-\half\bar\varphi} &\,
e^{i(\eta_3 H_3 +\eta_\su H_\su +\bar\eta_3 \bar H_3 + \bar\eta_\su \bar H_\su)} 
\\[-.3cm]
\times\Big(&
c_1(\etab,\bar\etab)
e^{\frac i2 H_\sl + \frac i2 \bar H_\sl} \,
\Phi^{(-1)}_{1-\frac k2;\frac k2+\frac14,\frac k2+\frac14}
+c_2(\etab,\bar\etab)
e^{\frac {3i}2 H_\sl + \frac i2 \bar H_\sl} \,
\Phi^{(-1)}_{1-\frac k2;\frac k2-\frac34,\frac k2+\frac14}
\\[.2cm]
&
+c_3(\etab,\bar\etab)
e^{\frac i2 H_\sl + \frac{3i}2 \bar H_\sl} \,
\Phi^{(-1)}_{1-\frac k2;\frac k2+\frac14,\frac k2-\frac34}
+c_4(\etab,\bar\etab)
e^{\frac{3i}2 H_\sl + \frac{3i}2 \bar H_\sl} \,
\Phi^{(-1)}_{1-\frac k2;\frac k2-\frac34,\frac k2-\frac34}
\Big)\Big]
}
with $\etab=(\eta_3,\eta_\su)$ the spinor weights $\eta_3,\eta_\su=\pm\hf$ and similarly for $\bar\etab$.  These operators have spacetime dimension $(\frac14,\frac14)$. BRST invariance on the left further relates $c_1$ to $c_2$, and $c_3$ to $c_4$, while on the right it relates $c_1$ to $c_3$, and $c_2$ to $c_4$.  All told there are sixteen solutions, which we can label by the sixteen coefficients $c_1(\etab,\bar\etab)$.  
The GSO projection selects from among these the eight $(R+,R+)\oplus(R-,R-)$ operators in type 0B, or the eight $(R+,R-)\oplus(R-,R+)$ operators in type 0A, and thus correlates the left and right spinor chiralities $\etab,\bar\etab$ in the four dimensions transverse to the long string.

In the spacetime theory, one can think of the operators \dimquarter\ as follows. The four left(right)-moving fermions $\chi^i(\bar\chi^i)$ introduced around equation~\sofourvec\ can be bosonized in terms of two left (right) moving scalars $H_i(\bar H_i)$. One can then construct $(\!R,\!R)$ sector operators of the form 
\eqn[spacetimerr]{e^{\frac{i}{2}(\pm H_1\pm H_2\pm \bar H_1\pm \bar H_2)}}
in the spacetime SW theory. This group of sixteen dimension $(\frac14,\frac14)$ operators corresponds to the sixteen operators \eqref{dimquarter} in the bulk string theory. 

In string theory we need to further impose a GSO projection on the worldsheet theory, due to the fact that the sixteen operators \dimquarter\ are not all mutually local. Similarly, in the spacetime CFT we need to further project \spacetimerr, for the same reason. In both, this leads to two inequivalent theories, 0A and 0B, as in critical string theory~\rcite{Polchinski:1998rr}. 

The analysis of low lying delta-function normalizable and non-normalizable operators on $AdS_3\times S^3$ can be summarized as follows. We have constructed the holographic map for the type 0 tachyon, \ebphivjmws, the gravity sector \dualcurr, \kkbar, \kkbarSW, and some low lying excited string states \dualbb, \eqref{onemoresu2}, \finalthird, \kkbarhh, \fermbil. We also mapped some $RR$ gauge fields \dimquarter\ to their analogs in the SW theory \spacetimerr. 

The resulting structure points to the SW theory being a bosonic sigma-model on $\bR_\phi\times\bS^3$ and four fermions $\chi^i$. The sum over spin structures for these fermions  is the standard non-chiral sum, that implements the projection $(-1)^F=(-1)^{\bar F}$. As usual, in the $(\NS,\NS)$ sector this means that only operators with an even number of fermions survive. In the $(\!R,\!R)$ sector there are two possibilities, the 0A and 0B theories, which differ in the way the left and right-moving spin fields are paired. The choice of GSO projection in the block of the symmetric product mirrors that of the worldsheet theory.

This conclusion about the GSO projection can be proven by studying the thermal partition sum of the theory, using worldsheet techniques. It is particularly simple to do that after deforming the theory by the single-trace $T\bar T$ deformation studied in~\rcite{Giveon:2017nie}. We discuss this deformation in section~\ref{sec:deflin}, and the thermal partition sum can be computed using the techniques of~\rcite{Hashimoto:2019wct}.

One interesting feature of our analysis is the presence in the theory of a large number of holomorphic operators. We have focused on dimension $(1,0)$ operators, such as $\partial\phi$ \dualbbzero\ and the three $SU(2)$ currents \dualcurr, \eqref{onemoresu2}, \finalthird, but we expect there to be an infinite tower of such operators, with dimensions of the form $(n,0)$ for $n>1$. We leave the construction of the corresponding vertex operators to future work.

From the point of view of the spacetime CFT, these operators correspond to the holomorphic operators in the SW theory $\bR_\phi\times\bS^3$. Since that theory is free at large positive $\phi$, it is natural that it has an infinite number of holomorphic operators in this limit. However, in the full theory we expect that the deviations from the free approximation lead to violations of holomorphy. 

Interestingly, in the worldsheet analysis we found that the vertex operators we constructed give operators that appear to be holomorphic in the full spacetime CFT. We will return to the question of the fate of these holomorphic operators in section~\ref{sec:wall} below. For now, we note that the operators that are expected to be conserved in the full theory, such as \dualcurr, come from the gravity sector of the bulk theory, while the ones that are expected to be broken, such as~\eqref{dualbbzero}, \eqref{onemoresu2}, and~\eqref{finalthird}, come from excited string modes (or oscillator states).


\subsection{Higher winding sectors }
\label{sec:higherwinding}

We have concentrated in this section on the unit winding sector, which corresponds to the block of the symmetric product. The analysis of~\rcite{Argurio:2000tb} then shows how the higher winding sectors build up the twisted sectors of the symmetric product.  The $L_0$ constraint in the winding $w$ sector was written in \eqref{L0excited} above. The corresponding spacetime scaling dimension  can be written in the form 
\eqna[windingdim]{
h_{\ST} &= -\Big(m + \frac{k}{2} w\Big) = -\Big[\frac{h_1}w + \frac k4\Big( w-\frac 1w \Big) \Big]~,
\\[.2cm]
h_1 &= -\frac{j(j-1)}{k} +\frac{j'(j'+1)}{n} + \big(N_\osc-\delta_0\big) + \frac{k}4  ~.
}
Here we restrict to negative $w$, since in our conventions vertex operators with $w<0$ correspond to in-states in spacetime, while their conjugates with $w>0$ describe out-states.  For example, the spacetime dimensions on the first line of~\eqref{windingdim} are given by the expression on the second at $w=-1$. The latter (given by $h_1$ on the second line) was discussed at some length above -- it is the spectrum of the SW theory. 

Interestingly, the first line of \windingdim\ with $w=-|w|$ is also the expression for the dimensions of operators in the  $\bZ_{|w|}$ twisted sector of a symmetric orbifold of a CFT whose building block is the SW theory~\rcite{Klemm:1990df,Argurio:2000tb}. Thus, we conclude that the scaling dimensions of delta-function normalizable and non-normalizable operators in sectors with $w\not=0$ in string theory on $AdS_3\times \bS^3$ agree with that of the symmetric product of the SW theory. In the case of delta-function normalizable operators this was discussed before \eg\ in~\rcite{Argurio:2000tb,Giveon:2005mi,Chakraborty:2019mdf}, but for the non-normalizable ones the statement is new. As we saw, non-normalizable operators play an important role in the theory. For example, they are the ones that give rise to the infinite set of holomorphic operators mentioned in the previous subsection.

In the discussion around equation \eqref{matchwave} we saw that the radial wavefunctions of vertex operators with winding $w=-1$ match those of the corresponding operators in the building block of the dual symmetric product. Another way of saying that is that the radial coordinate in $AdS_3$ can be identified with that in the SW block theory. 

This correspondence of radial profiles can be extended to other values of $\sltwo$ winding $w$, where the dual spacetime operator is in the $\bZ_{|w|}$ twisted sector of the symmetric product spacetime CFT.  Consider a cycle $\tau$ of length $|w|$ that cyclically twists together copies $\tau(1),...,\tau(|w|)$ of the block theory.
The ``center-of-mass'' of $\bR_\phi$ is the canonically normalized $\bZ_{|w|}$ invariant scalar
\eqn[phizero]{
\phi_0 = \frac{\phi_{\tau(1)}+...+\phi_{\tau(|w|)}}{\sqrt{|w|}}~,
} 
which has a linear dilaton 
\eqn[Qtilde]{\Qtil_\ell = \sqrt{|w|} \,Q_\ell~.}
The $\phi_0$ dependence of the twist operator takes the form
\eqn[twistphidep]{
\exp\big[-\hf\Qtil_\ell\, \phi_0\big]\,\Psi\big(\phi_0 \big)
~~,~~~~
\Psi\big(\phi_0 \big) = 
\exp\Big[ \frac{Q_\ell(j_\ST-\half)}{\sqrt{|w|}}\,\phi_0\Big]~,}
where $\exp[-\hf\medtilde Q_\ell\, \phi_0]$ is the usual string coupling dependence, and $\Psi(\phi_0)$ is the wavefunction for the center-of-mass coordinate \phizero.  
For further details on the construction of the twist operators, see Appendix~\ref{app:twistops}.
Furthermore, if we set
\eqn[jSTjWS]{Q_\ell\Big(j^{~}_{\ST}-\half\Big)=Q\Big(j^{~}_{\WS}-\frac{1}{2}\Big)~,}
then the spacetime scaling dimension~\eqref{hwbosLa} of the cyclic twist operator matches that of the worldsheet operator, written in~\eqref{windingdim}. 
Note that for $|w|=1$, the expressions above reproduce \eqref{matchwave}. 

One way of interpreting the wavefunction scaling~\eqref{twistphidep} and worldsheet/spacetime map \eqref{jSTjWS} is as follows. Plugging \jSTjWS\ into \twistphidep, we find that the wavefunction $\Psi$ can be naturally written as \eqn[newphi]{\Psi=\exp\Big[Q\left(j_\WS-\half\right)\,\phi_{\rm ave}\Big]~,
} 
where 
\eqn[phiave]{
|w|\,\phi_{\rm ave} \equiv \sqrt{|w|}\,\phi_0 = \phi_{\tau(1)}+...+\phi_{\tau(|w|) ~.}
}
The wavefunction \newphi\ describes an object in the symmetric product CFT, but we used the worldsheet/spacetime map \eqref{jSTjWS} to write it in terms of the worldsheet variables $(Q,j_\WS)$. 

The form of \newphi\ makes it natural to identify $\phi_{\rm ave}$ with the worldsheet coordinate $\phi$ describing a long string with winding $|w|$, since then the worldsheet and spacetime wavefunctions coincide, generalizing the discussion of the $|w|=1$ case above. As we point out next, this identification is also natural from the perspective of locality of string joining/splitting. 

Indeed, in the bulk theory one can consider a process in which two strings with winding $w_1$ and $w_2$ join to form a string of winding $w_1+w_2$. This process can occur locally in $\phi$, \ie\ both the two initial strings and the final one are located at the same value of $\phi$. 

From the point of view of the boundary theory, the above process corresponds to a merging of two operators in the $\bZ_{w_1}$ and $\bZ_{w_2}$ twisted sectors into a single operator in the $\bZ_{w_1+w_2}$ twisted sector. In this process, the center of mass coordinates of the two initial operators, $\phi_1+\dots+\phi_{w_1}$ and $\phi_{w_1+1}+\dots+\phi_{w_1+w_2}$ combine into the center of mass coordinate of the final one, $\phi_1+\cdots+\phi_{w_1}+\phi_{w_1+1}+\dots+\phi_{w_1+w_2}$. In terms of the definition of $\phi_{\rm ave}$ \phiave, this merging corresponds to 
\eqn[bulklocal]{
w_1 \phi_{\rm ave}^{(w_1)} + w_2 \phi_{\rm ave}^{(w_2)} = (w_1+w_2)\phi_{\rm ave}^{(w_1+w_2)} ~.
}
The identification of $\phi_{\rm ave}$ with the worldsheet $\phi$ is consistent with the locality of the process in $\phi$, since \bulklocal\ is satisfied if we set all three $\phi_{\rm ave}$ in it to the same value, $\phi$, and identify this value with the position in the bulk where this process takes place. 

To summarize, the symmetric product structure described above provides a natural description of non-normalizable and delta-function normalizable operators in sectors with non-zero winding of string theory on $AdS_3\times \bS^3$. An interesting question is what is the role of the sector with winding zero described in section~\ref{sec:type0short} in this picture. We will return to this question in section~\ref{sec:discussion}.

In the particular case of type 0 string theory, the $w=0$ sector contains a BF violating closed string tachyon~\VNS, \eqref{Lzerotach}, and therefore the theory is expected to be unstable.  To avoid this instability, we next turn to theories with chiral GSO projections, in which the tachyon can be eliminated.


\section{Type II theories}
\label{sec:typeIIshort}

As mentioned above, the type 0 string theories on $AdS_3\times\bS^3$ constructed in sections~\ref{sec:type0short}, \ref{sec:type0long} are unstable. One way to ensure stability is to impose spacetime supersymmetry, which involves a chiral GSO projection.  In this section, we construct the resulting type II theories and discuss aspects of their spectrum.


\subsection{Chiral GSO}
\label{sec:chiralGSO}

The standard chiral GSO projection used in critical string theory, $(-1)^F=1$, with $(-1)^F$ defined in~\eqref{Fnum}, is anomalous in non-critical dimensions.  In the attempt to construct the twisted, $(\NS,\!R)$ and $(\!R,\NS)$, sectors of such an orbifold, one is combining Ramond states of one worldsheet chirality with NS states of the other. Outside the critical dimension, the worldsheet conformal dimensions of the two sectors do not differ by an integer, as we see for example from~\eqref{L0excited}, where there is a mismatch by $\delta_{0,\NS}-\delta_{0,R}=\frac14$ that cannot be cancelled by the other terms. This leads to a (worldsheet) torus partition function that fails to be invariant under $\tau\to\tau+1$, and correlation functions of vertex operators that are not single-valued. The condition of equality of the fractional part of the left and right conformal dimensions of states and operators is known as level-matching~\rcite{Vafa:1986wx}. Satisfying this condition is sufficient to guarantee that a chiral GSO is anomaly-free~\rcite{Freed:1987qk}. 

One way to ensure level matching is to require spacetime supersymmetry. A spacetime supercharge is obtained by (contour) integrating a holomorphic dimension one Ramond sector worldsheet operator, the supersymmetry current. Starting with a type 0 theory, and demanding that all operators are mutually local with respect to the supersymmetry current can be thought of as a chiral orbifold that gives, by construction, a spectrum that satisfies level matching. This is obviously true in the untwisted sector of the orbifold, which is a subset of the original type 0 spectrum. It is also true, by construction, in the twisted sectors, which are obtained by applying the supercharge to the original $(\NS,\NS)$ and $(\!R,\!R)$ sector operators. This gives rise to $(\NS,\!R)$ and $(\!R,\NS)$ sectors, which consist of spacetime fermions. 

As reviewed in section~\ref{sec:GKP}, spacetime supersymmetry is also natural from the way the models we study appear in string theory. Before adding the strings, the near-horizon geometries \twodlst\ preserve $(2,2)$ supersymmetry with supersymmetry currents that were constructed in~\rcite{Kutasov:1990ua,Giveon:1999zm}. After adding the strings, the supersymmetry is enhanced to $(2,2)$ superconformal, with worldsheet supersymmetry currents that were constructed in~\rcite{Giveon:1999jg,Berenstein:1999gj,Giveon:2003ku}. They are given by: 
\eqna[susyops]{
S_r^\pm &= \exp\bigg[-\frac\varphi 2 +ir\big(H_\sl\mp H_3\big) \pm\frac {i\,a}2\, Z \pm \frac{i}{\sqrt{2k}}\,Y \bigg]~,
\\[.2cm]
J_R &= i\sqrt{2k}\, \partial Y~,
}
where $a=\sqrt{1-\frac2n}$ is the normalization of the $\frac\sutwo\uone$ R-current (see equation~\eqref{su2bos}).  
The following comments are useful for understanding \susyops.
\begin{enumerate}[1)]
\item 
The fields appearing in \susyops\ were defined in sections~\ref{sec:review}, \ref{sec:type0short}. In particular, the bosonized fermions were defined in 
\sltwobos, \Hdefs, while the scalars $Y$ and $Z$ were defined in 
\eqref{su2bos}.  
\item 
In this formula, the supersymmetric $SU(2)$ WZW model has been decomposed into the supersymmetric $\frac{SU(2)}{U(1)}$ coset and a $U(1)$, as in~\eqref{pfdecompsq}. This is useful, since, as we will see, the natural starting point for the chiral GSO projection that gives a supersymmetric theory is type 0 on a squashed $\bS^3$, denoted $\sqsphere$, with a particular value of the squashing parameter. One can generalize the construction to other values of the squashing parameter, including the unsquashed case, but we will be mainly interested in the supersymmetric case.
\item 
The parameter $r$ in \susyops\ takes two values, $r=\pm\hf$. Contour integrating the corresponding worldsheet supercurrents gives rise to the global superconformal charges in the spacetime theory, $\cG^\pm_{r}$. 
\item 
The zero mode of the worldsheet current $J_R$ in \susyops\ is the $U(1)_R$ that appears in the global spacetime $N=2$ superconformal algebra. It is proportional to the $SU(2)$ generator $J^3_{\rm su}$, but they are normalized in a different way, \eqref{su2bos} versus \susyops. The normalization of $J_R$ is fixed by the $N=2$ superconformal algebra, in particular the fact that the supercharges in \susyops\ must have charge $\pm 1$, while that of $J^3_{\rm su}$ is fixed by the $SU(2)$ algebra. 
\item 
From the expression for the global $N=2$ (super)charges \susyops, one can construct the spacetime supercurrents, following~\rcite{Kutasov:1999xu}. One finds 
\eqna[Gst]{
\cG^\pm(x) &\simeq \int\!d^2z\, 
\left[\left(S^\pm_\half-x S^\pm_{-\half}\right)\partial_x\Phi_1(x,\bar x;z,\bar z)-2 S^{\pm}_{-\half}\Phi_1(x,\bar x;z,\bar z)\right]
e^{-\bar\varphi}\,\bar\psi(\bar x;\bar z) ~,
\\
\cJ_R(x) & \simeq  \int\!d^2z\, 
e^{-\varphi-\bar\varphi}\, \psi^3_\su\, \bar\psi(\bar x;\bar z)\,
\Phi_1(x,\bar x;z,\bar z)  ~.
}
The operators $\cG^\pm(x)$ are (spacetime) dimension $\left(\frac32,0\right)$ holomorphic operators, corresponding to the $N=2$ superconformal generators. $\cJ_R(x)$ is the $U(1)_R$ current; it has dimension $(1,0)$. The operators \eqref{Gst} are normalized as in~\rcite{Kutasov:1999xu}, by requiring that they satisfy the $N=2$ superconformal algebra. The global supercharges \susyops\ are obtained by expanding the operators $\cG^\pm(x)$ in modes in the usual way,
\eqn[expandsuper]{\cG^\pm(x)=\sum_r \cG^\pm_r x^{-r-\frac32}}
and similarly for $\cJ_R(x)$.
\item An inspection of the currents $S_r^\pm$ above in comparison to~\eqref{pfdecomp}-\eqref{paramrel} reveals that they involve the squashed $\sutwo$ vertex operator $V^{(\eta_\su,\bar\eta_\su,w',\bar w')}_{j';m',\mbar'}$ of equation~\eqref{pfdecompsq} with 
\eqn[Sqnums]{
R=\sqrt{n+1}
~,~~~
j'=m'=\mbar'=0
~,~~~
\eta_\su = 1
~,~~~
\bar\eta_\su = \hf
~,~~~
w'=\bar w'=\hf ~, 
}
or its charge conjugate. In other words, the naive spin field built out of the bosonized worldsheet fermions has been decorated with a half unit of spectral flow in the total vector current $J^3_\su+\bar J^3_\su$, which flows simultaneously by a half unit each in the bosonic~\eqref{jbos} and fermionic~\eqref{sutwobos} contributions.  This extra spectral flow contribution results in a level-matched operator.
\end{enumerate}

\noindent 
The vertex operators of the supersymmetric theory are obtained by starting with the type 0 theory on $\sqsphere$ at the particular point $R=\sqrt{n+1}$ along the marginal line parametrized by the squashing parameter $R$ (see the discussion around eq.~\eqref{squonespec}), demanding that all states are mutually local with respect to $S_r^\pm$ \susyops, and adding the twisted sector operators. One way to exhibit the geometric action of this $\bZ_2$ orbifold is to consider a general vertex operator, which carries ghost picture, bosonized worldsheet fermion, $Y$ and $Z$ charges 
\begin{align}
\label{etaYZcharges}
&\exp\Big(-q_\varphi\varphi-\bar q_{\bar\varphi}\bar\varphi 
\tight+ i\eta_\sl H_\sl \tight+i\eta_3 H_3 \tight+   i\bar\eta_\sl \bar H_\sl \tight+i\bar \eta_3 \bar H_3 \Big) \,
\\
&\hskip 1.5cm
\times \exp\Big( ip_Y Y \tight+ i \bar p_Y \bar Y \tight+ ip_Z Z \tight+ i \bar p_Z \bar Z \Big)~,
\nn
\end{align}
and to analyze the condition of mutual locality with \susyops. The physical vertex operators have additional contributions, such as wavefunctions on $AdS_3$ and the parafermion disc, as well as oscillators, but those do not contribute to the monodromy properties of such operators with respect to the supercurrents $S_r^\pm$. In performing the calculation it is also useful to recall that the momenta $p_Z,\bar p_Z$ are given in~\eqref{bospf2}, while $p_Y,\bar p_Y$ are given in~\eqref{squonespec}, \eqref{paramrel}.

The phase generated by taking $S_r^\pm$ once around an operator $\cO$ of the form \eqref{etaYZcharges} is
\eqn[monodromy]{
\exp\Big[
2\pi i\Big(-\half q_\varphi +r(\eta_\sl\mp\eta_3) \pm\half\eta_\su \Big)
\pm i\pi \Big(m'\tight-\mbar'+\eta_\su\tight-\bar\eta_\su+\frac{n\tight-2}2(w'\tight-\bar w')\Big)
\Big] ~.
}
One can think of the first term in the exponent as being generated by the naive spin field in $S_r^\pm$ (the $(-1)^F$ projection), while the second term comes from the half-unit of $J^3_\su+\bar J^3_\su$ spectral flow we added to it, and involves the total $J^3_\su-\bar J^3_\su$ charge $\sqrt{\frac n2}(p_Y-\bar p_Y)$ of $\cO$.
Demanding the triviality of this phase gives the chiral GSO projection.  

As usual, mutual locality with respect to one of the $S_r^\pm$ guarantees that with respect to the others, since they are related by flipping an even number of signs in \susyops.  We choose as a convention to check mutual locality with respect to $S_{-\half}^+$, for which the condition is
\eqna[chiralGSO]{
(-1)^F\,\Omega &= 1 ~,
}
with the component contributions
\eqna[GSOphase]{
(-1)^F &= \exp\Big[i\pi\Big(-q_\varphi + \epsilon_\sl\eta_\sl + \epsilon_3\eta_3 + \epsilon_\su\eta_\su \Big)\Big]
\\[.2cm]
\Omega &=
\exp\Big[i\pi \Big( (m'\tight-\mbar') + (\eta_\su\tight-\bar\eta_\su) +\frac{n-2}2(w'\tight-\bar w') \Big) \Big] ~,
}
where $\epsilonb=(\epsilon_\sl,\epsilon_3,\epsilon_\su)=(-1,1,1)\equiv\epsilonb_2$ (see the discussion after equation~\Fnum).

In addition to the untwisted, $(\NS,\NS)$ and $(\!R,\!R)$, sectors of the orbifold \eqref{chiralGSO}, one has twisted sectors with $(\!R,\NS)$ or $(\NS,\!R)$ boundary conditions.  These also obey the chiral GSO constraint~\eqref{chiralGSO}, \eqref{GSOphase}, and thus carry a half unit of vector spectral flow in the total $\sutwo$ current.

The above approach fits within a general procedure for constructing worldsheet models of $N=2$ supersymmetric $AdS_3$ backgrounds laid out in~\rcite{Giveon:1999zm,Giveon:1999jg,Giveon:2003ku}.  There one considers a worldsheet model $\sltwo\times \cN$ where $\cM={\cN}/{\uone}$ is an $N=2$ superconformal theory.  The theories $\cN$ and $\cM$ are related by $\cN = \( \cM\times\uone \)/\bZ_n$ for some $n\in\bZ$, where $\bZ_n$ acts as a discrete R-symmetry rotation of $\cM$ combined with a $\bZ_n$ shift on the $\uone$. 
However this does not uniquely specify $\cN$ because one is free to dial the $\uone$ radius.  The appropriate choice (made implicitly in~\rcite{Giveon:1999jg}) is to set it according to the $\sltwo$ level so that it can form the R-current of spacetime supersymmetry.  The chiral GSO projection was not given explicitly, but again implied by the demand of a spacetime $N=2$ superconformal algebra, and the requirement of mutual locality of vertex operators with respect to the supersymmetry currents~\eqref{susyops}.  In particular, the fact that the $(\!R,\NS)$ and $(\NS,\!R)$ sectors carry $Y,\bar Y$ momentum is a result of the fact that spacetime fermions carry spacetime R-charge, which is measured by the spacetime R-current built out of the worldsheet $U(1)_Y$.  Our particular example sets $\cN$ to be a squashed $\bS^3$, with the chiral GSO projection generated by~\eqref{chiralGSO}.

While we motivated and derived the chiral GSO constraint above using spacetime supersymmetry, there is a second, non-supersymmetric, choice for a chiral GSO projection at the same value of the squashing parameter, obtained by flipping an odd number of signs in the vector $\epsilonb$ used to define $(-1)^F$.  We will adopt the choice $\epsilonb_0=(1,1,1)$ that appeared in section~\ref{sec:type0short}. This alternate GSO projection can be viewed as being due to the requirement of mutual locality with respect to the worldsheet operators
\eqn[GSO0currents]{
\Psi_r^\pm = \exp\bigg[-\frac\varphi 2 +ir\big(H_\sl\pm H_3\big) \pm\frac {i\,a}2\, Z \pm \frac{i}{\sqrt{2k}}\,Y \bigg]~,
}
which differ from $S_r^\pm$ in~\eqref{susyops} by the sign of $H_3$ in the exponent. To understand this GSO projection better, it is useful to construct the local operators in the spacetime CFT that correspond to \eqref{GSO0currents}. One finds
\eqna[defpsii]{\Psi^\pm(x) \simeq \int\!d^2z\, 
\left(\Psi^\pm_\half-x \Psi^\pm_{-\half}\right)e^{-\bar\varphi}\,\bar\psi(\bar x;\bar z)\Phi_1(x,\bar x;z,\bar z) ~.}
Thus, we see that while in the supersymmetric GSO projection \susyops\ the spectrum of local operators includes a pair of dimension $\left(\frac32,0\right)$ operators $\cG^\pm$, the generators of an $N=2$ superconformal algebra, in the alternative construction \eqref{GSO0currents} we have instead a holomorphic dimension $\left(\frac12,0\right)$ operator \defpsii\ -- a complex free fermion. Both theories contain a holomorphic $U(1)$ current (second line of \susyops, \eqref{Gst}), under which the fermionic operators have unit charge.

It is important to emphasize that the two GSO projections described above are mutually exclusive. In the first we have $N=2$ superconformal symmetry but no $\Psi$, while in the second we have $\Psi$ but no superconformal symmetry. Thus, we refer to them as (chiral) GSO~2 (for $N=2$ SUSY) and GSO~0 (for no SUSY), respectively. 

One has a choice of either GSO~0 or GSO~2 separately for left-movers and for right-movers, leading to four distinct theories, which we denote GSO ({\it a,b}), $a,b=0,2$, that have ({\it a,b}) superconformal symmetry in spacetime. It is tempting to interpret GSO~0 as describing a theory with spontaneously broken $N=2$ superconformal symmetry. We will return to this issue below in section~\ref{sec:discussion}.

The GSO (0,0) and (2,2) theories can be constructed by starting with the 0B theory, and applying to it the GSO projection \chiralGSO\ with $\epsilonb=\epsilonb_0$ and $\epsilonb=\epsilonb_2$, respectively. Note that we only need to do this for one worldsheet chirality, since the type 0 theory already has $(-1)^F=(-1)^{\bar F}$. We can think of the resulting theories as IIB string theories.

Similarly, the GSO (0,2) and GSO (2,0) theories can be obtained by starting with type 0A, and applying the chiral GSO projection \chiralGSO\ with $\epsilonb=\epsilonb_0$ and $\epsilonb=\epsilonb_2$, respectively, giving rise to IIA string theories. We will henceforth consider only the (2,0) theory; the (0,2) one is obtained from it by exchanging left and right-movers on the worldsheet.

The $\bZ_2$ eigenvalue~\eqref{GSOphase} that determines the chiral GSO projection is topological, and in particular invariant under the $\bS^3$ squashing deformation introduced in section~\ref{sec:currentalg}.  Thus, the type II theories introduced in this section can be constructed by starting with a type 0 theory on $AdS_3$ times a squashed $\bS^3$ with any value of the squashing parameter $R$, discussed in sections~\ref{sec:type0short} and~\ref{sec:type0long}, and orbifolding it by the $\bZ_2$ generated by $(-1)^F\Omega$.  The $(\!R,\NS)$ and $(\NS,\!R)$ sectors arise as the twisted sectors of the orbifold, and supersymmetry appears after one deforms along the moduli space from $R=1$ to $R=\sqrt{n+1}$. 

Note also that the charged $SU(2)$ currents, $J^\pm_{\rm su}$ are projected out by the orbifold for all values of $R$ -- they are invariant under $(-1)^F$, but go to minus themselves under $\Omega$.  Hence, the $SU(2)_L\times SU(2)_R$ symmetry is broken to $\uone_L\times\uone_R$ in the type II theories for all values of the squashing parameter $R$. This is also the case in the underlying type 0 models for all $R\not=1$.


\subsection{The spectrum}
\label{sec:typeIIspec}

We are now ready to describe the spectrum of the various type II theories. As in the type 0 case, we split the discussion into two parts. In this section we study the sector with winding zero, and in the next turn to non-zero winding, our primary focus. 

Since the $SU(2)$ symmetry is broken in the type II theory, it is convenient to describe the states in terms of their $\frac{SU(2)}{U(1)}$ and $U(1)$ components. It is also useful to remember that the chiral GSO projection~\eqref{chiralGSO} correlates worldsheet fermion parity with the axial $U(1)$ charge.

The type 0 tachyon operator is described in terms of the above decomposition by the vertex operator
\eqn[typeIItach]{
e^{-\varphi-\bar\varphi}\, e^{i\left(p_Y Y+\bar p_Y\bar Y\right)} \, 
\Lambda_{j';m',\mbar'}^{(0,0)} \, \Phi_j(x,\bar x;z,\bar z) ~.
}
Here $\Lambda_{j';m',\mbar'}^{(0,0)}$ is a primary in the supersymmetric $\frac{SU(2)}{U(1)}$ model, defined in \pfdecomp, whose dimension is given by \pfspec, and the momentum $\left(p_Y,\bar p_Y\right)$ takes the form \squonespec, which in this case is given by 
\eqn[pYtach]{
\left(p_Y,\bar p_Y\right)=\frac1{\sqrt{2n}}
\left[\frac{m'+\mbar'}{R} \pm \big(m'-\mbar' \big)R\right] ~.
}
For $R=1$, \typeIItach, \pYtach\ is just another way of writing \VNS, while for general $R$ it describes the type 0 tachyon at a general value of the squashing parameter. The mass shell condition for general $R$ is
\eqn[tachshell]{
-\frac{j(j-1)}{k} + \frac{j'(j'+1)-m'^2}{n} 
+ \frac{1}{4n}\Big[\frac{m'+\mbar'}{R}+(m'-\mbar')R\Big]^2 =\half ~.
}
This comes from the left-moving $L_0$ constraint. The right-moving sector gives a similar condition, with $m'\leftrightarrow\bar m'$, which is equivalent to \tachshell. Again, for $R=1$ \tachshell\ reduces to \Lzerotach.

The vertex operator \typeIItach\ belongs to the $(\NS-,\NS-)$ sector, where~$\pm$ refers to the $(-1)^F$ parity, as in the critical string~\rcite{Polchinski:1998rr}. Therefore, under the chiral GSO~\eqref{chiralGSO}, the modes with $m'-\bar{m}'\in 2\mathbb{Z}$ are projected out, while those with $m'-\bar{m}'\in 2\mathbb{Z}+1$ are kept. 

The spacetime scaling dimension of the operators \typeIItach\ is $(j,j)$, with $j$ given by \tachshell. Some of these operators are tachyonic for general squashing factor $R$. The most tachyonic operator that survives the GSO projection~\eqref{chiralGSO} has $j'=\pm m'=\mp\mbar'=\half$. The mass shell condition \tachshell\ leads in this case to
\eqn[jtach]{
j= \half + \frac{\sqrt{R^2-n+3}}{2\sqrt{n+1}} ~.
}
Thus as long as the squashing factor $R$ exceeds $\sqrt{n-3}$, there are no BF-violating tachyonic modes.  The supersymmetric point $R=\sqrt{n+1}$ lies safely within the stable regime. Of course, this had to be the case, since supersymmetry (together with unitarity) is inconsistent with the existence of BF-violating tachyons. In what follows, we will mostly restrict to this value of the squashing factor.

In section \ref{sec:type0short} we discussed vertex operators corresponding to the gravity sector, \eqref{WXxspace}, \hybridvertex. We did that at the $SU(2)$ point, but using the results of section~\ref{sec:review}, they can be extrapolated to any value of the squashing parameter $R$. For the operators in \eqref{WXxspace}, one finds 
\eqna[WXtypeII]{
\cW_{j';m',\mbar'} &= 
e^{-\varphi-\bar\varphi}\,\psi(x)\bar\psi(\bar x) \,
\Phi_j(x,\bar x) \,
e^{i\left(p_Y Y+\bar p_Y\bar Y\right)} \, 
\Lambda_{j';m',\mbar'}^{(0,0)} ~,
\\
\cX_{j';m',\mbar'} &= 
e^{-\varphi-\bar\varphi}\,
\Phi_j(x,\bar x) \,
e^{i\left(p_Y Y+\bar p_Y\bar Y\right)} \, 
\Lambda_{j';m',\mbar'}^{(1,1)} ~.
}
On the first line of \WXtypeII, the left and right-moving momenta $\left(p_Y,\bar p_Y\right)$  take the values \pYtach; the mass-shell condition is
\eqn[tachshellW]{
-\frac{j(j-1)}{k} + \frac{j'(j'+1)-m'^2}{n} 
+ \frac{1}{4n}\Big[\frac{m'+\mbar'}{R}+(m'-\mbar')R\Big]^2 =0 ~.
}
On the second line, \eqref{pfdecomp}-\eqref{paramrel} lead to 
\eqn[pYX]{
\left(p_Y,\bar p_Y\right)=\frac1{\sqrt{2n}}
\left[\frac{m'+\mbar'+2}{R} \pm \big(m'-\mbar' \big)R\right] ~,
}
while the mass-shell condition is
\eqn[Xshell]{
-\frac{j(j-1)}{k} + \frac{j'(j'+1)-\left(m'+1\right)^2}{n} 
+ \frac{1}{4n}\Big[\frac{m'+\mbar'+2}{R}+(m'-\mbar')R\Big]^2 =0~.
}
Since the operators \WXtypeII\ belong to the $(\NS+,\NS+)$ sector, the chiral GSO projection now allows $m'-\bar{m}'\in 2\mathbb{Z}$, and projects out states with $m'-\bar{m}'\in 2\mathbb{Z}+1$. The spacetime scaling dimensions of these operators are $(j-1,j-1)$ for the first line in \WXtypeII, and $(j,j)$ for the second. 

Another operator constructed in section \ref{sec:type0short} is $\cK_{j'}$ \hybridvertex. At the supersymmetric point it takes the form 
\eqn[hybridsusy]{\cK_{j';m',\mbar'} = 
e^{-\varphi-\bar\varphi}\,
\bar\psi(\bar x)\Phi_j(x,\bar x) \,
e^{i\left(p_Y Y+\bar p_Y\bar Y\right)} \, 
\Lambda_{j';m',\mbar'}^{(1,0)} ~.}
Its spacetime scaling dimension is $(j,j-1)$, where $j$ is determined by the mass-shell condition 
\eqn[masskj]{-\frac{j(j-1)}{k} + \frac{j'(j'+1)-\left(m'+1\right)^2}{n} 
+ \frac{1}{4n}\Big[\frac{m'+\mbar'+1}{R}+(m'+1-\mbar')R\Big]^2 =0~.
}
The momenta $\left(p_Y,\bar p_Y\right)$ take in this case the values
\eqn[momkj]{\left(p_Y,\bar p_Y\right)=\frac1{\sqrt{2n}}
\left[\frac{m'+\mbar'+1}{R} \pm \big(m'+1-\mbar' \big)R\right] .
}
The chiral GSO projection allows modes with $m'-\mbar'\in 2 \mathbb{Z}+1$, and projects out the modes $m'-\mbar'\in 2 \mathbb{Z}$.

We now turn to the $(\!R,\!R)$ sector. As discussed above, the type II theories are obtained by starting with the type 0 ones, and orbifolding by \chiralGSO. The action of $(-1)^F$ on the $(\!R,\!R)$ operators $\cO^{(i)}$ was described in \actionepsilon. Thus, to specify the action of \chiralGSO\ we need to compute the action of $\Omega$ \GSOphase on these operators. To do that it is convenient to write $\Omega$ as $\Omega_L\Omega_R$, with 
\eqna[omegalr]{&\Omega_L=\exp\Big[i\pi \Big( m' + \eta_\su +\frac{n-2}2w' \Big) \Big]~,
\cr 
&\Omega_R=\exp\Big[-i\pi \Big( \mbar'+\bar\eta_\su+\frac{n-2}2\bar w' \Big) \Big]~.}
One finds 
\eqna[OmegaLOi]{&\Omega_L ~{\mathcal O}^{(1)}=e^{i\pi\left(j'+\frac{1}{2}\right)}{\mathcal O}^{(1)}~,~~~~~\Omega_L ~{\mathcal O}^{(2)}=e^{i\pi\left(j'+\frac{1}{2}\right)}{\mathcal O}^{(2)}~,
\cr
&\Omega_L ~{\mathcal O}^{(3)}=e^{i\pi\left(j'-\frac{1}{2}\right)}{\mathcal O}^{(3)}~,~~~~~\Omega_L ~{\mathcal O}^{(4)}=e^{i\pi\left(j'-\frac{1}{2}\right)}{\mathcal O}^{(4)}~,}
The transformations under $\Omega_R$ is similar, with eigenvalues that are the complex conjugates of the ones in \OmegaLOi.

As discussed in section \ref{sec:type0short}, the $(\!R,\!R)$ operators in \eqref{RRcohomology} are the highest weight states in a given $SU(2)$ representation, and one needs to act on them with $J^-_{\su;0}$ to get the general state. Thus, we are also interested in the action of $\Omega_L$ on $J^-_{\su;0}$, which is 
\eqn[OmegaLJm]{\Omega_L~J^-_{\su;0}~\Omega_L^{-1}=-J^-_{\su;0}~,}
and a similar result for the right-movers. These results allow us to compute the action of the projection \chiralGSO\ on the type 0 $(\!R,\!R)$ spectrum in \GSOeven. 

To demonstrate this action consider, as an example, the (2,0) theory, obtained by applying the GSO projection \chiralGSO\ with $\epsilonb=\epsilonb_2$ to the 0A theory. Looking back at \GSOeven, we see that one of the operators in the 0A theory is ${\cal O}^{(2)} \bar{{\cal O}}^{(1)}$. From \actionepsilon\ (or, more precisely, the line under that equation) we see that this operator is odd under $(-1)^F$. From \OmegaLOi\ and its right-moving analog, we see that it is even under $\Omega$. Thus, this operator is odd under \chiralGSO, and does not survive the projection.

Using \OmegaLJm, we can act an odd number of times with the lowering operators $J^-_{\rm su;0}$, $\bar J^-_{\rm su;0}$ to obtain an operator that {\it is} invariant under the chiral GSO \chiralGSO. We summarize this information by the statement that the operator ${\cal O}^{(2)} \bar{{\cal O}}^{(1)}$ is in the spectrum of the IIA theory if\footnote{Here $m'$ and $\bar m'$ are the values of $J^3_{\rm su;0}$ and $\bar J^3_{\rm su;0}$ of the general operator in the representation whose highest component is the operator ${\cal O}^{(2)} \bar{{\cal O}}^{(1)}$ given in \eqref{RRcohomology}, \GSOeven.} $m'-\bar m'\in2\mathbb{Z}+1$. 

Note also that the calculation described above is done at the point in the moduli space of type 0 theories where the $\bS^3$ is unsquashed. However, as mentioned earlier in this section, the GSO projection is a topological operation, and we can perform it at any point in the moduli space labeled by the squashing parameter $R$. In the discussion above, we  performed the projection at $R=1$, and then changed $R$ to the supersymmetric value $R=\sqrt{n+1}$. 

Repeating this exercise for all the other operators in \GSOeven, we find that the spectrum of the $(2,0)$ theory includes the $(\!R,\!R)$ operators 
\eqna[RRIIA]{(2,0):&~~{\cal O}^{(1)} \bar{{\cal O}}^{(2)},{\cal O}^{(4)} \bar{{\cal O}}^{(2)},{\cal O}^{(1)} \bar{{\cal O}}^{(3)},{\cal O}^{(4)} \bar{{\cal O}}^{(3)}~,~~~~m'-\bar{m}'\in2\mathbb{Z}~,
\cr 
&~~{\cal O}^{(3)} \bar{{\cal O}}^{(1)},{\cal O}^{(2)} \bar{{\cal O}}^{(1)},{\cal O}^{(3)} \bar{{\cal O}}^{(4)},{\cal O}^{(2)} \bar{{\cal O}}^{(4)}~,~~~~m'-\bar{m}'\in2\mathbb{Z}+1 ~.
}
The (0,0) and (2,2) theories are obtained from the 0B theory by applying the GSO projection \chiralGSO\ with $\epsilonb=\epsilonb_0$ and $\epsilonb_2$, respectively. In this case we find that the following $(\!R,\!R)$ operators in~\eqref{GSOeven} survive the projection:
\eqna[RRIIB]{(0,0):&~~{\cal O}^{(2)} \bar{{\cal O}}^{(2)},{\cal O}^{(3)} \bar{{\cal O}}^{(3)},{\cal O}^{(2)} \bar{{\cal O}}^{(3)},{\cal O}^{(3)} \bar{{\cal O}}^{(2)}~,~~~~m'-\bar{m}'\in2\mathbb{Z}~,\cr
&~~{\cal O}^{(1)} \bar{{\cal O}}^{(1)},{\cal O}^{(4)} \bar{{\cal O}}^{(4)},{\cal O}^{(1)} \bar{{\cal O}}^{(4)},{\cal O}^{(4)} \bar{{\cal O}}^{(1)}~,~~~~m'-\bar{m}'\in2\mathbb{Z}+1~,\cr
(2,2):&~~{\cal O}^{(1)} \bar{{\cal O}}^{(1)},{\cal O}^{(4)} \bar{{\cal O}}^{(4)},{\cal O}^{(1)} \bar{{\cal O}}^{(4)},{\cal O}^{(4)} \bar{{\cal O}}^{(1)}  ~,~~~~m'-\bar{m}'\in2\mathbb{Z}~,\cr
&~~{\cal O}^{(2)} \bar{{\cal O}}^{(2)},{\cal O}^{(3)} \bar{{\cal O}}^{(3)},{\cal O}^{(2)} \bar{{\cal O}}^{(3)},{\cal O}^{(3)} \bar{{\cal O}}^{(2)}~,~~~~m'-\bar{m}'\in2\mathbb{Z}+1~.}
One can use our discussion above to find the mass shell condition for all the operators in \RRIIA, \RRIIB, but we will not do this here (we will study a particular example below). Also, some of the $(\!R,\!R)$ operators above are normalizable and some are not; we will not attempt a complete classification here, but will mention some examples later. 

So far, we have discussed the untwisted sector of the GSO $\bZ_2$ orbifold. The twisted sectors contain operators that belong to the $(\!R,\NS)$ and $(\NS,\!R)$ sectors. They can be obtained by acting on operators in the untwisted sectors by the spacetime supercharges \susyops, or the fermionic operators \eqref{GSO0currents}, for GSO~2 and GSO~0, respectively. We will not attempt to discuss the general case here. Instead, we will consider in the next subsection some examples of this action in special cases of interest.


\subsection{The BPS spectrum of the (2,2) theory}
\label{sec:BPS}

Earlier in this section we described theories with varying amounts of supersymmetry. In this subsection we will comment briefly on the spectrum of BPS operators in the theory with (2,2) superconformal symmetry.  We start with the zero winding chiral spectrum ~\rcite{Giveon:1999zm,Argurio:2000tb}.

In the $(\!R,\!R)$ sector, one finds that the operators ${\cal O}^{(1)} \bar{{\cal O}}^{(1)}$ in~\eqref{RRIIB}, with $m'=\bar{m}'=j'=0,\hf,\cdots, \frac{n-2}{2}$, are chiral on both left and right (\ie~$(c,c)$ operators).  We will denote them by  
\eqna[IIB22Y]{{\cal Y}_{j'}=
e^{-\frac{\varphi}{2}-\frac{\bar{\varphi}}{2}} \,
e^{\frac{i}{2}H_\sl+\frac{i}{2}H_3+\frac{i}{2}\bar{H}_\sl+\frac{i}{2}\bar{H}_3} \,
e^{i\frac{1+2j'}{\sqrt{2n(n+1)}}(Y+\bar{Y})} \, \Lambda^{\left(\frac{1}{2},\frac{1}{2}\right)}_{j';j',j'} \,
\Phi^{(0)}_{j;-j,-j} ~,}
where $j$ satisfies the on-shell condition
\eqna[jY]{j=\frac{1}{2}+\frac{1+2j'}{2(n+1)}~.}
The spacetime dimension of~\eqref{IIB22Y} is \eqn[chiraldim]{
\big(h_{\ST},\bar h_{\ST}\big) = 
\big(j-\hf,j-\hf\big)
=\Big(\frac{1+2j'}{2(n+1)},\frac{1+2j'}{2(n+1)}\Big) ~,
}
while the R-charge follows from \eqref{susyops}, and is given by $(R_\ST,\bar{R}_\ST)=\left(2h_{\ST},2\bar h_{\ST}\right)$ as expected for a $(c,c)$ operator.

The vertex operator \eqref{IIB22Y} is written in the $(m,\bar m)$ basis, and moreover it is normalizable, since $m=\bar m=-j$ and $j$ \eqref{jY} is in the unitary range $\frac{1}{2}<j<\frac{k+1}{2}$. Thus, this vertex operator describes the normalizable state associated via the state-operator correspondence to the corresponding $(c,c)$ operator.\footnote{The spacetime CFT we are dealing with does not have in general a state-operator correspondence, like Liouville theory, $\frac{\sltwo}{U(1)}$ CFT, etc, but it does have such a correspondence for a subset of the operators, a phenomenon that is familiar from $\frac{\sltwo}{U(1)}$ CFT~\rcite{Aharony:2004xn}.}

The operator \eqref{IIB22Y} is annihilated by the supercharges $\oint\! dz\, S^\pm_{\frac{1}{2}}$ and $\oint\! dz\, S^{+}_{-\frac{1}{2}}$ in \eqref{susyops}, the latter following from the fact that \eqref{IIB22Y} is chiral. The action of the supercharge $\oint\! dz\, S^{-}_{-\frac{1}{2}}$ produces the $(\NS,\!R)$ operator
\eqna[IIB22YRNS]{e^{-\varphi-\frac{\bar{\varphi}}{2}} \,
e^{\frac{i}{2}\bar{H}_\sl+\frac{i}{2}\bar{H}_3} \,
e^{i\frac{2j'-n}{\sqrt{2n(n+1)}}Y+i\frac{1+2j'}{\sqrt{2n(n+1)}}\bar{Y}} \, \Lambda^{\left(0,\frac{1}{2}\right)}_{j';j',j'} \,
\Phi^{(0)}_{j;-j,-j}~,}
where $j$ is again given by \eqref{jY}. The operator \eqref{IIB22YRNS} has spacetime dimension 
\eqn[Gmhalfdim]{
\big(h_{\ST},\bar h_{\ST}\big) = 
\big(j,j-\hf\big)
=\Big(\frac{2j'+n+2}{2(n+1)},\frac{1+2j'}{2(n+1)}\Big)~,
}
and R-charge
\eqn[RstRNS]{
\big(R_\ST,\bar R_\ST\big) 
=\Big(\frac{2j'-n}{n+1},\frac{1+2j'}{n+1}\Big) ~.
}
It is of course normalizable, since acting with a supercharge on a normalizable operator gives a normalizable operator.

Acting on \eqref{IIB22YRNS} with the right-moving supercharge $\oint\! d\bar{z}\, \bar{S}^{-}_{-\frac{1}{2}}$, we find a normalizable $(\NS,\NS)$ operator, with $h_{\ST}=\bar h_{\ST}=j$, $R_\ST=\bar R_\ST=\frac{2j'-n}{n+1}$. Note that while this operator is in the same sector, $(\NS,\NS)$, as some operators we constructed earlier in this section, \eg\ \WXtypeII, it is different from those operators. A quick way to see that is that it is normalizable, while the operators constructed earlier are not.

Another set of chiral operators corresponds to the operators ${\cal W}_{j'}$ in~\eqref{WXtypeII}, with $m'=\bar{m}'=j'$, which are given in the $m$ basis by
\eqna[IIB22W]{{\cal W}_{j'}=
e^{-\varphi-\bar{\varphi}} \,
 \,
e^{i\frac{2j'}{\sqrt{2n(n+1)}}(Y+\bar{Y})} \,
\Lambda^{\left(0,0\right)}_{j';j',j'} \,
\left(\psi_\sl\bar\psi_\sl\, \Phi^{(0)}_{j}\right)_{j-1;m=\mbar=-j+1}~,}
where the notation $\left(\psi_\sl\bar\psi_\sl\, \Phi^{(0)}_{j}\right)_{j-1;m=\mbar=-j+1}$ was introduced in \eqref{WX}, and $j$ now satisfies
\eqna[jW]{j=1+\frac{j'}{n+1}~.}
These operators have spacetime scaling dimension $\big(h_{\ST},\bar h_{\ST}\big)=\left(j-1,j-1\right)=
\big(\frac{j'}{n+1},\frac{j'}{n+1}\big)$, and $R$-charge $(R_\ST,\bar R_\ST)=\left(2h_{\ST},2\bar h_{\ST}\right)$. Thus, they are also $(c,c)$ operators in the spacetime SCFT, but since $j>\frac{k+1}{2}$, they are non-normalizable.

The $(c,c)$ operators~\eqref{IIB22Y}, \eqref{IIB22W} have analogues with non-zero $\sltwo$ winding~\rcite{Argurio:2000tb}. Using the parafermion decomposition~\eqref{pfdecompsq}, we find the $(\!R,\!R)$ operators
\eqna[IIB22 Ywneq0 susypt]{{\cal Y}^{w_\su,w_\sl,\eta_3}_{j'}=
e^{-\frac{\varphi}{2}}\,
e^{i\frac{Y}{\sqrt{2n(n+1)}}\left(2j'+n w_\su+1\right)}\,
\Lambda^{(\frac{1}{2})}_{j';m'=j'}e^{i\left(\frac{1}{2}-w_\sl\right)H_\sl+i\eta_3H_3}\,
\Phi^{(w_\sl)}_{j;m=-j}~,}
where for simplicity here and in some subsequent equations we write only the left-moving part of these left-right symmetric operators. 
These operators are BRST closed provided $j$ satisfies
\eqn[jYwneq0]{j-\frac{1}{2}=\frac{k}{2}\left(2\eta_3w_\su+w_\sl\right)+\frac{2\eta_3}{n+1}\left(j'+\frac{1}{2}\right)~.}
The spacetime scaling dimension and R-charge of \eqref{IIB22 Ywneq0 susypt} are
\eqna[YhSTwneq0gen]{&\left(h_{\ST},\bar{h}_{\ST}\right)=\Big(j-\frac{1}{2}-\frac{k}{2}w_\sl,j-\frac{1}{2}-\frac{k}{2}w_\sl\Big)~, 
\cr 
&\left(R_\ST,\bar R_\ST\right)=\left(\frac{2j'+nw_\su+1}{n+1},\frac{2j'+nw_\su+1}{n+1}\right)~.}
Demanding that \eqref{IIB22 Ywneq0 susypt} describe chiral operators with $w_\sl\leq -1$ and with $j$ in the range $\frac{1}{2}<j<\frac{k+1}{2}$, fixes $\eta_3=+\frac{1}{2}$, $w_\su=-w_\sl\equiv -w$. 
This gives $(c,c)$ operators which we denote by ${\cal Y}^{(w)}_{j'}$ for $w\leq -1$. They have spacetime scaling dimension
\eqn[YhSTwneq0]{\left(h_{\ST},\bar{h}_{\ST}\right)
=\Big(j-\frac{1}{2}-\frac{k}{2}w\,,\;j-\frac{1}{2}-\frac{k}{2}w\Big)~,}
where $j$ is again given by~\eqref{jY}.
Note that for $j'=0, w=-1$, this $(c,c)$ operator has spacetime scaling dimension $\left(\frac{1}{2},\frac{1}{2}\right)$, and therefore the top component of the same supermultiplet is a marginal $(\NS,\NS)$ operator in the spacetime CFT. This observation will be important in section~\ref{sec:wall} below.

Similarly, one has the $(\NS,\NS)$ operators
\eqna[IIB22 Wwneq0 susypt]{{\cal W}^{w_\su,w_\sl}_{j'}=
e^{-\varphi}  \,
e^{i\frac{Y}{\sqrt{2n(n+1)}}\left(2j'+n w_\su\right)} \,
\Lambda^{\left(0\right)}_{j';j'} \,
\left(\psi_\sl\Phi^{(0)}_{j}\right)^{\left(w_\sl\right)}_{j-1;m+1}~,}
where $\left(\psi_\sl\Phi^{(0)}_{j}\right)^{\left(w_\sl\right)}_{j-1;m+1}$ is the spin $j-1$ $\sltwo$ representation, with total $\sltwo$ winding $w_\sl$.  More explicitly, we start with the spin $j-1$ $\sltwo$ representation combination given by \eqref{WX}, and perform a supersymmetric $\sltwo$ spectral flow of $w_\sl$, see \eqref{superPhi}, \eqref{aass}. Note that the subscript $m+1$ in \eqref{IIB22 Wwneq0 susypt} denotes the total $J^3_{\sl;0}$ eigenvalue before the supersymmetric spectral flow.

Demanding that these operators are on-shell fixes
\eqn[mWneq0]{m=\bar{m}=\frac{1}{w_\sl}\left[-\frac{j(j-1)}{k}+\frac{j'}{n}+\frac{\left(2j'+n w_\su\right)^2}{4n(n+1)}\right]-\frac{k}{4}w_\sl -1 ~.}
These operators are in the $(c,c)$ ring for $w_\su=-w_\sl\equiv -w\geq1$, and $m=-j$. In this case \eqref{mWneq0} gives
\eqn[jWneq0]{j=1+\frac{j'}{n+1}~.}
The spacetime scaling dimension of \eqref{IIB22 Wwneq0 susypt} is then
\eqn[WhSTwneq0]{\left(h_{\ST},\bar{h}_{\ST}\right)=\left(-m-1-\frac{k}{2}w_\sl,-\bar{m}-1-\frac{k}{2}w_\sl\right)=\left(\frac{2j'-n w}{2(n+1)},\frac{2j'-n w}{2(n+1)}\right)~.}
Like \eqref{IIB22W}, these chiral ring operators $\cW^{(w)}_{j'}$ are non-normalizable, since $j>\frac{k+1}{2}$.  Note that perturbing the theory by the integrated top component of the operator $\cW_\half^{(-1)}$ is an $(\!R,\!R)$ marginal deformation of the spacetime CFT.

Thus we have a spectrum of $\half$-BPS single string states associated to the $(c,c)$ operators $\cY^{(w)}_{j'}$ with conformal weight and R-charge
\eqn[YwRchg]{
2h_{\rm ST}=R_\ST=\frac{2j'+1-wn}{n+1} \equiv \frac{\kappa}{n+1}
}
(recall $w\le 0$).  There is one single-string state at each value of R-charge $\kappa/(n+1)$, $\kappa\in\bN$ with the exception of $\kappa\in n\bN$.  Some further details of the BPS spectrum are the subject of Appendix~\ref{app:halfBPS}. 

The $(a,a)$ spectrum follows from conjugation in the R-charge~\eqref{susyops}.  More precisely, one employs the $\sutwo$ involution which sends $j^3_\su\to-j^3_\su$ and $j^\pm_\su\leftrightarrow j^\mp_\su$ (and similarly for the fermions $\psi^a_\su$), which when carried through the squashing deformation amounts to 
\eqn[Rconj]{
p_Y^{~}\to-p_Y^{~}
~~,~~~~
p_Z^{~}\to-p_Z^{~}
~~,~~~~
\eta_3^{~}\to-\eta_3^{~} 
}
(the last of these follows from the definition~\eqref{Hdefs} of $H_3$).

There are also $(c,a)$ and $(a,c)$ rings, for which $\bar p_Y\tight=- p_Y$, $\bar p_Z\tight=-p_Z$, and $\bar\eta_3\tight=-\eta_3$.  Equal and opposite left and right R-charge is achieved by setting $p\tight=P\tight=0$ in the $Y$ momenta~\eqref{squonespec}, and we also change the parafermion operator $\Lambda_{j';j'j'}$ to $\Lambda_{j';j',-j'}$ in~\eqref{IIB22Y}, \eqref{IIB22 Ywneq0 susypt} (for a $(c,a)$ operator).  
One has for instance the normalizable twisted chiral operators
\eqna[Ytwisted]{{\medhat\cY}^{(w)}_{j'}=
e^{-\frac{\varphi}{2}-\frac{\bar\varphi}2}\,
e^{i\sqrt{\frac{n+1}{2n}}\left(2j'+1+n w\right)(Y-\bar Y)}\,
\Lambda^{(\frac{1}{2},-\frac{1}{2})}_{j';j',-j'}\, 
e^{i\left(\frac{1}{2}-w\right)(H_\sl+\bar H_\sl)+\frac i2(H_3-\bar H_3)}\,
\Phi^{(w)}_{j;-j,-j}~.}
The spectrum of twisted chiral R-charges is $n\tight+1$ times sparser in the $(c,a)$ ring as compared to the $(c,c)$ ring
\eqn[caspec]{
R_\ST= \ell+n L  ~,
}
with $\ell\tight=2j'+1\tight=1,2,...,n-1$ and $L\in\bN_0$.

Beyond the $\hf$-BPS spectrum in the (2,2) theory, one has the $\frac14$-BPS spectrum (or the $\hf$-BPS spectrum in the (2,0) theory) counted by the elliptic genus (see~\rcite{Yamaguchi:2000dz,Eguchi:2004yi,Giveon:2015raa} for related discussions).  We will not give details here, but instead turn to a detailed discussion of long string states and operators in the type II theories.


\section{Long strings in type II}
\label{sec:typeIIlong}

Having described the short string sector of the type II theories, we are now ready to tackle the long string sector, along the lines of the analysis of the type 0 theory in section~\ref{sec:type0long}. In that case we saw that the spectrum of non-normalizable and delta-function normalizable operators in non-zero winding sectors gives rise in spacetime to a symmetric product, that describes the spacetime CFT at large positive $\phi$ (the radial direction in $AdS_3$, which becomes one of the target space dimensions of the spacetime CFT). The building block of this symmetric product is given by the $w=-1$ sector (see the discussion around equation~\eqref{windingdim}). Hence we begin by analyzing that sector.

The type II theories involve a squashed $\bS^3$ geometry~-- a particular squashing factor $R=\sqrt{n+1}$ is required in order to have supersymmetry or massless fermions.  Plugging the $\sqsphere$ spectrum~\eqref{pfspec}, \eqref{squonespec} into~\eqref{windingdim}, we find
\eqn[windingdimII]{
h_1 = -\frac{j(j-1)}{k} + \bigg[\frac{j'(j'+1)-(m'+\alpha)^2}{n} +\frac{\alpha^2}{2} + \frac{p_Y^{\,2}}{2}\bigg] + \big(N_\osc-\delta_0\big)+\frac k4 ~,
}
where $p_Y$ is determined in terms of the $\sutwo$ quantum numbers $j',m',\mbar'$, fermion charges $\eta_\su,\bar\eta_\su$, and spectral flow quantum numbers $w',\bar w'$ as in~\eqref{paramrel}, and $\alpha$ is defined after~\eqref{pfdecomp}. As in the type 0 theory, the first factor is the contribution to the conformal dimension from the vertex operator of the linear dilaton theory $\bR_\phi$ in the building block of the symmetric product
\eqn[expII]{
e^{\beta\phi} 
~~,~~~~
\beta=-\frac{Q_\ell}{2} + Q\Big(j-\half\Big)
=-\frac{Q_\ell}{2}+Q_\ell\Big(j_{\rm st}-\half\Big) ~,
}
where in the last equality we used~\eqref{jSTjWS}. The term in square brackets in~\eqref{windingdimII} suggests that the block of the symmetric product will have as a component the same squashed $\bS^3$ theory as the worldsheet theory. To verify these expectations and find the detailed structure of the block of the symmetric product, we next exhibit some low lying operators in the winding $w=-1$ sector. We will use extensively the results of section~\ref{sec:type0long}, which provides useful background for this analysis.

The $(\NS,\NS)$ and $(\!R,\!R)$ sectors of the type II theories correspond to the untwisted sector of the GSO orbifold. Thus, to analyze them we need to see which of the operators in section~\ref{sec:type0long} survive the chiral GSO projection.


\subsection{Low-lying operators}
\label{sec:block ops}

The first operator we discussed in section~\ref{sec:type0long} was \ebphiws. The vertex operator on the r.h.s. of that equation is invariant under the GSO projection \chiralGSO. In the block theory, it corresponds to the operator $e^{\beta\phi}$ on the l.h.s. of that equation, in agreement with the general picture above.

The same is true for the operator $\cI^{(-1)}$ in equation \eqref{wsidd}. As in the type 0 case, this operator corresponds to the $w=-1$ representation of the operator $\cI$, which plays the role of the identity operator in various current algebras, and hence must be present. 

The holomorphic operator $\partial_x\phi$ on the l.h.s. of~\dualbbzero, which is described by the worldsheet vertex operator on the r.h.s. of that equation, also survives the GSO projection and hence is present in all type II theories. This is again in agreement with expectations. 

All the operators discussed above have constant wavefunctions on the three-sphere. They are trivially invariant under $\Omega$ in \GSOphase, and take the same form throughout the moduli space of squashed $\bS^3$ described earlier in the paper. The first operator discussed in section~\ref{sec:type0long} that does not have this property is the block operator $v_{j';m',\bar m'} \, e^{\beta\phi}$, equation~\ebphivjmws. This operator is invariant under $(-1)^F$, like \ebphiws; therefore, the GSO projection imposes on it the constraint $m'-\mbar'\in 2\bZ$. 

The operator \ebphivjmws\ was written in the unsquashed theory (the one with $R=1$). To describe it at the supersymmetric point $(R=\sqrt{n+1})$, we need to perform the squashing that we discussed before (both on the l.h.s. and on the r.h.s.). The vertex operator on the r.h.s. now takes the form 
\eqn[blocktach]{
e^{-\varphi-\bar\varphi}\, e^{i\left(p_Y Y+\bar p_Y\bar Y\right)} \, 
\Lambda_{j';m',\mbar'}^{(0,0)} \, 
e^{iH_\sl+i\bar H_\sl} \,
\Phi^{(-1)}_{j;m,\mbar}~, 
}
where $m=\bar{m}$ is found from the on-shell condition,
\eqn[mblocktach]{m=\frac{j(j-1)}{k}+\frac{k+2}{4}-\frac{j'(j'+1)-m'^2}{n}-\frac{p_Y^2}{2}~,}
and the momenta $(p_Y,\bar p_Y)$ are given in \eqref{pYtach}. This operator has spacetime dimension $\left(h_{\ST},h_{\ST}\right)$, with
\eqn[hSTblocktach]{h_{\ST}=-m-1+\frac{k+2}{2}=-\frac{j(j-1)}{k}+\frac{k-2}{4}+\frac{j'(j'+1)-m'^2}{n}+\frac{p_Y^2}{2}~.}
The corresponding operator in the block of the spacetime CFT is \eqn[blocktachST]{e^{\beta\phi}\, e^{i\left(p_Y Y+\bar p_Y\bar Y\right)} \, 
\Lambda_{j';m',\mbar'}^{(0,0)}~,}
with $\beta$ related to $j$ by \eqref{expII}. Using this relation, it is straightforward to check that \eqref{hSTblocktach} agrees with the scaling dimension of \eqref{blocktachST}.

An interesting special case of the worldsheet operators~\eqref{blocktach} is given by
\eqn[WFZZgenj]{\cW_{j,j'}=e^{-\varphi-\bar{\varphi}} \,
 \,
e^{i\frac{2j'}{\sqrt{2n(n+1)}}(Y+\bar{Y})} \,
\Lambda^{\left(0,0\right)}_{j';j',j'} \,
e^{i H_\sl+i\bar{H}_\sl}\, \Phi^{(-1)}_{j;m,\mbar}~,}
where $m,\mbar$ are fixed by the on-shell condition,
\eqn[mWFZZ]{m=\mbar=\frac{j(j-1)}{k}+\frac{k+2}{4}-\frac{j'}{n}-\frac{j'^2}{n(n+1)}~.}
This operator has spacetime scaling dimension $\left(h_\ST,\bar{h}_\ST\right)$, where
\eqn[hST Wgenj]{h_\ST=\bar{h}_\ST=-m-1+\frac{k+2}{2}=-\frac{\left(j-\frac{1}{2}\right)^2}{k}+\frac{\left(k-1\right)^2}{4k}+\frac{j'}{n}+\frac{j'^2}{n(n+1)}~.}
When $j=1-\frac k2+\frac{j'}{n+1}$, the operator \eqref{WFZZgenj} is chiral, and has the same spacetime quantum numbers as \eqref{IIB22W}. The relation between \eqref{IIB22W} and \eqref{WFZZgenj} is the same as that observed above between \eqref{Kst} and \eqref{dualcurr}, and between \eqref{Ioperator} and \eqref{wsidd}.

The worldsheet operator \eqref{WFZZgenj} has the same quantum numbers as the operator
\eqn[Wblockop]{e^{\beta\phi}\Lambda^{(0,0)}_{j';j',j'}e^{i\frac{2j'}{\sqrt{2n(n+1)}}\left(Y+\bar{Y}\right)}~,}
in the block of the spacetime CFT. $\beta$ in \eqref{Wblockop} is related to $j$ in \eqref{WFZZgenj} by \eqref{expII}. When $j=1-\frac k2+\frac{j'}{n+1}$, \eqref{Wblockop} is a chiral operator in spacetime, as expected.

Note that our discussion above provides another example of a phenomenon mentioned in section~\ref{sec:type0long} (see the discussion after equation~\eqref{singletr}). The contribution of the squashed $\bS^3$ to the worldsheet vertex operator \blocktach\ is the same as its contribution to the spacetime operator \blocktachST, but the two live in different theories, and so should not be confused.


\subsection{Holomorphic operators}
\label{sec:holo ops}

In the type 0 discussion we found in the $w=-1$ sector (at the $SU(2)$ point, $R=1$) three sets of holomorphic $SU(2)$ generators given by \eqref{dualcurr}, \eqref{onemoresu2}, \eqref{finalthird}. All three are invariant under $(-1)^F$, but under the axial shift $\Omega$ of equation~\eqref{GSOphase}, the charged $SU(2)$ generators are odd, while the ones corresponding to the Cartan subalgebra are even. Thus, instead of three copies of $SU(2)$, we find three holomorphic $U(1)$ currents. Including the current \dualbbzero, we have four holomorphic currents, as well as four anti-holomorphic ones coming from the other worldsheet chirality. All these operators are present throughout the moduli space of squashed $\bS^3$ labeled by $R$.\footnote{Note that this is also the case in the type 0 theory, for any finite squashing of the $\bS^3$, \ie\ for $R\not=1$.}

To understand the role of these holomorphic currents, and more generally the structure of the spacetime SCFT, we would like to analyze the spectrum of holomorphic operators in the theory. Unlike the type 0 case, here we expect to find holomorphic operators of dimension $(\hf,0)$, coming from the $(\!R,\NS)$ sector. To find these operators, we proceed as follows. 

We discussed in section~\ref{sec:type0long} the worldsheet vertex operator dual to the operator $\exp(\beta\phi)$ in the spacetime SCFT, \ebphiws. This operator is independent of $R$, so we can act on it with the supersymmetry generators $\cG^\pm_{-\frac{1}{2}}$, which are described in the worldsheet theory by \susyops
\eqn[susygens]{{\cG^\pm_{-\frac{1}{2}}}=\oint\! \frac{dz}{2\pi i}\, e^{-\varphi/2}\,e^{-\frac{i}{2}H_\sl\pm\frac{i}{2}H_3\pm i\frac{a}{2}Z\pm i\frac{Y}{\sqrt{2k}}}~. }
Recall that $a=\sqrt{1-\frac2n}$.
We expect the commutator $\big[\cG^\pm_{-\frac{1}{2}},e^{\beta\phi}\big]$ to go to zero as $\beta\to 0$, so computing it for finite $\beta$ and isolating the leading term, that we expect to go like $\beta$, we find a complex fermion
\eqn[sppmm]{\cS^\pm=\big[\cG^\pm_{-\frac{1}{2}},\phi\big]~.}
Plugging in the vertex operators for the supercharges, \susygens, and $e^{\beta\phi}$, \ebphiws, we can calculate the vertex operator for the fermion,   
\eqna[GonExpWS]{\big[{\cG^{\pm}_{-\frac{1}{2}}} \, , \, e^{\beta\phi}\big]&~~\longleftrightarrow~~
\oint\! \frac{dz}{2\pi i}\, e^{-\varphi/2}
e^{-\frac{i}{2}H_\sl\pm\frac{i}{2}H_3\pm i\frac{a}{2}Z\pm i\frac{Y}{\sqrt{2k}}}(z) \; 
e^{-\varphi}\,e^{i H_\sl}\,\Phi^{(-1)}_{j;m}(0)
\\[.2cm]
&\hskip 2cm
=e^{-\frac{3}{2}\varphi} 
e^{\frac{i}2 H_\sl\pm \frac{i}2 H_3\pm i\frac{a}{2}Z\pm i\frac{Y}{\sqrt{2k}}}\,\Phi^{(-1)}_{j;m}~,}
where $m$ is given by \massshell, and $j$ is related to $\beta$ by \expII. 
Here and in some of the following equations, we suppress the right-moving dependence as it is unaffected by the manipulations.

The operator \eqref{GonExpWS} is written in the $-\frac32$ picture. We can move it to the canonical $-\half$ picture by acting with $e^{\varphi}G(z)$ with $G(z)$ given by \eqref{Gsu} and~\eqref{Gsl}. The non-trivial terms come from $j^-_\sl\psi^+_\sl$, $\psi^3_\sl J^3_\sl$, and $\psi^3_\su J^3_\su$. The result is
\eqn[GonExpWS2]{\big(m+j-1\big)e^{-\frac{\varphi}{2}+i\frac{3}{2}H_\sl\pm \frac{i}2 H_3 \pm i\frac{a}{2}Z\pm i\frac{Y}{\sqrt{2k}}} \,
\Phi_{j;m-1}^{(-1)}+ \Big(m-\frac{k}{2}\Big)
e^{-\frac{\varphi}{2}+\frac{i}2 H_\sl\mp \frac{i}2 H_3\pm i\frac{a}{2}Z\pm i\frac{Y}{\sqrt{2k}}}\,
\Phi_{j;m}^{(-1)}~.}
One can check that, as expected, as $\beta\to 0$ both terms in \eqref{GonExpWS2} go to zero. Comparing the terms of order $\beta$, we arrive at the vertex operator for the fermion
\eqn[blockferm]{\cS^\pm ~\longleftrightarrow~e^{-\frac{\varphi}{2}-\bar{\varphi}}e^{\pm i\frac{a}{2}Z\pm i\frac{Y}{\sqrt{2k}}+i\bar{H}_\sl}\left(
\frac{1}{1-k}e^{\frac{3i}{2}H_\sl\pm \frac{i}2 H_3}\,\Phi_{1-\frac{k}{2};\frac{k}{2}-1,\frac{k}{2}}^{(-1)}+ e^{\frac{i}2 H_\sl\mp \frac{i}2 H_3 }\,\Phi_{1-\frac{k}{2};\frac{k}{2},\frac{k}{2}}^{(-1)}\right)~,}
where we restored the right-moving part of the vertex operator.

Some comments are in order at this point:
\begin{enumerate}[1)]
\item The fermions $\cS^\pm$ should not be confused with the ones appearing at winding zero in the non-supersymmetric GSO~0 described in section~\ref{sec:typeIIshort}, $\PsiST^\pm$, \eqref{defpsii}. As explained there, those fermions do not exist in the supersymmetric GSO~2. 
\item Another way to obtain the vertex operator for the fermions $\cS^\pm$ is to start with that of $\partial_x\phi$ \dualbbzero, and apply the lowering operators $\cG^\pm_{\frac{1}{2}}$. One can check that this gives the same answer, \blockferm, as one expects from the supersymmetry algebra.
\item The operator $\partial_x\phi$ is one of four holomorphic dimension $(1,0)$ operators that we found above that are independent of the squashing deformation, and thus have the same form at the supersymmetric point. One can apply this procedure to the other three, and check that one does not find any additional dimension $(\hf,0)$ operators. In more detail, the $a=3$ component of the operator \eqref{dualcurr} is annihilated by $\cG^\pm_{\frac{1}{2}}$, in agreement with the fact that it is the $w=-1$ representation of the vertex operator of the spacetime $U(1)_R$ current, which is the bottom component of the spacetime superconformal multiplet. Acting with $\cG^\pm_{\frac{1}{2}}$ on the two operators \eqref{onemoresu2}, \eqref{finalthird}, one finds that one combination is annihilated, while the other gives the operators $\cS^\pm$ in \eqref{blockferm}, which again agrees with expectations. Indeed, our discussion above leads one to expect that out of the four holomorphic $(1,0)$ operators, two are the bottom components of two independent supercurrent multiplets in the block (which we exhibit in Appendix~\ref{app:blocksusy}), and the other two belong to a chiral superfield, whose fermionic components are $\cS^\pm$. 
\item One can find higher dimension holomorphic operators by acting with the raising operators \susygens\ on the dimension one operators. For example, \sppmm\ implies that acting with \eqref{susygens} on $\partial_x\phi$ gives the dimension $(\frac32,0)$ operators $\partial_x\cS^\pm$. Acting with \eqref{susygens} on \dualcurr\ (with $a=3$) gives the supercurrents $\cG^\pm$ in the block of the symmetric product.  We analyze these operators in appendix~\ref{app:blocksusy}.
\end{enumerate}

\noindent
To summarize, in our analysis so far we found that the spectrum of non-normalizable vertex operators of the (2,2) superconformal theory dual to string theory on $AdS_3\times \sqsphere$ in the sector with $w=-1$ contains a complex dimension one half fermion $\cS^\pm$ \blockferm, and four dimension one currents. The latter were already present in the type~0 theory that gave rise to the type~II one via chiral GSO, but here they belong to superfields described in comment (3) above. These superfields also contain the fermions $\cS^\pm$.


\subsection{The block theory}
\label{sec:blocktheory}

To understand this spectrum better, it is useful to recall that in theories of the sort we are studying here (string theory on $AdS_3$ with $(\NS,\NS)$ B-flux), the structure of the spacetime CFT closely mirrors that of the worldsheet one. This was a major theme in the literature from the late 1990's on this subject (\eg\ ~\rcite{Giveon:1998ns,Kutasov:1999xu}), and we have seen examples of this here as well.

In the worldsheet theory, we obtained the type II theories by demanding that the spacetime supercharges \eqref{susygens} be in the spectrum, \ie\ that all operators be local with respect to them. Thus, it is natural to do the same in spacetime. We start with the spacetime CFT constructed in section~\ref{sec:type0long} for the type 0 case, $\bR_\phi\times\sqsphere$, and demand that the operators $\cS^\pm$ be in the spectrum. Since these operators are charged under the various $U(1)$'s present in the type 0 model (the left-moving momentum on the $Y$ circle, the $U(1)_R$ in the $N=2$ minimal model, and the fermion number in $\bR_\phi\times\bS^1_Y$), we can write them in terms of the natural variables in the type 0 theory as
\eqn[Spmcandidate]{
\cS^\pm = \exp\Big[ \pm \frac i2 \Big( -H_3 + a\, Z + \sqrt{\frac2k}\, Y\Big) \Big] ~.
}
Note that here, as in a number of other examples above, while the notation looks similar to that in (for instance) equation~\susygens, the interpretation is different. There, the fields live on the worldsheet, and operators like \susygens\ are vertex operators in the worldsheet theory, while in \Spmcandidate\ the fields $\cS^\pm$ are operators in the spacetime SCFT.

The operators \Spmcandidate\ look like the worldsheet supercurrents $S^\pm_{-\frac12}$ in \susyops\ after stripping off the ghost $(\varphi)$ and longitudinal $(H_{\rm sl})$ parts, 
in a kind of light cone gauge construction.  The reason for their similarity is that the worldsheet vertex operator~\eqref{blockferm} is a spinor in the six dimensional target space $AdS_3\times \bS^3_\flat$; thus the corresponding operator in the spacetime CFT is simultaneously a spinor in $x$-space (the directions along the long string) as well as in $\bR_\phi\times \bS^3_\flat$ (the directions transverse to the string).  The former is accounted for by the fact that the operator is of dimension $h_{\ST}-\bar h_{\ST}=\half$, and the latter by the fact that it is a spin field for the four fermions $\chi^i$ introduced in section~\ref{sec:type0long}.%
\footnote{
In the $\sutwo$ representation theory underlying the squashed $\bS^3$ discussed in section~\ref{sec:currentalg}, one can see that the $\chi^i$ are in a vector representation while the $\cS^\pm$ are spinors.}
Finally, the operator must also have unit R-charge under the spacetime $N=2$ superconformal algebra, which accounts for the $Y$ dependence.

Thus, our proposal for the theory of the block in the supersymmetric type II theory is a chiral orbifold of the one we found in section~\ref{sec:type0long} for type 0, defined by requiring mutual locality of all operators with respect to \Spmcandidate. The resulting theory must be $N=2$ superconformal (since the corresponding string theory on $AdS_3$ is). The superconformal generators pair the bosons $\phi\pm iY$ with the fermions $\cS^\pm$, see equation~\eqref{sppmm}.  These supersymmetry generators are somewhat complicated by the fact that $\cS^\pm$ is a composite operator~\eqref{Spmcandidate} in our construction.  There is however a convenient trick to deriving them.

The exotic realization of the $N=2$ superconformal algebra on $\bR_\phi\times \bS^3_\flat$ involving $\cS^\pm$ is related to the standard one 
\eqna[stdsusy]{
\cG^{\pm}_{\rm free} = \pm e^{\pm i\medhat H_3}(\partial \phi\tight\mp i\partial \medhat Y)\pm Q_\ell\,\partial e^{\pm i \medhat H_3}
~~&,~~~~
\cJ_{R}^{\rm free} = i\partial \medhat H_3 +iQ_\ell\,\partial \medhat Y~,
\\[.2cm]
\cG^\pm_{\LG} = \psi^\pm_{\pf} \, \exp\Big[\pm \frac ia \medhat Z \Big]
\qquad\quad~~&,~~~~
\cJ_{R}^{\LG} = i\,a\,\partial \medhat Z~,
}
by a rotation in field space
\eqn[fieldredef]{
\big(  H_3, Z, Y \big) = \cR\cdot \big( \medhat H_3, \medhat Z, \medhat Y \big) ~,
}
as we demonstrate in appendix~\ref{app:blocksusy}.  
This standard supersymmetry pairs the scalars $\phi\pm i\medhat Y$ parametrizing $\bR_\phi\times \bS^1$ with the vector fermions $\chi^\pm=e^{\pm i\medhat H_3}$ of section~\ref{sec:type0long}. 
A feature of this field redefinition is that it maps the total R-current $\cJ_R=\cJ_R^{\rm free}+\cJ_R^{\LG}$ of the standard realization of supersymmetry to the spacetime R-current $\cJ_R=i\sqrt{2k}\,\partial Y$, realizing the R-symmetry as geometrical translations along $\bS^1_Y$.

Because this rotation is simply a linear field redefinition, one can work with the standard $N=2$ supersymmetry on $\bR_\phi\times\bS^3_\flat$ (employing the relation $\bS^3_\flat = \big(\bS^1\times \frac\sutwo\uone\big)/\bZ_n$), and then apply the transformation~\eqref{fieldredef}.
Rather than working with the exotic supersymmetry in the field space frame that includes $\cS^\pm$, one can simply transform all quantities to the frame of the standard supersymmetry and work there.
In what follows we will find it much more convenient to do so, for instance in constructing the twist fields of the symmetric orbifold that build the wall deformation of the symmetric orbifold.
Note that the R-symmetry in the standard realization of supersymmetry is not realized geometrically as translations on the $Y$ circle, but rather measures charge along some linear combination of $(\medhat H_3,\medhat Z,\medhat Y)$ given by $\cJ_R=\cJ_R^{\rm free}+\cJ_R^{\rm pf}$ of~\eqref{stdsusy}; thus the marginal line in this standard realization, while still generated by the deformation $\cJ_R\bar\cJ_R$, no longer corresponds to simply changing the radius of the $Y$ circle as in~\eqref{susyops}.

A similar analysis can be done in the type II theories without supersymmetry. In this case, instead of the supercharges $\oint\! dz\, S^{\pm}_{-\frac{1}{2}}$ we act with the fermionic modes $\oint\! dz\, \Psi^{\pm}_{-\frac{1}{2}}$, where $\Psi^\pm_r$ is given by \eqref{GSO0currents}. Comparing \eqref{susyops} and \eqref{GSO0currents}, we see that they only differ in the sign in front of the $H_3$ term. 

Once again there is a collection of holomorphic operators.  As in section~\ref{sec:type0long}, we can deduce some of them using spacetime symmetries.   
Starting again with the exponential operator $e^{\beta\phi}$, equation~\eqref{ebphiws}, and acting with the $-\frac{1}{2}$ mode of the fermion operators $\Psi^\pm(x)$, equation~\eqref{GSO0currents}, \eqref{defpsii}, one obtains an operator of dimension $(\hf,0)$ in the limit $\beta\to0$ 
\eqn[blockfermzero]{e^{-\frac{\varphi}{2}-\bar{\varphi}}e^{i\bar{H}_\sl\pm i\frac{a}{2}Z\pm i\frac{Y}{\sqrt{2k}}+\frac{i}2 H_\sl\pm \frac{i}2 H_3 }\,\Phi_{1-\frac{k}{2};\frac{k}{2},\frac{k}{2}}^{(-1)} ~.
}
In this case, the coefficients don't vanish in the limit, as expected, since $\Psi^\pm_{-\frac{1}{2}}$ acting on the identity is not zero. In fact, \blockfermzero\ has a simple interpretation -- it is the $w=-1$ representation of the fermion vertex operator. 

Following the same logic for the GSO~0 projection that we employed in the GSO~2 case, one arrives at an expression for this operator in the block theory 
\eqn[Spmhat]{
\PsiST^\pm = \exp\Big[ \pm \frac i2 \Big( H_3 + a\, Z + \sqrt{\frac2k}\, Y\Big) \Big] ~.
}
Note that under the GSO~2 projection, the theory in the block in spacetime has both the fermions $\cS^\pm$ as well as the supersymmetry currents $\cG^\pm$.  Similarly, under the GSO~0 projection, there are the fermions $\PsiST^\pm$, and also the same procedure can be used to derive a spin-$\frac32$ supersymmetry current $\cG^\pm$ in the block (see Appendix~\ref{app:blocksusy}), which again differs from the one for GSO~2 by a flip in the sign of the $H_3$ contributions.  Thus, although the full theory with GSO~0 is not supersymmetric, the asymptotic structure at large $\phi$ has such a supersymmetry, which we can think of as being softly broken by the wall.  This softly broken supersymmetry is again related to the standard $N=2$ supersymmetry~\eqref{stdsusy} in the block by a field space rotation analogous to~\eqref{fieldredef}, which we also exhibit in Appendix~\ref{app:blocksusy}.

Thus even though in the worldsheet theory one must choose between having a non-normalizable holomorphic spin-1/2 operator in spacetime (GSO~0) or instead a holomorphic spin-3/2 operator (GSO~2), and these options are mutually exclusive, it seems that the operator spectrum in the winding one sector always has both.  This result extends a theme we encountered already in the type 0 theory in section~\ref{sec:type0long}, that the unit winding sector contains more holomorphic operators than the zero winding sector, leading to apparent symmetries that we expect to be absent in the full theory.  Here the issue is brought home even more forcefully~-- it seems that at large $\phi$ there is an essentially unique theory that corresponds to a particular symmetric product (the block fermions and supersymmetry currents differ only by a trivial parity flip in $H_3$).  The two chiral GSO projections differ only in the structure of the wall that regulates the strong coupling region.

We can now lay out the GSO projection of the spacetime theory.  As in the type 0 theory, once again the GSO projection in the block of the symmetric product (or more generally in a given cycle of the symmetric product) parallels that of the worldsheet theory~\eqref{etaYZcharges}-\eqref{GSOphase}: 
\eqna[blockGSOphase]{
(-1)^F &= \exp\Big[i\pi\Big( \epsilon_3\eta_3 + \epsilon_\su\eta_\su \Big)\Big]~,
\\[.2cm]
\Omega &=
\exp\Big[i\pi \Big( (m'\tight-\mbar') + (\eta_\su\tight-\bar\eta_\su) +\frac{n-2}2(w'\tight-\bar w') \Big) \Big] ~,
}
with $\epsilonb_2=(\epsilon_3,\epsilon_\su)=(-1,1)$ and $\epsilonb_0=(1,1)$.
The fermions $\cS^\pm$ and $\PsiST^\pm$ have the underlying $\sutwo$ quantum numbers~\eqref{Sqnums}
\eqn[Sqnums2]{
j'=m'=\mbar'=0
~,~~~~
w'=\bar w'=\bar\eta_\su=\half
~,~~~~
\eta_\su=1~,
}
and thus $\Omega=i$, which compensates $(-1)^F=-i$ to allow the operators $\cS^\pm$ in the GSO~2 spectrum, and $\PsiST^\pm$ in the GSO~0 spectrum.  On the other hand, the fermions $\chi^i$ (introduced above~\eqref{sofourvec}) all have $(-1)^F\Omega=-1$, and are thus projected out.

The type II symmetric product can thus have $(\!R,\NS)$ or $(\NS,\!R)$ cycles, allowed by the GSO projection.  The cycles themselves are fermionic, and have $h_\ST-\bar h_\ST=\half$, but so long as the number of such cycles is even one has an allowed state/operator in the full CFT.


\section{The wall}
\label{sec:wall}


\subsection{The story so far and the strategy going forward}
\label{sec:sofar}

Before continuing the development of our holographic duality, it is worthwhile to pause and summarize the picture obtained so far. We showed that in string theory on $AdS_3$ with NS B-field and $R_{AdS}<\lstr$ (or $k<1$, \kval), the spacetime CFT can be described in some region by a symmetric product of $\bR_\phi\times\sqsphere$. Here $\bR_\phi$ is the real line with linear dilaton $Q_\ell$ \longss, while $\sqsphere$ is the squashed three-sphere CFT, obtained from $SU(2)_n$ supersymmetric WZW by the marginal deformation described in section~\ref{sec:currentalg}. 

The region in which this description is valid is large positive $\phi$. From the point of view of the bulk theory  (string theory on $AdS_3$) this is the region near the boundary of $AdS_3$, while from the point of view of the spacetime CFT it is the region where the string coupling on $\bR_\phi$, $\exp(- Q_\ell\phi/2)$, goes to zero. What the restriction to large $\phi$ means in practice is that we can only compare non-normalizable and delta-function normalizable operators in string theory on $AdS_3$ with their analogs in the symmetric product. The reason is that, as is familiar from CFT's such as Liouville theory, the spectrum of such operators is insensitive to any modifications of the background at finite $\phi$. On the other hand, the spectrum of normalizable operators, as well as correlation functions, are sensitive to such modifications, and are not expected to be well described by the above symmetric product.     

We arrived at the above description of the large $\phi$ structure of the spacetime CFT by systematically analyzing the worldsheet theory on $AdS_3$. We showed that the sector of long strings with $\sltwo$ winding $w\tight=-1$ in that theory gives precisely the spectrum of non-normalizable and delta-function normalizable operators in the building block of the symmetric product, $\bR_\phi\tight\times\sqsphere$, while sectors with higher winding give rise to the twisted sectors of the symmetric product orbifold. The (radial dependence of the) wavefunctions of the bulk and boundary operators agree as well. In this sense, the large $\phi$ symmetric product form of the spacetime CFT was {\it derived} from the bulk $AdS_3$ description. 

It is clear that the picture of the previous sections must receive corrections at finite $\phi$. There are many ways of seeing that. For example, if the spacetime CFT was given by a symmetric product involving the linear dilaton theory for all $\phi$, all correlation functions of non-normalizable and delta-function normalizable operators would be singular, which does not agree with what one finds in the worldsheet theory. Also, the symmetric product theory does not contain any normalizable states while, as described in previous sections, the bulk theory does contain such states. 

What kind of modification of the symmetric product structure do we expect? In other examples of theories of this sort, \eg\ those mentioned in section~\ref{sec:introsummary}, such a modification takes the form of an infrared wall, a modification of the background that prevents the field $\phi$ in the spacetime CFT from exploring the region $\phi\to-\infty$, and thus leads to a non-singular theory. It is natural to expect such a wall to make an appearance here as well. For example, the wall is presumably responsible for the reflection relation \reflection\ that the spacetime CFT inherits from the worldsheet theory. The goal of this section is to show that this expectation is realized, and construct the wall.

We will do this by following the strategy we used in studying the asymptotic structure in previous sections. We will use the bulk theory as a tool to see how the symmetric product is modified at finite $\phi$, and deduce the form of the wall from this analysis. 

A sensitive probe of such modifications is the properties of the holomorphic operators we found in sections \ref{sec:type0long}, \ref{sec:typeIIlong}. As discussed there, we expect most of these operators to only be holomorphic in the large $\phi$ approximation that gives rise to the symmetric product. The deviations from the symmetric product structure at finite $\phi$ are expected to break holomorphy, and thus studying them seems like a good strategy for probing the structure of the wall. We next turn to this study. 

\subsection{Deviation from holomorphy}
\label{sec:nonholo}
The first operator we will consider is $\partial_x\phi$ in the spacetime CFT. This operator is non-normalizable, since the corresponding wavefunction diverges like $\exp(Q_\ell\phi/2)$. It exists in all the theories we considered (type 0, type II with all possible GSO's, and for all values of the squashing parameter), and corresponds in the bulk worldsheet description to the vertex operator on the r.h.s. of \dualbbzero. 

As mentioned in section \ref{sec:type0long}, this operator is naively holomorphic, but its holomorphy is expected to be broken in the full theory. We now examine its properties more closely. 

The vertex operator for $\partial\phi$ was found in section \ref{sec:type0long} to be given by \dualbbzero. In order to compute $\bar\partial\partial\phi$, we need to find the commutator of this vertex operator with $\bar J_0^-$. This is the computation we turn to next. 

We start with the commutator of $\bar J_0^-$ with the first term on the r.h.s. of \dualbbzero ,

\eqna[ddbarphiWS]{
&~~\left[\bar{J}_0^-,(\partial\varphi+i\partial H_\sl)e^{-\varphi-\bar\varphi} \, e^{i(H_\sl+\bar H_\sl)} \, 
\Phi^{(-1)}_{1-\frac{k}{2};\frac{k}{2}-1,\frac{k}{2}}\right]
\\
&\hskip 2cm 
= e^{-\varphi-\bar\varphi}\,
e^{iH_\sl}\, (\partial\varphi+i\partial H_\sl)\,
\Big(\sqrt2\,\bar\psi^3_\sl\, \Phi^{(-1)}_{1-\frac k2;\frac k2-1,\frac k2}
+ e^{i\bar H_\sl}\bar j^-_{\sl;0}\, \Phi^{(-1)}_{1-\frac k2;\frac k2-1,\frac k2} \Big) 
\\
&\hskip 2cm 
= \big\{\bar{Q}_{\it BRST}, (\partial\varphi+i\partial H_\sl)e^{-\varphi-2\bar\varphi}\bar\partial\bar\xi e^{iH_{\sl}}\Phi^{(-1)}_{1-\frac k2;\frac k2-1,\frac k2}\big\}~.
}
In the last line we used \eqref{Lmbda}, and the fact that $\bar m+j-1=\bar m-\frac k2=0$ for the operator on the r.h.s. of \dualbbzero. Similarly, the commutator of $\bar J_0^-$ with the second term on the r.h.s. of \dualbbzero is BRST exact. Hence, the operator $\bar\partial\partial\phi$ corresponds in the bulk theory to a BRST exact worldsheet operator. This seems to suggest that $\partial_x\phi$ is holomorphic in the full theory. 

It turns out that this conclusion is premature. One way to see that is to note that when using \eqref{Lmbda},\footnote{With left and right-movers exchanged, since there we were computing the $x$ derivative of an operator, while here we are interested in the $\bar x$ derivative.} the factor $\bar m+j-1$, that vanishes for the operator \dualbbzero, multiplies the operator 
$\Phi^{(-1)}_{1-\frac k2;\frac k2-1,\frac k2-1}$, which diverges. This operator is an example of the discussion around equation \modphiop. As explained there, approaching the particular values of $(j,m,\bar m)$ needed here corresponds to approaching an LSZ pole, and one has to take the limit more carefully, as in \modphiop.

In order to do that, we move slightly away from \dualbbzero, by turning on a small non-zero $\beta$  in \dualbb. In this limit, $(j,m,\bar m)$ in \dualbb\ shift by an amount of order $\beta$ from their values at $\beta=0$, which are $j=1-\frac k2$, $m=\bar m=\frac k2$. For $\beta=\epsilon Q_\ell$, one has, to first order in $\epsilon$,
\eqn[polevals2]{
j = 1-\frac k2+\epsilon
~~,~~~~
m= \mbar= \frac k2+\frac{1-k}{k}\epsilon
~~,~~~~
m+j-1=\mbar+j-1 = \frac{\epsilon}{k}
~,
}
Looking back at \dualbb\ we see that to leading order in $\epsilon$ we can cancel the prefactors on the left and right hand sides, both of which go like $\epsilon$. Thus, we have
\eqn[dualbb2]{\partial_x\phi \, e^{\beta\phi} ~~\longleftrightarrow~~ 
\frac{1}{\sqrt{2}(1-k)}(\partial\varphi+i\partial H_\sl)e^{-\varphi-\bar\varphi} \, e^{i(H_\sl+\bar H_\sl)} \, \Phi_{j;m-1,\bar m}^{(-1)}-e^{-\varphi-\bar\varphi}\psi^3_\sl e^{i\bar{H}_\sl}\Phi_{j;m,\bar m}^{(-1)}~.
}
Note that the operators on both sides of \eqref{dualbb2} are non-normalizable, both at finite $\epsilon$ and in the limit $\epsilon\to 0$. 

Next we compute the $\bar x$ derivative of \eqref{dualbb2} using the techniques of section~\ref{sec:type0long}. We start by considering the first term on the r.h.s. of \eqref{dualbb2}, while the second term will be discussed shortly. We find
\eqna[dualbb3]{&~ \left[\bar{J}_0^-,(\partial\varphi+i\partial H_\sl)e^{-\varphi-\bar\varphi} \, e^{i(H_\sl+\bar H_\sl)} \, 
\Phi^{(-1)}_{j;m-1,\bar{m}}\right]
\\
&\hskip 2cm 
= e^{-\varphi-\bar\varphi}\,
e^{iH_\sl}\, (\partial\varphi+i\partial H_\sl)\,
\Big(\sqrt2\,\bar\psi^3_\sl\, \Phi^{(-1)}_{j;m-1,\bar{m}}
+ e^{i\bar H_\sl}\bar j^-_{\sl;0}\, \Phi^{(-1)}_{j;m-1,\bar{m}} \Big)
\\
&\hskip 2cm
= (\bar{m}+j-1)e^{-\varphi-\bar\varphi}\,
e^{iH_\sl+i\bar{H}_\sl}\, (\partial\varphi+i\partial H_\sl)(\bar{\partial}\bar\varphi+i\bar{\partial} \bar{H}_\sl)\,
\Phi^{(-1)}_{j;m-1,\bar{m}-1}
\\
&\hskip 3cm
-\sqrt{2}\left(\bar{m}-\frac{k}{2}\right)e^{-\varphi-\bar\varphi}\,
e^{iH_\sl}\, (\partial\varphi+i\partial H_\sl)\bar{\psi}^3_\sl\,
\Phi^{(-1)}_{j;m-1,\bar{m}}~,
}
where in the last equality we used \eqref{Lmbda} and dropped a BRST exact term. 

In equation \eqref{dualbb3}, the quantum numbers $(j,m,\bar m)$ take the values \eqref{polevals2}. The last step is to take $\epsilon\to 0$. In this limit, the coefficients of both terms in the last expression of \eqref{dualbb3} go to zero. However, while the first term sits on a LSZ pole, the second term does not. Thus, only the first term in this last expression gives a finite contribution, which is equal to the {\it normalizable} operator
\eqn[holoviol]{
e^{-\varphi-\bar\varphi}\,
e^{i(H_\sl+\bar H_\sl)}\, 
(\partial\varphi+i\partial H_\sl) 
(\bar\partial\bar\varphi+i\bar\partial \bar{H}_\sl)\,
\Phi^{(-1)}_{1-\frac k2;\frac k2-1,\frac k2-1}~. 
}

To compute the $\bar x$ derivative of \eqref{dualbb2}, we also need to include the contribution from the commutator of $\bar{J}^-_0$ with the second term on the r.h.s. of \eqref{dualbb2}. This can be computed in a similar way to \eqref{dualbb3}. However, in this case the operator does not sit on top of an LSZ pole, and hence this contribution vanishes as $\epsilon\to0$. 

Thus, we find that the l.h.s. of \eqref{dualbb3}, which goes in the limit $\epsilon\to0$ to $\partial_{\bar x}\partial_x\phi$, is equal in the bulk theory to the operator \eqref{holoviol}. This implies that the holomorphy of the operator $\partial_x\phi$ is violated -- $\partial_{\bar x}\partial_x\phi$ does not vanish, but instead is given by the operator in the spacetime CFT that corresponds to \holoviol.

Coming back to \ddbarphiWS, which seemed to suggest that $\partial\phi$ is holomorphic, we can now state more clearly the status of this equation. The analysis done there omitted the contribution to $\partial_{\bar x}\partial_x\phi$ of the r.h.s. of \eqref{dualbb3}. The reason for the omission was that the coefficient $\bar m+j-1\to 0$ in the limit $\epsilon\to 0$. This is justified at the level of non-normalizable operators, but the limit is finite for the normalizable operator \holoviol. 

Thus, the precise statement that follows from our analysis is that $\partial_{\bar x}\partial_x\phi$ {\it vanishes at the level of non-normalizable operators}, but it receives a {\it non-vanishing normalizable contribution}. This contribution must be due to the wall, and in fact we deduce from the calculation that the wall is described by a normalizable operator, the CFT dual of \holoviol. 

\subsection{Identifying the wall}
\label{sec:whatsthewall}

Having derived the form of the operator that provides the wall from the worldsheet point of view, \holoviol, it remains to identify it in the spacetime symmetric product theory. The fact that \holoviol\ belongs to the winding $-1$ sector suggests that it is an operator in the building block of the symmetric product. The results of section~\ref{sec:review} further imply that it has spacetime scaling dimension $(1,1)$, as expected (since it is equal to $\bar\partial\partial\phi$ that has this dimension, and the full theory is conformal). Thus, it is a marginal operator. We can compute its (leading) radial profile at large $\phi$ by using~\eqref{jSTjWS} with $j_\WS=1-\frac k2$. This gives $j_\ST=1$, and therefore $\beta=-Q_\ell$.%
\footnote{Note that we are now discussing normalizable wavefunctions, for which we need to take $j_\ST\to1-j_\ST$ in the spacetime wavefunction~\eqref{twistphidep}, and similarly for the wavefunction on the worldsheet.  Equation~\eqref{jSTjWS} is invariant under this substitution on both sides.}

A quick way to identify the operator that corresponds to \holoviol\ in the spacetime CFT is the following. We saw in section \ref{sec:type0long} that the operator $\partial_x e^{\beta\phi}$ in the block of the spacetime CFT corresponds on the worldsheet to the vertex operator on the r.h.s. of~\eqref{dualbb}. This vertex operator differs from the one for $e^{\beta\phi}$, given by the r.h.s. of \ebphiws, only in its left-moving structure (on the worldsheet).

As we have noted before, a recurring theme in string theory on $AdS_3$ with NS $B$-field is that chirality on the worldsheet is mapped to chirality in spacetime. Thus, it is natural to expect that the operator $\partial_x\phi\,\partial_{\bar x}\phi\, e^{\beta\phi}$ in the spacetime theory is described on the worldsheet by a vertex operator whose right-moving part is the same as the left-moving part in~\eqref{dualbb}. The resulting operator map is valid for non-normalizable operators, but if we formally continue $\beta\to -Q_l$, we find that the worldsheet operator \holoviol\ corresponds to the normalizable spacetime operator 
\eqn[ddbarphiRHS]{
\partial_x\phi\,\partial_{\bar x}\phi \, e^{-Q_\ell\phi}  ~.
}
Further support for this identification will be presented in the next subsection.

The wall inferred from the worldsheet theory corresponds to adding the operator 
\ddbarphiRHS\ to the Lagrangian in the block,
\eqn[Lblock]{
\LL_{\rm block}=\LL_0+\lambda\,\partial_x\phi\,\partial_{\bar x}\phi \, e^{-Q_\ell\phi} 
}
where $\LL_0$ is the Lagrangian of the sigma model on $\bR_\phi\times\sqsphere$, $\LL_0=\partial\phi\bar\partial\phi+\cdots$, and $\lambda$ is a coupling that we will omit below; for example, it can be absorbed in a shift of $\phi$. The interaction term in \Lblock\ is the leading correction to the free Lagrangian $\LL_0$. As is clear from the general form~\eqref{Phiasymp} of the operators involved in the derivation, there are further corrections that are higher order in $\exp\left(-Q_\ell\phi\right)$. 

The Lagrangian \Lblock\ leads to a modified equation of motion for the field $\phi$, 
\eqna[eqnphi]{
\partial_{\bar x}\partial_x\phi  \simeq
\partial_x\phi\,\partial_{\bar x}\phi \, e^{-Q_\ell\phi}~,
}
where $\simeq$ means that on the r.h.s. we are omitting a multiplicative constant and keeping only the leading term at large $\phi$.
Superficially, it looks like \eqnphi\ violates the holomorphy of $\partial\phi$, as expected from the worldsheet analysis, but in fact one can remove this apparent violation of holomorphy by a field redefinition,
\eqna[phiredef]{
\phi\to\phi+A\,e^{-Q_\ell\phi}+O\big(e^{-2Q_\ell\phi}\big)
}
with $A$ tuned such that $\partial\phi$ remains holomorphic to order $\exp\left(-Q_\ell\phi\right)$. A simple way to see this is to note that the Lagrangian \Lblock\ can be written as $\LL_{\rm block}=G(\phi)\partial\phi\bar\partial\phi$, with the field space metric $G(\phi)=1+\lambda e^{-Q_\ell\phi}+O\left(e^{-2Q_\ell\phi}\right)$. One can set the metric $G(\phi)$ to one by a reparametrization~\phiredef.

Thus, it looks like the deformation \Lblock\ does not actually achieve the goal of keeping the field $\phi$ away from the strong coupling region. Interestingly, the worldsheet theory contains the resolution of this problem. The operator \holoviol\ belongs to the principal discrete series, and as such has an FZZ dual representation~\eqref{FZZduality}, which in this case involves an operator in the $w=-2$ sector, given by
\eqn[weq2wall]{
e^{-\varphi-\bar\varphi}\,
e^{i(H_\sl+\bar H_\sl)}\, 
(\partial\varphi+i\partial H_\sl) 
(\bar\partial\varphi+i\bar\partial \bar{H}_\sl)\,
\Phi^{(-2)}_{k;k,k}  ~.
}
As dicussed earlier in the paper, the operators \eqref{holoviol} and \eqref{weq2wall} are not distinct. FZZ duality states that there is one operator whose large $\phi$ expansion contains a component with a radial profile $\exp[-Q(1-\frac k2)\phi]$ and another component with radial profile $\exp(-Qk\phi)$ which, using \Qtilde-\jSTjWS, correspond to $\exp(-Q_\ell\phi_{\rm ave})$ and $\exp(-Q_\ell\frac{n+1}2\phi_{\rm ave})$, respectively, in the spacetime CFT ($\phi_{\rm ave}$ is defined in~\eqref{phiave}, and is identified with $\phi$ on the worldsheet). The former gives rise to the deformation \Lblock\ in the block of the symmetric product. The latter, being an operator in the sector with winding $w=-2$, is expected to live in the $\bZ_2$ twisted sector of the orbifold. Our next goal is to construct this operator. 

\subsection{Supermultiplet structure in the (2,2) theory}
\label{sec:susystruc}

In order to identify the operator \eqref{holoviol}, \eqref{weq2wall} in the spacetime theory, it is useful to consider it in the theory with $(2,2)$ superconformal symmetry (corresponding to GSO~2 for both the left and right-movers on the worldsheet, in the analysis of section \ref{sec:typeIIshort}). As is familiar from general studies of $(2,2)$ SCFT's, a large class of moduli in such theories is obtained by starting with (anti)chiral operators of dimension $\left(\half,\half\right)$ and applying the appropriate supercharges. There are four classes of such operators~-- the $(c,c)$ operators, with R-charge $(1,1)$, their adjoints the $(a,a)$ operators, with R-charge $(-1,-1)$, as well as $(c,a)$ and $(a,c)$ operators with R-charges $(1,-1)$ and $(-1,1)$, respectively.  For instance, in the context of Calabi-Yau sigma models, the $(c,c)$ and $(a,a)$ operators describe complex structure deformations, while $(a,c)$ and $(c,a)$ operators correspond to K\"ahler moduli. 

In our case, the operator \eqref{holoviol}, \eqref{weq2wall} is expected to be the highest component of a chiral superfield.  It is a normalizable operator in the $(\NS,\NS)$ sector of the worldsheet theory, so the bottom components are expected to be normalizable operators in the worldsheet $(\!R,\!R)$ sector. 

We have constructed such operators in section \ref{sec:typeIIshort}. In particular, the $(c,c)$ operator $\cY_0^{(-1)}$ of~\eqref{IIB22 Ywneq0 susypt}, its $(a,a)$ conjugate, and the lowest dimension operators in the $(c,a)$ and $(a,c)$ spectrum~\eqref{caspec} with $\ell\tight=1,L\tight=0$, have just the right properties.  From~\eqref{GonExpWS}, \eqref{blockferm} and \eqref{ebphiws}, we have that these $\hf$-BPS operators correspond to the spacetime operators
\eqn[SSbar]{
\cS^\pm \bar\cS^{\pm} e^{-Q_\ell\phi} 
~~\longleftrightarrow~~
e^{-\frac\varphi2-\frac{\bar\varphi}2}\, e^{\frac{3i}2(H_\sl+\bar H_\sl)}\, 
e^{\pm\frac i2 (H_3+aZ+\sqrt{\frac2k}\,Y) \pm \frac i2 (\bar H_3+a\bar Z+\sqrt{\frac2k}\,\bar Y)} \,
\Phi^{(-1)}_{1-\frac k2,\frac k2-1,\frac k2-1} ~,}
where the $\pm$ on the left and right are independent,
and on the l.h.s. we only wrote the leading behavior of the operator at large $\phi$. 

Acting with the supersymmetry generators~\eqref{susygens} on the operators on the r.h.s. of \SSbar, we find the highest components (writing only the left-moving component for simplicity)
\eqna[NSwall]{
\big\{ \cG^\mp_{-\half} , \, \cS^\pm  e^{-Q_\ell\phi} \big\}  
&\longleftrightarrow \oint\! \frac{dz}{2\pi i}\, e^{-\frac{\varphi}{2}}
e^{-\frac{i}{2}H_\sl \mp( \frac{i}{2}H_3 + i\frac{a}{2}Z + i\frac{Y}{\sqrt{2k}})}(z) \; 
e^{-\frac{\varphi}{2}} \,
e^{\frac{3i}{2}H_\sl \pm (\frac i2H_3 + i\frac{a}{2}Z + i\frac{Y}{\sqrt{2k}})} \,
\Phi^{(-1)}_{1-\frac k2;\frac k2-1}  
\\[.2cm]
&= -\frac 12\Big[ \partial\varphi+i\partial H_\sl \pm \Big(i\partial H_3+ ia\,\partial Z +i\sqrt{\frac2k}\,\partial Y\Big)\Big] \,
e^{-\varphi}\, e^{i H_\sl}\, \Phi^{(-1)}_{1-\frac k2;\frac k2-1} 
\\[.2cm]
&\equiv\cV^\pm~,}
where the contour integral is taken around the location of the second operator.  Combining left- and right-movers, the sum of all four $\hf$-BPS operators
\eqn[BPSsum]{
\big(\cV^+ + \cV^-\big)\big(\bar\cV^+ + \bar\cV^-\big) = 
 \,e^{-\varphi-\bar\varphi}\,
e^{i(H_\sl+\bar H_\sl)}\, 
(\partial\varphi+i\partial H_\sl) (\bar\partial\varphi+i\bar\partial \bar{H}_\sl)\,
\Phi^{(-1)}_{1-\frac k2;\frac k2-1,\frac k2-1} }
gives~\eqref{holoviol}. A calculation of the l.h.s. of~\eqref{BPSsum} directly in the spacetime CFT reproduces the operator~\eqref{ddbarphiRHS}, that we argued in the previous subsection is the spacetime counterpart to the worldsheet operator on the r.h.s. of \BPSsum.

Thus, we conclude that the ``wall operator'' \holoviol\ is a particular combination of $(c,c)$, $(c,a)$, $(a,c)$ and $(a,a)$ moduli. 
FZZ duality implies that the operator \eqref{weq2wall} must also have this property. Therefore, to find it in the symmetric orbifold, we need to look for a chiral operator of dimension $1/2$ in the $\bZ_2$ twisted sector. This is the problem we turn to next.

\subsection{The \texorpdfstring{$\bZ_2$}{} twist deformation}
\label{sec:Z2twist}

The worldsheet calculations above indicate that the wall has a component in winding sector two, equation~\eqref{weq2wall}. The dual spacetime field is expected to live in the $\bZ_2$ twisted sector of the symmetric orbifold. The goal of this subsection is to construct that operator. 

For this purpose, it is convenient to describe the block of the symmetric product, $\bR_\phi\times\sqsphere$, as (a $\bZ_n$ orbifold of)  
\eqn[rphiy]{
\bR_\phi\times\bS^1_Y\times LG_n ~.
}
The superconformal generators acting on this space are the standard ones \eqref{stdsusy}, that pair $\phi$, $Y$, and two fermions $\chi^\pm=\chi_\phi\pm i\chi_Y$ into a chiral superfield, and a separate superconformal generator that acts on the Landau-Ginsburg superfield. 

We are interested in studying the sector of the theory that is twisted by the $\bZ_2$ that exchanges two copies of the block theory. For the Landau-Ginzburg model the (anti)chiral $\bZ_2$ twist field $\Sigma^\pm_{\LG}$ that creates the BPS twisted ground states has the standard dimension~\rcite{Klemm:1990df,Fuchs:1991vu}
\eqn[SigLGdim]{h\left[\Sigma_\LG^\pm\right]=\frac{c_\LG}{12}=\frac14-\frac1{2n}} 
(for details, see Appendix~\ref{app:twistops}).

For the $(\phi,Y)$ theory one can proceed as follows. We decompose the two copies into   
symmetric and antisymmetric combinations
\eqna[SAfields]{
\phi_S=\frac1{\sqrt2}\big(\phi_{(1)}+\phi_{(2)}\big)
~&,~~~~
\phi_A=\frac1{\sqrt2}\big(\phi_{(1)}-\phi_{(2)}\big)~,
\\[.2cm]
Y_S=\frac1{\sqrt2}\big(Y_{(1)}+Y_{(2)}\big)
~&,~~~~
Y_A=\frac1{\sqrt2}\big(Y_{(1)}-Y_{(2)}\big) ~,
}
and similarly for the corresponding fermions $\chi^\pm_{S,A}$ built out of (anti)symmetric combinations of $\chi^\pm=(\chi_\phi^{~}\pm i\chi_Y^{~})/\sqrt2$ in each block. 

The fields $\phi_{(i)}$ are described by a linear dilaton CFT with slope $Q_\ell$. After the change of variables \SAfields, we find one field, $\phi_S$, with linear dilaton slope $\sqrt2\,Q_\ell$, and another, $\phi_A$, with no linear dilaton.

The antisymmetric fields $\phi_A$, $Y_A$, $\chi^\pm_A$ are all $\bZ_2$ twisted. The twist operators for the bosons $\sigma^{~}_\phi$ and $\sigma^{~}_Y$ have dimension $\frac{1}{16}$; there is a single such twist operator because there is a single fixed locus. 
For the fermions $\chi^\pm_A$, the appropriate twist operators 
$\sigma^{\pm}_{\chi_A^{~}}$
have dimension $\frac18$ and carry a half unit of fermion charge (they act as spin fields).

Putting together all the ingredients, we find $\bZ_2$ twisted ground states created by the operators 
\eqn[justtwist]{\sigma_{\phi^{~}_A} \sigma^{~}_{Y_A} \sigma^{\pm}_{\chi_A}\Sigma^\pm_{\LG}}
(where the $\pm$ are correlated), whose dimension is 
\eqn[excess]{2\times \frac{1}{16}+\frac18+\frac14-\frac{1}{2n}=\frac12-\frac{1}{2n} ~.
}
In \justtwist\ we explicitly wrote only the left-moving part of the operator. We will put left and right-movers together shortly.

The operator \justtwist\ is (anti)chiral,\footnote{The R-charge of \justtwist, $\pm\left[\half+\left(\half-\frac1n\right)\right]$, is equal to plus or minus twice its scaling dimension \excess.}   
but its dimension is smaller than $\half$. We need to modify it so that its dimension is exactly $\half$, while maintaining chirality.  
A simple way of doing that is to multiply \justtwist\ by a contribution from the untwisted fields, of the form 
$$e^{\beta(\phi_S\mp i Y_s)}$$
with $\beta$ determined such that the dimension of this operator is precisely equal to the amount we need to cancel in \excess, $\frac{1}{2n}$. 

Recalling that the linear dilaton slope for $\phi_S$ is $\sqrt2\,Q_\ell$, we thus conclude that 
$$-\half\beta Q_\ell\sqrt2=\frac{1}{2n}~,$$
\ie\ $\beta=-\frac{1}{2\sqrt k}$. 
Putting all the elements together, we find that the $\bZ_2$ sector of the symmetric orbifold contains (anti)chiral operators of dimension $1/2$, given by
\eqn[twist2]{
\Sigma^\pm = \Sigma^\pm_{\rm free}\,\Sigma^\pm_\LG = \exp\Big[ -\frac{1}{2\sqrt{k}}\big( \phi_S \mp iY_S\big) \Big]\,
\big( \sigma_{\phi^{~}_A} \sigma^{~}_{Y_A} \sigma^{\pm}_{\chi_A} \big) \, \Sigma^\pm_{\LG}  ~.
}
In \eqref{twist2} we exhibited explicitly only the left-moving structure. Adding the right-movers we find four dimension $(\half,\half)$ operators $\Sigma^{\pm,\pm}$ that belong to the four sectors $(c,c)$, $(c,a)$, $(a,c)$, $(a,a)$, like their worldsheet counterparts \SSbar.  For more details on the construction of these operators, see Appendix~\ref{app:twistops}.

As a check of the correspondence between \SSbar\ and \eqref{twist2}, the former has 
$j_\WS^{~}=k$ (which can be read off from \eqref{weq2wall}), while the latter has 
$j_\ST^{~}=\frac n2$, which is a consequence of~\eqref{phizero}, \eqref{twistphidep} (remembering that we are now discussing the normalizable profile, related to the non-normalizable one via $j_\ST\to 1-j_\ST$, and similarly for $j_\WS$). 
The matching between the worldsheet and spacetime quantum numbers was discussed in section~\ref{sec:type0long}, and is in general given by equation~\eqref{jSTjWS}. The values of $j_\WS$ and $j_\ST$ given above satisfy this relation, consistent with our identification of the corresponding operators.  

Of course, \eqref{twist2} is the bottom component of the spacetime superfield. In order to construct the modulus \eqref{weq2wall} we need to act with the spacetime supercharges, as was done on the worldsheet in \NSwall, \BPSsum.
Also, in comparing to the worldsheet theory, one needs to remember to reinterpret the construction of the twist operators using the field redefinition of Appendix~\ref{app:blocksusy} that relates the frame of the fermions $\cS^\pm$ to that of the fermions $\chi^\pm$ of standard supersymmetry employed above.

\subsection{Deviation from holomorphy for general operators}
\label{sec:genops}

In our discussion above we have focused on the properties of the operator $\partial\phi$ in the spacetime theory. We saw that while it is holomorphic at large $\phi$, its holomorphy is broken at finite $\phi$ by the contribution of a normalizable mode~-- the wall that cuts off the region $\phi\to-\infty$. 

As explained in earlier sections, the operator $\partial\phi$ is just one of an infinite number of operators with similar properties. Examples include the $\sutwo$ currents \eqref{onemoresu2}, \finalthird, and the fermions $\cS^\pm$ \blockferm. We expect the fate of these operators to be similar to that of $\partial\phi$~-- their holomorphy at large $\phi$ should be violated by their interaction with the wall. In this subsection we will demonstrate this by studying the fermions \blockferm, and explain how we expect the general case to work. 

The worldsheet vertex operator for the spacetime fermion $\cS^\pm$, \eqref{blockferm}, contains the operator $\Phi^{(-1)}_{1-\frac k2;\frac k2-1,\frac k2}$ whose $\bar x$ derivative can be treated similarly to our discussion of $\partial\phi$ above. In particular, it sits on an LSZ pole and gives a non-zero result that includes the normalizable operator $\Phi^{(-1)}_{1-\frac k2;\frac k2-1,\frac k2-1}$. Repeating the previous analysis, we find 
\eqn[Sholoviol]{
\partial_{\bar x}\cS^\pm \simeq  
e^{-\varphi/2-\bar\varphi}\, e^{\frac{3i}2 H_\sl+ i\bar H_\sl} \,
e^{\pm(\frac i2 H_3+i\frac a2 Z +\frac{Y}{\sqrt{2k}})}
(\bar\partial\varphi+i\bar\partial \bar{H}_\sl)\,
\Phi^{(-1)}_{1-\frac k2;\frac k2-1,\frac k2-1} ~.
}
Note that the second term in~\eqref{blockferm} doesn't contribute because its $\bar x$ derivative is not sitting on an LSZ pole. 

Following closely our discussion of the breakdown of holomorphy for $\partial\phi$, we first ask what is the leading behavior at large $\phi$ of the operator in the spacetime CFT dual to the vertex operator on the r.h.s. of \Sholoviol. Since the left-moving part of the vertex operator looks like that of $\cS^\pm$, the r.h.s. looks like that of $\bar\partial\phi$, and the scaling of the operator with $\phi$ is $\exp\left(-Q_\ell\phi\right)$, as before, it is natural to identify the operator \Sholoviol\ with 
\eqn[ddbarS RHS]{
\cS^\pm \partial_{\bar x}\phi \,e^{-Q_\ell\phi}
}
in the spacetime theory. 

Thus, the equation of motion of $\cS^\pm$ that we found at large $\phi$ is modified by the presence of the wall to 
\eqna[holoviolSS]{
\partial_{\bar x}\cS^\pm & \simeq 
\cS^\pm \partial_{\bar x}\phi \,e^{-Q_\ell\phi} ~.
}
As in our discussion of $\partial\phi$ above, around equation \phiredef, this naive violation of holomorphy can be undone by a field redefinition, 
\eqna[fieldredefS]{
\cS^\pm\to\cS^\pm\Big[1+B\,e^{-Q_\ell\phi}+O\big(e^{-2Q_\ell\phi}\big)\Big] ~,
}
with $B$ a constant. Thus, \holoviolSS\ does not lead to a real violation of holomorphy. 

However, again as in the discussion of $\partial\phi$, taking into account the FZZ correspondence implies that there is another contribution to the operator on the r.h.s. of \Sholoviol, obtained by replacing 
$\Phi^{(-1)}_{1-\frac k2;\frac k2-1,\frac k2-1}$ on the r.h.s. by $\Phi^{(-2)}_{k;k,k}$. This corresponds to an operator in the $\bZ_2$ twisted sector, which leads to a real violation of the holomorphy of the operators $\cS^\pm$. 

One can read off the holomorphy violating effect directly from the spacetime theory. Recall that we found before that the wall in the symmetric product spacetime CFT corresponds to the top component of the superfield whose bottom component is the chiral operator \eqref{twist2}, or more precisely a combination of such operators \BPSsum. 

As discussed below~\eqref{fieldredef}, after a rotation in field space, the superconformal generators that need to be applied to $\Sigma^\pm$ \eqref{twist2} are the standard $N=2$ superconformal generators on $\bR_\phi\times \bS^1$ and the $N=2$ minimal model, and the fermions $\chi^\pm$ are the superpartners of $\phi\pm iY$ in the block of the symmetric product.

Thus, to compute $\bar\partial\cS^\pm$, we need to compute $\bar\partial\chi^\pm$ in the symmetric product CFT with standard supersymmetry, deformed by the top component of \eqref{twist2}. Acting on this operator with the superconformal generators we find a few terms. First, we can act with the superconformal generators on the minimal model contribution or on that of $\bR_\phi\times \bS^1$. The former does not contribute to the holomorphy violation of $\chi^\pm$. Therefore we focus on the latter. 

The terms that give rise to violation of holomorphy are the ones that involve the action of the superconformal generators on the untwisted fields $(\phi_S, Y_S)$. This gives a deformation of the Lagrangian of the form 
\eqn[deformL]{
\delta\LL=\chi_S^+\bar\chi_S^+\,\exp\Big[ -\frac{1}{2\sqrt{k}}\big( \phi_S - iY_S\big) \Big]\,\sigma+\cdots~,
}
where $Y_S=Y_S^L+Y_S^R$ (the sum of the left and right-moving scalars), $\sigma$ is the contribution of the twisted fields to \eqref{twist2}, and the ellipsis stands for the contribution of the other three terms, associated with the operators $\Sigma^{+-}$, $\Sigma^{-+}$ and $\Sigma^{--}$ in the discussion after \eqref{twist2}.  

The equation of motion for $\chi_S^+$ that follows from this Lagrangian is 
\eqn[deformeomS]{\bar\partial\chi_S^+=
\bar\chi_S^+\,\exp\Big[ -\frac{1}{2\sqrt{k}}\big( \phi_S + i\widetilde Y_S\big) \Big]\sigma
+\bar\chi_S^-\,\exp\Big[ -\frac{1}{2\sqrt{k}}\big( \phi_S + i Y_S\big) \Big]\sigma~,
}
where $\widetilde Y_S=Y_S^L-Y_S^R$ and $Y_S=Y_S^L+Y_S^R$, as above. Thus, we conclude that: (1) the holomorphy of $\cS^\pm$ is violated, and (2) the operator on the r.h.s. of \deformeomS\ is the spacetime CFT dual of the worldsheet operator on the r.h.s. of \Sholoviol.

As mentioned in the beginning of this subsection, there are many other operators that are holomorphic at large $\phi$ (in fact, formally, an infinite number of them in the $g_s\to 0$ limit) and are expected to follow the same pattern as that exhibited above for $\partial_x\phi$ and $\cS^\pm$. In the rest of this subsection we will briefly comment on two related issues:
\begin{itemize}
\item What is the difference between the operators whose holomorphy is violated at finite $\phi$, and those that remain holomorphic in the full theory, such as the currents constructed in~\rcite{Kutasov:1999xu}.
\item What is the mechanism for the violation of holomorphy for generic operators.
\end{itemize}
Starting with the first issue, consider, for example, the $SU(2)$ currents constructed in the winding $w=-1$ sector in section~\ref{sec:type0long}. While the holomorphy of the currents \eqref{onemoresu2}, \finalthird\ is expected to be violated at finite $\phi$ that of the current \dualcurr\ is not. 

At a technical level, the origin of the difference is that the latter current contains the $\sltwo$ operator $\Phi^{(-1)}_{1-\frac{k}{2};\frac{k}{2},\frac{k}{2}}$. When we differentiate w.r.t. $\bar x$, the value of $\bar m$ is shifted down by one unit, and we get an operator that contains $\Phi^{(-1)}_{1-\frac{k}{2};\frac{k}{2},\frac{k}{2}-1}$. This operator is non-normalizable and does not sit on top of an LSZ pole, because of the value of $m=\frac{k}{2}$. Therefore, the limit $\bar m+j-1\to 0$ that appeared in our calculation is trivial, and we do not find any violation of holomorphy. 

On the other hand, in both examples we have studied explicitly in which the holomorphy was violated, the value of $m$ was shifted down, to $m=\frac{k}{2}-1$, and after differentiating w.r.t. $\bar x$ we got an operator that does sit on an LSZ pole, and thus a non-zero normalizable contribution. 

Since the above shift of $m$ led to such dramatic consequences as violation of holomorphy, it is natural to ask what is the qualitative reason for it. Looking back at the vertex operators for (say) $\partial_x\phi$, \dualbbzero, and for the holomorphic current \dualcurr, we see that the shift of $m$ is due to an important physical difference between the two: while the current \dualcurr\ corresponds in the bulk to a {\it gravity} mode, that of $\partial\phi$ corresponds to an {\it excited} (oscillator) mode. 

Thus, it is natural to conjecture that the examples we analyzed in detail are a special case of a more general phenomenon: 
\begin{quote}
{\it Operators in the spacetime CFT that are holomorphic at large $\phi$ and are dual to supergravity sector bulk fields remain holomorphic in the full theory, while those that are dual to excited string modes are not holomorphic in the full theory.}
\end{quote}
As an example, the $SU(2)$ currents \eqref{onemoresu2}, \finalthird\ are not expected to be holomorphic following the analysis above, and their vertex operators indeed correspond to oscillator states in string theory. It should be clear from the above discussion that the effect is quite general. Adding left-moving oscillators to the worldsheet vertex operator, forces us to shift the value of $m$ down, to satisfy the worldsheet mass-shell condition. Once we do that, we enter the region where a downward shift of $\bar m$ by one unit, such as the one that is implemented by taking the $\bar x$ derivative, puts us on top of an LSZ pole, and leads to violation of holomorphy.


\section{Discussion
\label{sec:discussion}}


\subsec{Recap}
\label{sec:recap}

In this work, we have identified a new species in the zoology of holographic dualities, and analyzed its anatomy using the tools of worldsheet string theory.  

The bulk background involves string theory on $AdS_3\times \bS^3_\flat$, where $\bS^3_\flat$ is a squashed $\bS^3$, described on the worldsheet by a current-current deformation of an $SU(2)$ WZW model. The AdS radius is in this case in the region $R_{AdS}<\lstr$. Therefore, this theory does not contain BTZ black holes and the $SL(2,\bR)$ invariant vacuum is not normalizable~\rcite{Giveon:2005mi}. 

The boundary theory is a deformed symmetric product of the form \ourcft. The symmetric product captures the target space of the spacetime CFT for large values of the coordinate $\phi$ that parametrizes the factor $\bR_\phi$ in~\ourcft. This region corresponds to the vicinity of the boundary of $AdS_3$. The spacetime CFT is deformed away from the symmetric product form by adding to the Lagrangian a marginal $\bZ_2$ twist operator (constructed in section \ref{sec:wall} and appendix~\ref{app:twistops}) whose strength runs with $\phi$, vanishing as $\phi\to\infty$ and forming a wall in the strong coupling region $\phi\to-\infty$. This deformation also leads to a non-trivial metric on $\bR_\phi$ in the building block of the symmetric product.

Most of our analysis focused on the case where the theory is $(2,2)$ superconformal. In that theory, the aforementioned $\bZ_2$ twist operator belongs to a superfield whose bottom component is a half-BPS operator of dimension $\left(\half,\half\right)$, which is a linear combination of operators that belong to the $(c,c)$, $(c,a)$, $(a,c)$, and $(a,a)$ rings. The operator that is turned on in the Lagrangian of the deformed symmetric product is the top component of this superfield.  

The derivation of this duality involved the following logical chain. First, we established a map between the non-normalizable and delta-function normalizable operators in the winding sectors of the worldsheet theory on $AdS_3\times \bS^3_\flat$ and the operators in the symmetric product \ourcft. We focused on the winding one sector, which maps to the building block $\bR_\phi\times \bS^3_\flat$ of the symmetric product. 

This map relied in an important way on the fact mentioned above, that the weak-coupling region in $\bR_\phi$, where operators are defined in asymptotically linear dilaton theories, maps under the duality to the vicinity of the boundary of $AdS_3$, where operators are defined in string theory on asymptotically $AdS$ spacetimes. This is only true for models with $k<1$, such as the ones studied in this paper.

In constructing the above map, we found that the worldsheet analysis gives rise to a large set of dimension $(r,0)$ operators, $r\in\half\bN$, which are described by non-normalizable vertex operators in string theory on $AdS_3$. These operators were found to be holomorphic at large $\phi$, a statement that was made precise in sections \ref{sec:type0long}, \ref{sec:typeIIlong}, and \ref{sec:wall}. They were identified with holomorphic operators in the building block of the symmetric product \ourcft. 

Some of these operators were found to be holomorphic in the full theory (\eg\ the currents associated to isometries of $AdS_3\times \bS^3_\flat$), while others have the property that their holomorphy is violated at finite $\phi$. More precisely, we showed (in section \ref{sec:wall}) that these currents have the property that $\bar\partial J= V_{\rm norm}$, where $V_{\rm norm}$ is a normalizable vertex operator on $AdS_3$, \ie\ one whose wavefunction decays exponentially at large $\phi$. This is the sense in which the (non-normalizable) operator $J$ is holomorphic at large $\phi$. We also showed that the operators that are holomorphic in the full theory come from the supergravity sector of the bulk theory \ourads, while those that are only asymptotically holomorphic are associated with excited string states in this background. 

We used the violation of holomorphy seen in the worldsheet calculation to deduce the corresponding deformation in the spacetime symmetric product \ourcft. This involved identifying the worldsheet operator $V_{\rm norm}$ for different $J$'s in the spacetime CFT. 

We found that the spacetime deformation corresponds to an operator in the spacetime CFT that has two components. One belongs to the building block of the symmetric product, and has the form \Lblock. We argued that this deformation corresponds to a modification of the metric on $\bR_\phi$, and thus is redundant on its own~-- it can be removed by a reparametrization of the form \phiredef~-- although it might have the effect of bounding the range of $\phi$ at small $\phi$, like the cigar deformation discussed in section \ref{sec:intro}. This issue requires further study. 

The second component of the interaction is a $\bZ_2$ twist field in the spacetime CFT, that was constructed in section \ref{sec:wall} and appendix~\ref{app:twistops}.  From the worldsheet point of view, it is related to \Lblock\ by FZZ duality. Its radial profile behaves at large $\phi$ like $\exp(-\frac{1}{2\sqrt{2k}} \phi)$. The corresponding wavefunction goes rapidly to zero at large positive $\phi$, and therefore the symmetric product picture \ourcft\ is valid in that region. Conversely, as $\phi$ decreases, the effect of the $\bZ_2$ twist increases, and it modifies the symmetric product picture \ourcft\ significantly, as depicted in figure~\ref{fig:bagpipes}.

From the spacetime CFT point of view, the above deformation prevents the field $\phi$ from exploring the region of large negative $\phi$, where the coupling on $\bR_\phi$ is strong. Thus, we can think of it as an analog of the wall in other asymptotically linear dilaton CFT's, such as the ones mentioned in section \ref{sec:intro}. Interestingly, in this case the wall is primarily in the $\bZ_2$ twisted sector of the symmetric product, and one can wonder how effective it is in performing this duty. 


\subsec{Comments on the proposed duality}
\label{sec:whatisit}

While we have collected a lot of information about the structure of the boundary CFT that is dual to string theory on $AdS_3\times\sqsphere$, equation~\ourads, much remains to be understood. In this subsection we comment on some of the outstanding questions. 

In string theory on the Euclidean version of \ourads, the natural observables are correlation functions of operators in the spacetime CFT, such as those constructed in sections \ref{sec:type0short}-\ref{sec:typeIIlong}. For operators whose dimensions scale like $p^0$ in the limit $p\to\infty$, these correlation functions can be computed using the standard worldsheet formalism, in a power series in $g_s^2\sim 1/p$. 

A natural question is whether these correlation functions agree with those of the corresponding operators in the deformed symmetric product.  We expect the answer to this question to be yes.  More precisely, we expect the CFT obtained by deforming the symmetric product~\ourcft\ in the manner described in section~\ref{sec:wall} to be unique given the features of the model that have already been deduced from the bulk analysis. It would be interesting to compute such correlators and test this expectation. 

Another interesting question concerns non-perturbative effects in our models. As usual in string theory, we expect the perturbative series discussed in the beginning of this subsection to have the property that the genus $g$ contribution goes like $(2g)!$~\rcite{Shenker:1990uf}.  This growth is usually related to the presence of D-branes in the theory, that give rise to effects that go like $\exp(-1/g_s)$. If the perturbative series in $1/p$ is reproduced by the boundary theory we have proposed, such effects should go like $\exp(-\sqrt p)$. It would be interesting to understand their origin in this theory.

The worldsheet analysis of Euclidean correlation functions of non-normalizable operators in the bulk theory exhibits LSZ poles.  Our duality implies that such poles must also be present in the deformed symmetric product theory.  These poles owe their existence to the wall discussed in section \ref{sec:wall}.  It would be interesting to establish their existence and properties directly in the spacetime CFT. 

Another interesting question concerns the high energy spectrum of our theories. In~\rcite{Giveon:2005mi} it was argued that they exhibit Cardy growth, with the central charge $c=6kp$ replaced by 
\eqn[ceff]{c_{\rm eff}=6\left(2-\frac1k\right)p~.} 
It is at first sight surprising that we can say anything precise about the spectrum of states whose energies grow like $p$, where the perturbative analysis breaks down. The following line of argument makes such a claim more plausible.

String theory on $AdS_3$ is believed to always give rise to a modular invariant spacetime CFT. In any such theory, the high energy density of states is related to the dimension of the low lying operators. In particular, the quantity $c_{\rm eff}$ that determines the high energy density of states, is given by 
$c_{\rm eff}=c-24 h_{\rm min}$, where $h_{\rm min}$ is the lowest dimension in the theory \rcite{Kutasov:1990sv}. 

In our case, taking the lowest dimension operator to be the bottom of the continuum of delta-function normalizable states in each factor of the symmetric product \ourcft, leads to \ceff. One can ask whether this calculation is reliable, since it involves states with dimensions that grow like $p$. 

From the bulk point of view it seems reasonable that this is indeed the case, since the states in question involve $p$ strings that may be widely separated in the radial direction. For example, we can consider a configuration where $p-1$ of the strings are close together, making a space that is asymptotically $AdS_3$, and the last string is widely separated from them in the radial direction. Then the difference of dimensions of the state with and without this last string can be computed using the perturbative techniques used in this paper. 

This argument suggests that in the Lorentzian theory, to leading order in the $1/p$ expansion, states in the theory take the form of a symmetric product of single string states.  The total dimension of these states is a sum of the dimensions of the single string states, and is typically of order $p$. The $1/p$ corrections due to $g_s$ effects can change the energies of these states by an $O(1)$ amount. It would be interesting to substantiate this picture by further calculations, generalizing those of~\rcite{Giveon:2005mi}.  
 
In the boundary theory that we proposed, the deformed symmetric product, the fact that $c_{\rm eff}$ is given to leading order in the $1/p$ expansion by \ceff\ is built in. The torus partition sum of the model receives a contribution proportional to the (infinite) length of the $\phi$ direction. This contribution is due to the delta-function normalizable states; it can be computed in the (undeformed) symmetric product, and thus, the corresponding $c_{\rm eff}$ is given by the free result \ceff.


\subsec{Variations on a theme}
\label{sec:variants}

In this subsection we discuss several offshoots of the above construction which are worth further investigation.  

The type II theories studied in sections \ref{sec:typeIIshort}, \ref{sec:typeIIlong} allowed two types of GSO projection for each worldsheet chirality. In section \ref{sec:wall} we constructed the wall operator for the theory with chiral GSO~2 on both left- and right-movers, where the analysis was facilitated by the fact that the theory has $(2,2)$ supersymmetry in spacetime. One might then inquire about the status of the wall when one or both chiralities involve the non-supersymmetric chiral GSO~0 projection.

At large $\phi$, where the spacetime CFT is described by a symmetric product orbifold, it was shown in section \ref{sec:typeIIlong} that the two chiral GSO projections give the same block theory. Because of this, the difference between the two GSO projections should come from the wall operator, which deforms the symmetric product structure.  The wall deformation is a normalizable operator, and thus represents a vacuum expectation value in the theory rather than changing the theory by turning on a coupling (which would be the case if the wall deformation involved a non-normalizable operator).  This implies that the GSO~0 wall \textit{spontaneously} breaks supersymmetry in the full theory.  A consequence of this spontaneous breaking is the appearance of a pair of massless fermions in theories with left-moving GSO~0, and/or similarly on the right, which are the Goldstinos associated with the spontaneously broken supersymmetries.  Indeed, we find such a pair of massless fermions $\PsiST^\pm$ under GSO~0.  It is an interesting open problem to find the spacetime wall operator for chiral GSO~0 with the requisite properties.

Similarly, it would be interesting to understand whether there is a candidate for a wall in the theory with type 0 GSO projection. In this case, the worldsheet theory has a BF-violating tachyon and is unstable, so it's not clear if the spacetime CFT makes sense. A tachyon also appears for the type II theories as the squashing parameter $R$ is varied, see the discussion below \eqref{jtach}. For these theories, the wall operator for generic $R$ can be obtained by starting with the theory at the supersymmetric point, with the wall described in section~\ref{sec:wall}, and then changing the squashing of the $\bS^3$. We leave for future work the question of what happens to the type II theory when it is deformed to a value of $R$ for which tachyons appear.
The situation is rather reminiscent of the Kosterlitz-Thouless transition, where a marginal line ends due to a condensation instability (in this case, a condensation of unwound strings instead of vortices).

Most of the analysis of this paper focused on operators in string theory on \ourads\ that live in sectors with non-zero winding. One of the remaining open problems is the role played by worldsheet operators in the zero winding sector. We found that non-normalizable operators in this sector appear to constitute a subset of the ones that appear in sectors with non-zero winding. 

For example, in the sector with winding $-1$ we found the operators \ebphiws\ with any real $\beta>-\frac{Q_l}2$, while in the sector with winding zero only the operator with $\beta=0$ made an appearance, \Ioperator\ (which is equivalent to the winding $-1$ operator \eqref{wsidd}). Similarly, in the $w=-1$ sector we found the operators \eqref{suws1}, which correspond in the boundary theory to \eqref{totsu2}, whereas in the zero winding sector we only found the operators \Kst, which correspond to \eqref{totsu2} with $\beta=0$, (and thus the $w=-1$ operators \dualcurr). A similar relation holds between the operators \WFZZgenj, which are dual to \Wblockop, and their $w=0$ analogs \WXtypeII, and other examples. It would be interesting to understand the organizing principle of these findings.

Another interesting question has to do with the four (exactly) marginal deformations found in section 7 for the type IIB (2,2) theory. The worldsheet operators corresponding to these deformations are given by $\cV^\pm\bar{\cV}^\pm$ (with independent choice of signs for left- and right-movers), with the left-moving part of the operator defined in \eqref{NSwall}. They are exactly marginal since they are top components of dimension $(\hf,\hf)$ chiral superfields. A linear combination of these deformations corresponds to the $AdS_3$ wall operator discussed in section 7, equation~\eqref{BPSsum}, while the remaining three combinations give a moduli space of theories. It would be interesting to understand the structure of this moduli space better. 

A familiar way to regularize the symmetric product \ourcft\ is to replace the building block of the symmetric product by~\rcite{Giveon:1999px,Brennan:2020bju}
\eqn[blockwall]{
\bR_\phi\times\sutwo_n^\flat
~~\longrightarrow~~
\bigg(\frac{\sltwo_{k_\ell}\times\sutwo_n^\flat}{\uone_L\times\uone_R}\bigg)
=\bigg(\frac{\sltwo_{n(n+1)}}{U(1)\times \bZ_{n+1}}\times\frac{SU(2)_n}{U(1)}\bigg)/\bZ_n ~.
}
The resulting symmetric product theory probably appears somewhere in the moduli space mentioned in the previous paragraph, but it is separated from the theory that corresponds to string theory on \eqref{ourads} by a large distance in moduli space. Roughly speaking, to get to it, one has to turn off the deformations described in section \ref{sec:wall}, and turn on the ones that lead to \blockwall. 

Another set of exactly marginal deformations of our theory is given by the  non-normalizable $(\!R,\!R)$ operators,
\eqn[VR]{
\cV_R^{\pm\pm} = \cG^\mp_{-\half}\bar\cG^\mp_{-\half}\cW_{j'=\half;\pm\half,\pm\half}^{(-1)}~,
}
where the operators $\cW_{j'=\half;\pm\half,\pm\half}^{(-1)}$ were introduced in section \ref{sec:BPS} (the $\pm$ signs can be chosen independently on left and right, so there are four such operators). These operators have the property that they grow exponentially as $\phi\to\infty$. Therefore, they give rise to a wall in the block of the symmetric product theory that keeps the long strings from exploring this region.

Deforming the theory by the $(\!R,\!R)$ operators \eqref{VR} thus lifts the long string continuum, and turns it into a discretuum. In this sense, the situation is similar to that in the critical theory of NS5-branes on $\bS^1\times \bT^4$ with fundamental strings wrapping $\bS^1$, where the continuum of long strings studied in~\rcite{Seiberg:1999xz,Maldacena:2000hw} can be lifted by turning on any of a quartet of $(\!R,\!R)$ moduli.

While we have focused on the $k<1$ models as theories in and of themselves, they can arise as endpoints of holographic RG flows from $k>1$.  The broader little string theory landscape described in section~\ref{sec:GKP} allows us to start with a CY singularity of the form~\rcite{Giveon:1999zm}
\eqn[UVsing]{
F(z_1,z_2)+z_3^2+z_4^2+z_5^2=0 ~,
}
where $F(z_1,z_2)$ is any quasihomogenous polynomial in two variables, and add to it $p$ strings, leading in the infrared to an $AdS_3$ background, which for general $F(z_1,z_2)$ has $k>1$.  Perturbing the polynomial $F$ by a relevant operator triggers a holographic RG flow, which can be arranged to lead at low energies to any of the $k<1$ theories~\eqref{cccnnn}.  

In the UV of this RG flow, the spectrum is controlled by the dual of string theory on $AdS_3$ with $k>1$, corresponding to \UVsing. The generic high energy states are described by BTZ black holes. The IR theory has $k<1$ and its spectrum consists of long strings, as discussed in this paper. Thus, such RG flows allow one to study the correspondence transition as a function of energy, as originally envisioned in~\rcite{Horowitz:1996nw}, in a setting where one may have more control as one approaches it from below.


\subsec{\texorpdfstring{$AdS_3$}{} string theory as a function of \texorpdfstring{$k$}{} }
\label{sec:ktheory}

In this paper, we discussed an infinite sequence of theories labeled by the discrete parameter $n$ defined in subsection \ref{sec:GKP} (around equations \cccnnn-\twodlst). The level $k$ of $SL(2,\bR)$ in these models is given by \kntwod. It is smaller than one, but as $n\to\infty$ it approaches the critical value $k=1$ from below. It is natural to ask what happens to our duality in this limit, and how the structure we find relates to the one for $k>1$. In this subsection we briefly comment on these questions. 

The first issue we would like to address is the behavior as $k\to 1$ of the deformation that modifies the symmetric product \ourcft\ and provides a wall that keeps the $\phi$ coordinate away from the strong coupling region $\phi\to-\infty$ in the spacetime CFT. We saw in section \ref{sec:wall} that this deformation has two components. One is a deformation in the block of the symmetric product, whose leading behavior is given in \Lblock. As mentioned after eq. \phiredef, this corresponds to a modification of the metric on $\phi$ space, $G(\phi)$. It can be removed by a field redefinition, but the important point is that this field redefinition is trivial for $\phi\gg 1/Q_\ell$, and is non-trivial otherwise. Thus, as $Q_\ell\to 0$ (which is the case for $n\to\infty$ \longss), the region where the deformation \Lblock\ can be neglected is pushed to infinity. 

As explained in section \ref{sec:wall}, the deformation \Lblock\ itself is rather benign. The non-trivial component of the wall is associated with the $\bZ_2$ twisted modulus constructed in section \ref{sec:wall} and appendix~\ref{app:twistops}. This modulus decays exponentially as $\phi\to\infty$, and formally that decay persists as $n\to\infty$. However, the exponentially decaying form of this operator is obtained by taking the metric in the $\phi$ direction to be $G(\phi)=1$, and this metric is modified at $\phi\sim 1/Q_\ell$. For $\phi$ below this value, the profile of the $\bZ_2$ twisted modulus is expected to change significantly. In particular, it might grow faster with decreasing $\phi$ than would be expected from the asymptotic form found in section \ref{sec:wall} and appendix~\ref{app:twistops}.

Thus we conclude that the region where the spacetime CFT is guaranteed to be well described by the symmetric product \ourcft\ is $\phi\gg 1/Q_\ell$. Beyond this region, one has to analyze the more complicated problem with a non-trivial $G(\phi)$, and the symmetric product may be significantly modified. As $k\to 1$, $Q_\ell\to 0$ and the regime of validity of the symmetric product is pushed to infinity. 

Another way to probe the limit $n\to\infty$ (or, equivalently, $k\to 1$) of our construction, is to study the behavior of the operators of dimension $(r,0)$ constructed in sections \ref{sec:type0long}, \ref{sec:typeIIlong}, \eg\ those given in equations~\eqref{dualbbzero}, \eqref{dualcurr}, \eqref{onemoresu2}, \eqref{finalthird}, \eqref{blockferm}, \eqref{blockfermzero}. These non-normalizable operators all have $j=1-\frac k2$, and therefore their wavefunctions grow exponentially towards the boundary of $AdS_3$, 
\eqn[growthhol]{\Psi(\phi)\sim e^{Q\left(j-\half\right)\phi}=e^{\frac{1-k}{2}Q\phi}=e^{\frac{Q_\ell\phi}{2}} ~.}
As explained in section \ref{sec:wall}, for a fixed $n$, \ie\ fixed $k<1$, these operators are holomorphic in the region in which the symmetric product description is valid, which was argued above to be $Q_\ell^{~}\phi\gg1$. Interestingly, this is precisely the region in which the wavefunction \growthhol\ starts growing exponentially. 

As $k\to 1$, the wavefunction \growthhol\ becomes more and more flat, and the region in which it grows exponentially is pushed to infinity. It is natural to interpret this as the statement that the region in which the corresponding operators are holomorphic also disappears in this limit. The fact that this argument gives the same answer for the size of the region where the symmetric product description \ourcft\ is valid, provides a consistency check on the picture.   

All our models have $k$ strictly smaller than one, but it is natural to ask what happens when $k$ is exactly equal to one.\footnote{For example, this is the case in string theory on $AdS_3\times \bS^1\times \bT^2$ studied in~\rcite{Giribet:2018ada}.} This case is particularly subtle from our perspective. All the dimension $(r,0)$ operators have $j=\half$ in this case, \ie\ they lie at the threshold of the continuum of delta-function normalizable states.\footnote{A related observation was made in section 5.1.1 of~\rcite{Ferreira:2017pgt}.} Their wavefunctions are formally constant \growthhol, though, as is familiar from other contexts (see \eg~\rcite{Seiberg:1990eb}), in this case there is a second solution to the mass-shell condition for which $\Psi(\phi)\sim\phi$.  Formally, the size of the region in which the holomorphy of these operators is broken goes to infinity in this case, but further study is necessary to decide what really happens.\footnote{A particular model with $k=1$ was studied in~\rcite{Eberhardt:2019ywk} (and references therein). The bulk model considered by these authors does not belong to the class of models studied in this paper and in~\rcite{Giribet:2018ada}. Thus, the relation of their work to ours is unclear.}    

It is interesting to contrast the behavior we found for $k<1$ with the one for $k>1$. One important difference between the two cases is that for $k>1$ the spectrum does not contain any dimension $(r,0)$ operators, other than the ones that are holomorphic in the full theory.  By repeating the analysis of sections \ref{sec:type0long}, \ref{sec:typeIIlong}, one can show that there are no asymptotically holomorphic BRST invariant operators with these scaling dimensions. For example, if we try to use the construction of the holomorphic operators of these sections, and formally plug in $k>1$, we find that $j=1-\frac k2<\half$. To obtain a non-normalizable operator with $j>\half$ we have to use the reflection $j\to 1-j$, after which we obtain $j=\frac k2$, but then the details of the calculation change. 

For $k<1$, we saw that the existence of asymptotically holomorphic dimension $(r,0)$ operators is directly related to the fact that at large $\phi$ the spacetime CFT is described by the symmetric product \ourcft. The above operators are the holomorphic operators in the building block of the symmetric product, the Seiberg-Witten (SW) theory $\bR_\phi\times \sqsphere$. Since for $k>1$ there are no such holomorphic operators, it is natural to ask what is the status of the SW theory in that case. 

As we mentioned in the introduction, for $k>1$ the SW theory is at best valid in some finite range of $\phi$. If $\phi$ is too small, the notion of long strings that gives rise to this theory breaks down, while when $\phi\to\infty$ the effective string coupling of the SW theory goes to infinity, so it cannot provide a good description of that region (without introducing a wall that prevents the field $\phi$ from exploring it). 

Thus, the SW theory breaks down near the boundary of $AdS_3$, and cannot be used to describe non-normalizable operators, whose wavefunctions are exponentially growing in that region. The asymptotically holomorphic operators constructed in this paper are precisely such operators, and therefore their absence in the spectrum of the worldsheet theory on $AdS_3$ is compatible with the fact that the SW theory does not provide a good description of the region near the boundary. 

Another aspect of the theory that differs between $k<1$ and $k>1$ is that in the latter case the $SL(2,\bR)$ invariant vacuum and BTZ black holes are normalizable states in the theory, while in the former case they are not. As $k$ approaches $1$ from above, the nature of these states changes. For example, the (Euclidean) BTZ black hole involves a condensate of a fundamental string winding around the thermal circle at infinity~\rcite{Kutasov:2005rr}, and this condensate spreads out further into the large $\phi$ region the closer $k$ is to $1$. This effect was interpreted in~\rcite{Kutasov:2005rr} as a stringy smearing of the Euclidean horizon that grows without bound near the string/black hole transition at $k=1$. The loss of normalizability of the BTZ black hole solution for $k<1$ is due to this condensate. 

Thus, we are led to a picture where as $k\to1^+$, the horizon of black holes exhibits large fluctuations. For all $k<1$ we find that the theory contains an infinite number of asymptotically holomorphic operators, but the region in which they are holomorphic is pushed towards $\phi=\infty$ as $k\to1^-$. One can heuristically think of the situation as being (very loosely) similar to the Kosterlitz-Thouless transition in statistical mechanics, with the region $k<1$ reminiscent of the massless phase of the model, and $k>1$ reminiscent of the massive phase. 
Of course, there are important differences as well. In particular, in our case changing $k$ corresponds to changing the theory, whereas there changing the radius (or temperature) corresponds to changing a parameter in the theory. Also, in that case the massless phase includes one holomorphic current, whereas in our case there is an infinite number of them.


\subsec{Comparison to \texorpdfstring{$c<1$}{} string theory}
\label{sec:clt1}

The theories discussed in this paper exhibit some analogies to a class of theories that was studied about thirty years ago, sometimes referred to as the old matrix model (see \eg~\rcite{DiFrancesco:1993cyw,Marshakov:1993au,Morozov:1994hh} for reviews). In this subsection we briefly discuss some of these analogies. 

The old matrix model is in a sense the first example of a holographic duality. In the version of it that is relevant for our purposes, the bulk theory is the bosonic string on a background that consists of a minimal model CFT (which has $c<1$) and a Liouville field $\phi$.%
\footnote{The $N=1$ supersymmetric minimal models coupled to $N=1$ supersymmetric Liouville theory also fit into this framework.}
If one takes the minimal model to be unitary, these string theories are labeled by a single integer, $n=2,3,\dots$, and the central charge of the minimal model is given by $c=1-\frac6{n(n+1)}$. 

The boundary theory is a (multi-)matrix model, with a potential that couples the matrices in a particular way. The duality to string theory involves a double scaling limit, where the rank of the matrices, $N$, is sent to infinity, and the potential for the matrices is tuned to a critical point. 

These models naturally live in Euclidean signature -- there is no time coordinate. One can take the limit $n\to\infty$, or equivalently $c\to 1$, in which the minimal model is replaced by a free scalar field, and Wick rotate that scalar field to obtain matrix quantum mechanics,\footnote{See \rcite{Klebanov:1991qa,Ginsparg:1993is,Jevicki:1993qn,Martinec:2004td} for reviews, and \rcite{Balthazar:2017mxh} for some recent results.} but we will not discuss that case here. 

The Euclidean nature of the old matrix model seems different from our construction, where the (Lorentzian) $AdS_3$ contains a timelike direction on the boundary. However, our models with $k<1$ are also more naturally thought of as Euclidean, with $AdS_3$ replaced by the hyperbolic space $\bH_3^+=\frac\sltwoc\sutwo$. Indeed, in Lorentz signature, the model does not contain an $SL(2,\bR)$ invariant state, so the isometry of $AdS_3$ is not a symmetry of any state in the theory. Moreover, the dimensions of the lowest states in the theory grow linearly with the number of strings, $p$. In general, string perturbation theory on $AdS_3$ is not well suited for studying such states, since they have energies that go like $1/g_s^2$. 

If instead one studies our theories on $\bH_3^+$, one can compute correlation functions of delta-function normalizable and non-normalizable operators whose dimensions scale like $p^0$ in the limit $p\to\infty$. These correlation functions satisfy the Ward identities that follow from two dimensional conformal symmetry. The results of this paper are perhaps best viewed as constructing these Euclidean theories.  

Another analogy between our models and the old matrix model involves the physical degrees of freedom of the model. In our models, due to the fact that $k<1$, there are no black holes in the spectrum, and the physical degrees of freedom are the ones seen in string perturbation theory, namely fundamental strings and D-branes. This seems to lie at the heart of the fact that we were able to identify the boundary CFT by analyzing the worldsheet string theory. 

Similarly, in the old matrix model, the only degrees of freedom that seem to play a role are the ones that are seen in string perturbation theory. Black holes are irrelevant in that case, since there is no time coordinate,\footnote{In the $c=1$ case, where one can study the theory in Lorentz signature, one can formally write black hole solutions, but they are not in the spectrum for reasons similar to our case; see \eg~\rcite{Giveon:2005mi}.} and the parameters on which the partition sum depends (which correspond to couplings in a certain integrable hierarchy, see \eg~\rcite{DiFrancesco:1993cyw,Marshakov:1993au,Morozov:1994hh} for reviews), correspond to coefficients of non-normalizable operators in the worldsheet Lagrangian. 

In the old matrix model, the underlying reason for the simplicity of the theory is often said to be the fact that it corresponds to string propagation in $D\le 2$ dimensions, where there are no transverse directions for the string to oscillate in. This leads to the absence of a Hagedorn spectrum in these theories, and presumably is also related to the absence of black holes. 

In the theories studied in this paper, something similar happens. As discussed in section~\ref{sec:review}, these theories can be viewed as describing the dynamics of strings placed at the tip of a singular cone. In our case, the only directions available for such a string to oscillate in are transverse to the tip, while for $k>1$ there are also directions along the singularity that the string can oscillate in. For example, in the case \ardougads, the string can oscillate along the $\bT^2$ without leaving the singularity. It is likely that the presence of such directions is necessary for the existence of black holes and the complexity they bring to the problem. Conversely, it is likely that the absence of such directions in our case is responsible for the simplicity of the dynamics at large $\phi$, where the theory is well described by the symmetric product \ourcft.

The $c<1$ models are solvable due to their large symmetry. In particular, correlation functions are severely constrained by a structure known as the {\it ground ring}~\rcite{Witten:1991zd,Kutasov:1991qx,Ginsparg:1993is,Douglas:2003up}. The analog of this structure in our case remains to be understood, but it is likely related to the fact that the boundary CFT is asymptotically free -- it approaches at large $\phi$ a symmetric product CFT, which contains a tower of (spontaneously broken) holomorphic operators.   

Another analogy between the two constructions involves the limit $c\to 1$ in the old matrix model, and the corresponding limit $k\to 1$ in our case. In the old matrix model, the area operator $e^{\gamma\phi}$ approaches in this limit the critical value at the bottom of the continuum of delta-function normalizable operators, $\gamma_c= -Q/2$. In our case, something similar happens with the operator $\cI$, which appears in the worldsheet Lagrangian of string theory on $AdS_3$~\rcite{Giveon:2001up}.  In both theories, one can work either at fixed chemical potential $\mu$, which corresponds to the cosmological constant in $c<1$ string theory, and a chemical potential for the number of strings in our case, or Legendre transform, to fixed area there and to fixed number of fundamental strings here~\rcite{Porrati:2015eha}. 

Also, in both cases, when one goes beyond $c=1$ there, or $k=1$ here, the nature of the theory changes significantly.  There, for $c>1$ a Hagedorn spectrum appears; here, for $k>1$ BTZ black holes become normalizable, and the entropy goes from that of fundamental strings to that of black holes \rcite{Giveon:2005mi}.

Given these analogies between the $c<1$ and $k<1$ string models, it might be interesting to revisit some of the calculations that were done in the $c\le1$ models to see whether they can teach us something about our models.  These include
\begin{itemize}
\item
{\it Bulk amplitudes:} Correlation functions or scattering amplitudes where $\phi$ momentum is conserved modulo an integer power of $\beta_{\rm wall}$ can be evaluated in conformal perturbation theory in the wall operator (see for instance~\rcite{Goulian:1990qr,DiFrancesco:1991daf}).
\item
{\it The ground ring}:  
The correlation functions of $c\le1$, and the S-matrix at $c=1$, are governed by a symmetry structure known as the {\it ground ring}~\rcite{Witten:1991zd,Kutasov:1991qx,Ginsparg:1993is,Douglas:2003up}.  The natural candidate for a symmetry structure underlying the $k<1$ models is the tower of holomorphic operators in the block of the symmetric product, whose holomorphy is softly broken by the wall deformation.  This situation is also reminiscent of the SYK model, where an underlying spontaneously broken Virasoro symmetry governs the effective dynamics~\rcite{Maldacena:2016hyu}. 
\item
{\it Holographic RG flows}:
The $c\le1$ models exhibit a rich structure of holographic RG flows~\rcite{DiFrancesco:1993cyw,Hsu:1992cm}, which on the worldsheet can be thought of as gravitationally dressed flows between the minimal models.  One has a similarly rich set of RG flows between $N=2$ LG models~\rcite{Kastor:1988ef,Martinec:1988zu,Vafa:1988uu}, which can be dressed by the worldsheet $AdS_3\times\bS^1$ CFT, as discussed around eq. \UVsing. 
\item
{\it The thermal partition function:}
The partition function of the $c=1$ model compactified on a Euclidean circle is another interesting probe of the system (see \eg~\rcite{Klebanov:1991qa} for a review); it would be interesting to compute the thermal partition function of the deformed symmetric product, and compare it to a worldsheet computation on thermal $AdS_3$, \ie\ $\big(\frac\sltwoc\sutwo\big)/\bZ$, along the lines of~\rcite{Maldacena:2000kv,Ashok:2021vww}.
\item
{\it D-branes:}
The leading non-perturbative effects in $c\le1$ are given by D-brane amplitudes;  comparing the worldsheet and spacetime theories provides a sensitive probe of the structure of the wall in $c\le1$~\rcite{McGreevy:2003kb,Martinec:2003ka,Douglas:2003up,Seiberg:2003nm,Seiberg:2004at,Balthazar:2019rnh,Balthazar:2019ypi}. One might hope that a similar study will shed additional light on the wall in the $k<1$ models. 
\end{itemize}


\subsec{Deformation to linear dilaton asymptotics and \texorpdfstring{$T\bar T$}{}}
\label{sec:deflin}

As described in section \ref{sec:review}, the $AdS_3\times \bS^3_\flat$ background studied in this paper can be obtained by starting with the two dimensional LST background 
\eqn[2dlst]{\bR_t\times \bS^1\times\bR_\phi\times \bS^3_\flat~,} 
and adding to it a large number of fundamental strings wrapping the $\bS^1$ (see the discussion around equation \twodlst). The $AdS_3$ background is obtained by taking the infrared limit, \ie\ by studying the near-horizon region of the strings.  This limit is equivalent to taking the radius $R_x$ of $\bS^1$ (in string units) to infinity.

\begin{figure}[ht]
\centering
\includegraphics[width=.55\textwidth]{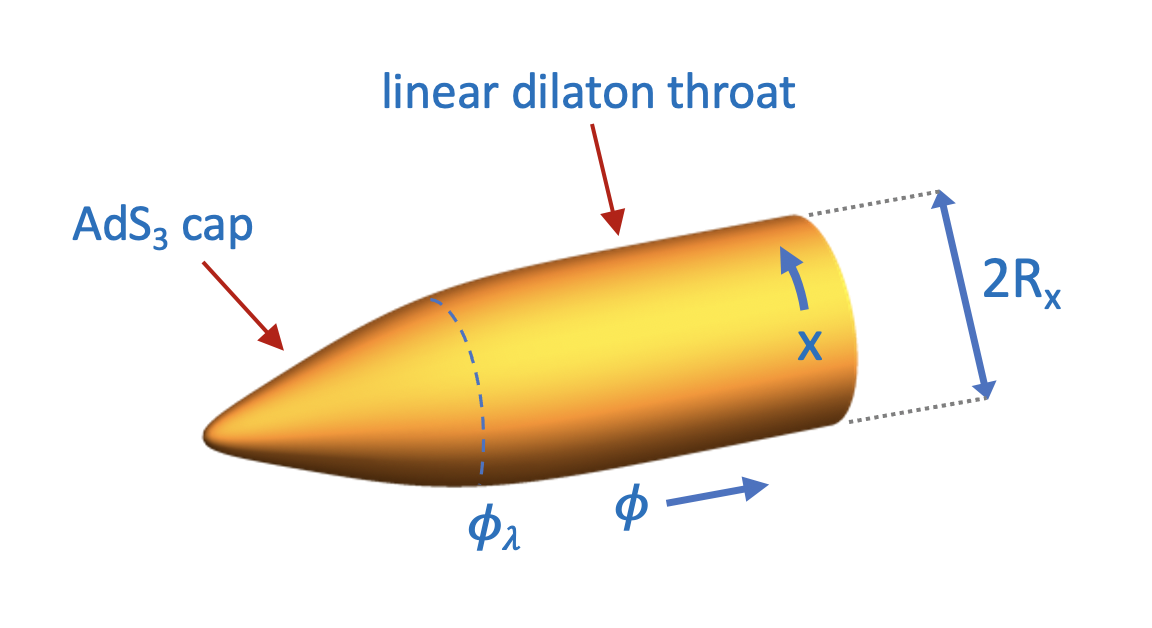}
\caption{\it Sketch of the deformed geometry, showing the spatial geometry on a fixed time slice (suppressing the $\bS^3_\flat$ directions).  The growth of the azimuthal circle in $AdS_3$ saturates at the size $R_x$ at a radial scale $\phi\approx\phi_\lambda$, where the geometry rolls over to a constant size circle and linear dilaton in $\phi$.}
\label{fig:TTbargeom}
\end{figure}
\vspace{1mm}

As discussed in~\rcite{Giveon:1999zm,Martinec:2017ztd,Giveon:2017nie,Chakraborty:2019mdf,Brennan:2020bju}, one can extend the bulk theory to the full LST geometry that interpolates between the linear dilaton geometry~\eqref{2dlst} in the UV and $AdS_3\times\bS^3_\flat$ in the IR (see figure~\ref{fig:TTbargeom}). This extension is most naturally done in the $(\!R,\!R)$ sector of the spacetime CFT. This CFT has multiple $(\!R,\!R)$ ground states, but for the purpose of computing the spectrum of delta-function normalizable long string states in the full theory, it is believed that all of them can be replaced by the massless BTZ background.\footnote{This belief is supported by the fact that the bulk descriptions of other $(\!R,\!R)$ ground states, for which the worldsheet theory is also exactly solvable~\rcite{Brennan:2020bju}, have an identical spectrum of such states.} The linear dilaton extension of this background is sometimes denoted by $\MM_3$, and we will use that notation below. It is described by a current-current deformation of $SL(2,\bR)$ CFT, which is exactly solvable.

In the boundary description, it corresponds to an irrelevant deformation of the ${\it CFT_{\rm 2}}$ dual to string theory on $AdS_3$. This deformation is closely related to the $T\bar T$ deformation~\rcite{Smirnov:2016lqw,Cavaglia:2016oda}, and is usually referred to as the single-trace $T\bar T$ deformation. As we will see, this name is particularly appropriate for the theories discussed in this paper.

For $k>1$, this deformation of the dual ${\it CFT_{\rm 2}}$ is not well understood, mainly because the ${\it CFT_{\rm 2}}$ itself is not well understood either. In our case, the $AdS_3/CFT_2$ duality is better understood, and thus it is natural to expect that we will have a better handle on the deformation as well. 

As shown in references~\rcite{Giveon:2017nie,Chakraborty:2019mdf}, the spectrum of delta-function normalizable (and non-normalizable) operators in the resulting theory is that of a symmetric product, whose building block is the $T\bar T$ deformed CFT $\bR_\phi\times \bS^3_\flat$. As is familiar from studies of $T\bar T$ deformed CFT, the spectrum of the resulting block theory on a spatial circle interpolates continuously between that of the CFT $\bR_\phi\times \bS^3_\flat$, which describes the region $\cE\ll \frac1\lambda$ (where $\cE$ and $\lambda$ are the energy and $T\bar T$ coupling in units of the radius of the $\bS^1$, $R_x$, respectively), and that of two dimensional LST, which describes the region $\cE\gg \frac1\lambda$. In particular, the entropy interpolates between Cardy entropy at energies $1\ll \cE\ll \frac1\lambda$ and Hagedorn entropy for $\cE\gg \frac1\lambda$.

From the bulk point of view, the fact that this deformation is irrelevant (in the RG sense) is reflected in the fact that the deformation keeps the infrared $AdS_3$ geometry untouched at large negative $\phi$, but changes the geometry at large positive $\phi$ to the linear dilaton one~\eqref{2dlst} (see figure~\ref{fig:TTbargeom}). The transition between the two regimes happens at some particular location $\phi_\lambda$ along $\bR_\phi$, which can be thought of as the holographic representation of the energy scale~$\frac1\lambda$. In terms of the string theory parameters, the dimensionless coupling $\lambda$ is mapped to $\alpha'/R_x^2$  \rcite{Giveon:2017nie}.

The discussion of this subsection raises a number of questions. One concerns the form of the deformed boundary theory. As explained earlier, if we think of the spacetime CFT as the symmetric product \ourcft, the irrelevant deformation that corresponds to turning on a finite coupling $\lambda$ is naturally identified with the $T\bar T$ deformation in the building block of the symmetric product, $\bR_\phi\times\bS^3_\flat$. 

However, as explained in the previous sections, the spacetime theory is actually described by the above symmetric product only at large $\phi$. The full theory has in addition a $\bZ_2$ twist deformation turned on. This deformation decays exponentially at large positive $\phi$, but has a significant effect on the dynamics at finite $\phi$. In particular, it destroys the symmetric product structure in that regime (see figure~\ref{fig:bagpipes}). 

It is thus natural to ask whether the single-trace $T\bar T$ deformation described in this subsection is compatible with the $\bZ_2$ twisted marginal deformation that is turned on in the spacetime CFT.  In other words, can one turn on both of them at the same time, despite the fact that the full theory is not a symmetric product? One answer to this question is that the bulk-boundary duality constructed in this paper predicts that this is indeed the case, in a particular sense we explain next.  

For $\phi\ll\phi_\lambda$, the deformed background of figure~\ref{fig:TTbargeom} looks like $AdS_3$, while in the opposite limit, $\phi\gg\phi_\lambda$, it looks like the linear dilaton background \eqref{2dlst}. Another natural scale that appears in this model is the radial size of low lying normalizable states in string theory on \ourads. We can denote this scale by $\phi_{AdS}$. The situation is cleanest if the hierarchy of scales in the deformed model of figure~\ref{fig:TTbargeom} is $\phi_{AdS}\ll\phi_\lambda$. One can think of $\phi_\lambda$ as a UV cutoff in the low energy theory (string theory on $AdS_3$), and in this regime it is large, in the sense that the spectrum of states below the cutoff exhibits Cardy growth. 

The deformation that changes the geometry from $AdS_3$ to $\MM_3$ can be described by adding an irrelevant operator to the Lagrangian of the dual CFT. Since the UV cutoff $\phi_\lambda$ is high in this case, we can identify this deformation with a particular non-normalizable operator in string theory on $AdS_3$. That operator is the single-trace $T\bar T$ operator in the symmetric product theory \ourcft. The requirement that $\phi_{AdS}\ll\phi_\lambda$ maps in the boundary theory to $\lambda\ll 1$, or more physically to the requirement that the irrelevant single-trace $T\bar T$ coupling be small at the compactification scale $R_x$.   

As one increases the coupling $\lambda$, the form of the operator in the spacetime CFT that implements the deformation from $AdS_3$ to $\cM_3$ may receive corrections due to the effects of the $\bZ_2$ twist field. These deformations are expected to become large as $\phi_\lambda$ approaches $\phi_{AdS}$. However, it is not unreasonable to expect that these corrections do not influence the non-normalizable and delta-function normalizable states, whose properties are determined at large $\phi$, $\phi\gg\phi_\lambda,\phi_{AdS}$. Indeed, one finds that the energy formula for such states in $\cM_3$ agrees with that of $T\bar T$ deformed block CFT for all $\lambda$.

The above picture is only applicable for states that can be studied at large $\phi$, where the symmetric product descrition \ourcft\ is valid, namely for the delta-function normalizable and non-normalizable operators. In particular, the energies of normalizable states in the presence of the above deformation need not have anything to do with $T\bar T$ deformed CFT. From the point of view of the spacetime CFT, these states owe their existence to the $\bZ_2$ twisted wall, and therefore for them the deformation of~\rcite{Giveon:1999zm,Giveon:2017nie} need not affect their spectrum according to the $T\bar T$ formula of~\rcite{Smirnov:2016lqw,Cavaglia:2016oda}. It is an interesting remaining question how to think about their energies at finite $\lambda$ in the CFT dual.\footnote{Of course, highly excited bound states, which in this context correspond to states with $m+j=-l$ with a large integer $l$, behave in many respects like scattering states, and thus for them the $T\bar T$ energy formulae of~\rcite{Smirnov:2016lqw,Cavaglia:2016oda} are approximately valid.} 

Another question raised by the above discussion concerns the nature of the RG in the bulk and on the boundary. We mentioned above that in the boundary theory the transition between the CFT and LST regimes happens at some particular energy, $\cE\approx\frac1\lambda$. In the bulk theory, a similar transition takes place in the geometry of figure~\ref{fig:TTbargeom} at a particular value of the radial coordinate, $\phi\approx\phi_\lambda$. It is natural to ask how these two facts are related, and in particular what is the precise map between $\phi$ and energy for perturbative string states. 

We will leave a more detailed discussion of these issues to future work. Here, we just note that a similar issue arises when studying black holes in these geometries (for $k>1$).  As the horizon moves out to larger $\phi$ in the geometry of figure~\ref{fig:TTbargeom} and crosses the scale $\phi_\lambda$, the black hole entropy transitions from Cardy to Hagedorn growth.  We expect a similar relation to govern the physics of perturbative string states in our case.


\subsec{Little string theory interpretation}
\label{sec:LSTinterp}

As discussed in section \ref{sec:review} and in the previous subsection, one can view the backgrounds \ourads\ studied in this paper in the broader context of Little String Theory (LST). From that point of view, the starting point of the discussion is a two dimensional vacuum of LST, that is described near the boundary at large $\phi$ by the linear dilaton spacetime~\eqref{2dlst}. 

We considered a particular superselection sector in that theory, that contains $p$ fundamental strings placed in the linear dilaton throat \eqref{2dlst}. At low energies, this system is described by the ${\it CFT_{\rm 2}}$ dual to the $AdS_3$ background \ourads. If we do not take the low energy limit, we find the single-trace $T\bar T$ deformed theory described in the previous subsection. 

The LST perspective may shed additional light on the difference between the $k<1$ models discussed in this paper and the ones with $k>1$ extensively studied in the last twenty years. In this subsection we briefly comment on this issue.

In the original example of six dimensional LST compactified on $\bT^4$ (or $K3$) and the associated string theory on $AdS_3\times \bS^3\times \bT^4$, the high energy behavior of the entropy of the LST exhibits Hagedorn behavior with $\beta_H=\sqrt{k}\,\lstr$~\rcite{Maldacena:1996ya}, and the string theory on $AdS_3$ exhibits Cardy behavior with central charge $c=6kp$, where $k$ is the number of $NS5$-branes and $p$ the number of strings~\rcite{Strominger:1997eq}. The LST entropy is due to the contribution of $\frac\sltwo\uone$ black holes, while that of the $AdS_3$ theory is due to the contribution of BTZ black holes. 

In the 1990's it was proposed that the above entropies can be thought of in terms of {\it little strings}.  These strings arise as fractionated constituents of fundamental strings confined to the fivebranes; as a consequence they have a tension that is smaller by a factor of~$k$~\rcite{Maldacena:1996ya,Dijkgraaf:1997ku}.  This fractionation is in particular responsible for the fact that the Hagedorn temperature of LST is lower by a factor of $\sqrt k$ from the one familiar from fundamental string theory. 

In the $AdS_3$ context, the worldsheet of the little strings fills an $\bR_t\times \bS^1$ inside the fivebranes. Their (transverse) oscillations take place in the remaining four directions along the fivebranes, \ie\ the $\bT^4$. This explains the formula for the central charge above, $c=6kp$. The six is due to the fact that the strings can explore four bosonic and four fermionic directions, and the $kp$ is due to the fact that the $p$ original fundamental strings behave like $kp$ fractional strings.  Essentially, one can {\it define} the little strings as the microscopic degrees of freedom of the spacetime CFT.

In both cases (the Hagedorn entropy of the asymptotically linear dilaton theory, and the Cardy spectrum of the $AdS_3$), the little strings are the microscopic degrees of freedom that explain the entropy of the corresponding black holes~\rcite{Martinec:1999gw}. Thus, it is not surprising that a detailed understanding of their dynamics is a non-trivial problem. Solving it is tantamount to providing a microscopic understanding of the states of these black holes.

When we move away from the original system of flat fivebranes and strings, to curved fivebranes, or equivalently the dynamics near conical singularities (see section \ref{sec:review}), the situation becomes more involved. The number of fivebranes $k$ is replaced in the above considerations by the level of $SL(2,\bR)$. This level is in general non-integer. For example, for the background \ardougads\ we find that $k$ is given by \knad, and in our case \klessone, it is given by \kntwod. 

It would be interesting to extend the little string description of the critical theories to the general non-critical case with $k>1$. Such a description would presumably involve the notion of little strings oscillating in some transverse directions while remaining inside the curved fivebranes, or at the tip of the dual cone, described in section \ref{sec:review}. The fact that $k$ is in general fractional is presumably due to the fact that the little strings are in general interacting in a curved geometry.  

From the perspective of the above discussion, the difference between the theories with $k>1$ and those with $k<1$ appears to be that in the latter case the little strings do not have any transverse directions in which they can oscillate, and therefore, the full theory becomes just that of fundamental strings located at the tip of the cone \cccnnn, whose transverse oscillations take them away from the tip. The fact that the little strings are effectively absent is responsible in this picture for the fact that BTZ black holes (and in the full LST $\frac\sltwo\uone$ black holes) are not in the spectrum. It would be interesting to make this picture more precise. 

One route toward this goal is to study D-branes in these geometries. In related settings, such as $c\le 1$ string theory~\rcite{McGreevy:2003kb,Martinec:2003ka,Douglas:2003up,Seiberg:2003nm,Seiberg:2004at,Balthazar:2019rnh,Balthazar:2019ypi}, 
and Double Scaled Little String Theory%
~\rcite{Giveon:1999px,Elitzur:2000pq,Eguchi:2000cj,Martinec:2019wzw}, 
the study of such D-branes provides important insights into the theory, \eg\ the appearance of non-abelian gauge symmetry as type IIB NS fivebranes approach each other, and the appearance of a non-trivial IR CFT (the $(2,0)$ theory) when IIA fivebranes coincide. It is possible that a similar study in our case will provide a better understanding of the stringy geometry seen by the little strings, and in particular clarify the difference between the cases $k>1$ and $k<1$.

\vskip 1cm


\ack{
\vskip .3cm
We thank O. Aharony, S. Minwalla, E. Rabinovici and S. Sethi for useful discussions.
The work of BB, DK, and EJM is supported in part by DOE grant DE-SC0009924.
The work of AG and DK is supported in part by BSF grant number 2018068.
The work of AG is also supported in part by a center of excellence supported
by the Israel Science Foundation (grant number 2289/18). DK thanks the Weizmann Institute for hospitality during part of the work on this paper.
}

\begin{appendices}

\newsec[app:cocycles]{Spin field conventions}

The supersymmetric WZW model $\sltwo_k\times SU(2)_n$ CFT includes 6 free fermions $\psi^a_\sl,\psi^a_\su$, discussed in section~\ref{sec:currentalg}. It is convenient to bosonize these free fermions as $H_\sl,H_\su,H_3$, as done in \eqref{sutwobos}, \eqref{sltwobos}, \eqref{Hdefs}. The spin fields of the free fermions can be written in terms of the bosonized fields as
\eqn[spinfields]{e^{\pm\frac{i}{2}H_\sl\pm\frac{i}{2}H_\su\pm\frac{i}{2}H_3}~.}
The OPE's between the free fermions and spin fields can then be obtained from the OPE's between the free scalars $H_\sl,H_\su,H_3$.

However, there are subtleties in these OPE's associated to $\pm$ signs, called cocycle factors \rcite{Polchinski:1998rr}. In practice, we write the 6 fermions zero modes in terms of $\Gamma$ matrices, and the spin fields as elements in this 8-dimensional vector space. To be explicit, we write the fermion zero modes as
\eqna[zzaa]{&\sqrt{2}\,\psi^3_{\sl,0}=\Gamma^0~,~~~~~\sqrt{2}\,\psi^3_{\su,0}=\Gamma^1~,~~~~~\sqrt{2}\,\psi^1_{\sl,0}=\Gamma^2~,\cr &\sqrt{2}\,\psi^2_{\sl,0}=\Gamma^3~,~~~~~\sqrt{2}\,\psi^1_{\su,0}=\Gamma^4~,~~~~~\sqrt{2}\,\psi^2_{\su,0}=\Gamma^5~,}
where we use the $\Gamma$-matrix conventions (from appendix B of~\rcite{Polchinski:1998rr}):
\eqna[zzbb]{&\Gamma^0=
\left(\begin{matrix}
i \sigma^2 & 0 &0 &0 \cr
0& -i \sigma^2 & 0 &0 \cr
0&0& -i \sigma^2 & 0 \cr
0&0&0&i \sigma^2 
\end{matrix}\right)
,~~~\Gamma^1=
\left(\begin{matrix} \sigma^1 & 0 &0 &0 \cr
0& -\sigma^1 & 0 &0 \cr
0&0& - \sigma^1 & 0 \cr
0&0&0&\sigma^1 
\end{matrix}\right)
\\[.3cm]
&\Gamma^2=
\left(\begin{matrix}
0 &-\sigma^0  &0 &0 \cr
-\sigma^0& 0 & 0 &0 \cr
0&0&0 & \sigma^0 \cr
0&0&\sigma^0&0 
\end{matrix}\right)
~\quad,~~~\Gamma^3=
\left(\begin{matrix}
0 & i \sigma^0 &0 &0 \cr
 -i \sigma^0 &0 & 0 &0 \cr
0&0& 0&-i \sigma^0 \cr
0&0&i \sigma^0 &0 
\end{matrix}\right)
\\[.3cm]
&\Gamma^4=
\left(\begin{matrix}0 & 0 &\sigma^0 &0 \cr
0& 0 & 0 &\sigma^0 \cr
\sigma^0&0& 0 & 0 \cr
0&\sigma^0&0&0 
\end{matrix}\right)
~~\quad\quad,~~~\Gamma^5=
\left(\begin{matrix}
0 & 0 &-i \sigma^0 &0 \cr
0& 0 & 0 &-i \sigma^0 \cr
i \sigma^0 &0& 0 & 0 \cr
0&i \sigma^0&0&0 
\end{matrix}\right),}
where $\sigma^a$ are the usual Pauli matrices, and $\sigma^0$ is the identity matrix. The spin fields are given by the following basis elements
\eqna[zzcc]{
e^{+\frac{i}{2}H_\sl+\frac{i}{2}H_3+\frac{i}{2}H_\su}&=v_1
~~,~~~~
e^{+\frac{i}{2}H_\sl-\frac{i}{2}H_3+\frac{i}{2}H_\su}=v_2~,\cr
e^{-\frac{i}{2}H_\sl+\frac{i}{2}H_3+\frac{i}{2}H_\su}&=v_3
~~,~~~~
e^{-\frac{i}{2}H_\sl-\frac{i}{2}H_3+\frac{i}{2}H_\su}=v_4~,\cr
e^{+\frac{i}{2}H_\sl+\frac{i}{2}H_3-\frac{i}{2}H_\su}&=v_5
~~,~~~~
e^{+\frac{i}{2}H_\sl-\frac{i}{2}H_3-\frac{i}{2}H_\su}=v_6~,\cr
e^{-\frac{i}{2}H_\sl+\frac{i}{2}H_3-\frac{i}{2}H_\su}&=v_7
~~,~~~~
e^{-\frac{i}{2}H_\sl-\frac{i}{2}H_3-\frac{i}{2}H_\su}=v_8~,}
where $v_i$ denotes the $8$-dimensional vector that is equal to 1 in the $i$-th entry and equal to zero otherwise.

To perform the OPE between the fermions and spin fields, we write the spin field in terms of the corresponding basis element $v_i$, and act with the corresponding $\Gamma$ matrix associated to the fermion zero mode. For example, using \zzaa-\zzcc, the singular terms in the OPE $\psi^3_\su(z)\,e^{-\frac{i}{2}H_\sl+\frac{i}{2}H_3+\frac{i}{2}H_\su}(0)$ are given by
\eqn[zzdd]{\frac{1}{\sqrt{2z}}\psi^3_{\sl;0}\cdot e^{-\frac{i}{2}H_\sl+\frac{i}{2}H_3+\frac{i}{2}H_\su}(0)= \frac{1}{\sqrt{2z}}\,\Gamma_1 \cdot v_3=
-\frac{1}{\sqrt{2z}}\,v_4=
-\frac{1}{\sqrt{2z}}\,e^{-\frac{i}{2}H_\sl-\frac{i}{2}H_3+\frac{i}{2}H_\su}(0)~.}
Note that this has the opposite sign to what would be obtained by doing the naive OPE using the bosonized free-field $H_3(z)$, namely
\eqna[zzee]{\psi^3_\su(z)\,e^{-\frac{i}{2}H_\sl+\frac{i}{2}H_3+\frac{i}{2}H_\su}(0)
&=\frac{1}{\sqrt{2}}\left(e^{i H_3(z)}+e^{-i H_3(z)}\right) e^{-\frac{i}{2}H_\sl+\frac{i}{2}H_3+\frac{i}{2}H_\su}(0)
\\
&\sim\frac{1}{\sqrt{2z}}\,e^{-\frac{i}{2}H_\sl-\frac{i}{2}H_3+\frac{i}{2}H_\su}(0)~.}
This has the opposite sign to \eqref{zzdd}, and therefore it does not give consistent results. However, if we do the OPE's using the $\Gamma$ matrices as in \zzdd\ , then the cocycle factors are automatically taken into account and the signs are consistent.


\newsec{Supersymmetry currents in the block theory \texorpdfstring{$\bR_\phi\times\bS^3_\flat$}{} }
\label{app:blocksusy}

The block theory $\bR_\phi\times \sqsphere$ has a natural $N=(2,2)$ superconformal symmetry, which makes use of the parafermionic decomposition 
\eqn[sqdecomp]{
\sqsphere=\bigg[\bS^1_Y\times\frac{\sutwo_n}\uone\bigg]/\bZ_n~.
}
Recombining the factors as $\bR_\phi\times \bS^1_Y$ and the super-parafermion theory $\frac\sutwo\uone$, one has the component $N=2$ supercurrents,
\eqna[susyNS]{
\cG^{\pm}_{\rm free} = \pm e^{\pm iH_3}(\partial \phi\tight\mp i\partial Y)\pm Q_\ell\,\partial e^{\pm iH_3}
~~&,~~~~
\cJ_{R}^{\rm free} = i\partial H_3 +iQ_\ell\,\partial Y
\\[.2cm]
\cG^\pm_{\LG} = \psi^\pm_{\pf} \, \exp\Big[\pm \frac ia\, Z \Big]
\qquad\quad~~&,~~~~
\cJ_{R}^{\LG} = i\,a\,\partial Z
~,
}
where we have used the bosonized representation $e^{\pm iH_3}=\frac1{\sqrt2}(\chi_\phi^{~}\pm i\chi_Y^{~})$ built from the $\chi^i$ of section~\ref{sec:type0long}, and $Z$ bosonizes the R-current of the parafermion theory as in~\eqref{su2bos}.  The operator $\psi^\pm_{\pf}$ is the eponymous parafermion of the $\frac\sutwo\uone$ coset model, which can be thought of as $\lambda_{j'=0,m'=1,\mbar'=0}$ in the notation of~\eqref{bospf} (\ie\ it arises from the currents $j^\pm_\su$ by stripping off the exponential that carries the $j^3_\su$ charge).

At the supersymmetric point $R=\sqrt{n+1}$, there exists a second $N=2$ supersymmetry realized on the composite fermions $\cS^\pm$ of~\eqref{Spmcandidate}  
\eqn[Spmapp]{
\cS^\pm = \exp\Big[ \pm \frac i2 \Big( -H_3 + a\, Z + \sqrt{\frac2k}\, Y\Big) \Big] ~.
}
Conceptually, we want this other supersymmetry to pair these fermions $\cS^\pm$ with some bosonic currents in the block theory, rather than pairing $\chi_\phi^{~}\tight\pm i\chi_Y^{~}$ with $\partial\phi\tight\pm i\partial Y$.  In addition, from the worldsheet construction, we know that the R-current of the block theory measures momentum along $Y$, see equation~\eqref{susyops}.  These two criteria are enough to determine the desired field redefinition.  Note also that once we demand that~\eqref{Spmapp} is allowed in the spectrum, the supercurrents $\cG^\pm_{\rm free}$ and $\cG^\pm_{\LG}$ must be projected out since they are not mutually local with respect to $\cS^\pm$; thus only this exotic second supersymmetry can be present.

In order to preserve the conformal dimensions of exponential operators
\eqn[blockexptls]{
\exp\big[ip_3^{~} H_3+ip_Z^{~} Z+ip_Y^{~} Y\big]
}
(writing for simplicity only the left-moving part of the operator),
the field redefinition should be a rotational automorphism $\cR$ of the three-dimensional lattice of charges $(p_3,p_Z,p_Y)$. 
This rotation is specified by the two criteria above
\eqna[GandSrotation]{
&\cR\cdot \Big(1\;,\;0\;,\;0\Big) =
\Big(-\half\;,\;\frac a2\;,\;\frac Q2\Big)
\\[.2cm]
&\cR\cdot \Big(1\;, \; a\;, \; Q_\ell\, \Big) =
\Big(0\;,\;0\;,\;\sqrt{2k}\,\Big) ~.
}
The first specifies that the complex fermions $\chi_\phi^{~}\tight\pm i\chi_Y^{~}$ in the supercurrents $\cG^\pm_{\rm free}$ rotate into $\cS^\pm$, while the second says that the total R-current~\eqref{susyNS} rotates into the $Y$ direction.
The solution of these constraints is
\eqn[Rsoln2]{
\cR = 
\left(\begin{matrix}
-\half & \half\sqrt{\frac{n-2}{n}} & \sqrt{\frac{n+1}{2n}} \\[.3cm]
\half\sqrt{\frac{n-2}{n}} & -\frac{n+2}{2n} & \sqrt{\frac{(n-2)(n+1)}{2n^2}} \\[.3cm] 
\sqrt{\frac{n+1}{2n}} & \sqrt{\frac{(n-2)(n+1)}{2n^2}} & \frac1n
\end{matrix}\right)  ~.
}
Using this transformation, one can now build dimension $(\frac32,0)$ supercurrents as the image of the supersymmetry currents~\eqref{susyNS} under $\cR$
\eqna[susyR]{
\medtilde\cG^{\pm}_{\rm free} &=-\cS^\pm \bigg[i\bigg(
\sqrt{\frac{n+1}{2n}} \,\partial H_3 + \sqrt{\frac{(n-2)(n+1)}{2n^2}} \,\partial Z + \frac{1}{n} \,\partial Y
\bigg)\pm\partial\phi\bigg]\mp Q_\ell\,\partial\cS^\pm
\\[.2cm]
\medtilde\cJ_R^{\rm free} &= -i\frac{n-2}{2n}\,\partial H_3 + i \frac{(n+2)\sqrt{n-2}}{2n^{3/2}} \,\partial Z +i\frac{n^2+n+2}{n^{3/2}\sqrt{2(n+1)}}\, \partial Y
\\[.2cm]
\medtilde\cG^\pm_{\rm pf} &= \psi^\pm_{\pf}\;\exp\bigg[\pm\frac i2 \bigg( H_3 - \frac{(n+2)}{\sqrt{n(n-2)}}\, Z + {\sqrt{\frac2k}}\,Y\bigg) \bigg] 
\\[.2cm]
\medtilde\cJ_R^{\rm pf} &=  i\frac{n-2}{2n}\,\partial H_3 - i \frac{(n+2)\sqrt{n-2}}{2n^{3/2}} \,\partial Z + i\frac{(n-2)\sqrt{n+1}}{\sqrt2 n^{3/2}} \,\partial Y
}
(the somewhat more complicated expression on the first line compared to $\cG^\pm_{\rm free}$ arises because $\partial Y$ in $\cG^\pm_{\rm free}$ also transforms under $\cR$).
Because we are simply performing a field redefinition, these operators are guaranteed to also satisfy the $N=2$ superconformal algebra. Combined, they form the total supercurrent
\eqn[GtotR]{
\cG^\pm = \medtilde\cG^{\pm}_{\rm free} + \medtilde\cG^\pm_{\rm pf}
}
of the block theory, with the total R-current given by
\eqn[RcurrentR]{
\cJ_{R} = \medtilde\cJ_R^{\rm free} + \medtilde\cJ_R^{\rm pf} = i\sqrt{2k}\, \partial Y 
}
since this was one of the conditions that defined the transformation.

We can apply the same logic to the GSO~0 projected theory with its spacetime fermion $\Psi^\pm_{-\half}$, which has more or less the same expression as a fermion in the block theory as we had for $\cS^\pm$ in the GSO~2 projection, equation~\eqref{Spmapp}, but with a sign flip for $H_3$ in the exponent 
\eqn[Shatpmapp]{
\PsiST^\pm = \exp\Big[ \pm \frac i2 \Big( H_3 + a\, Z + \sqrt{\frac2k}\, Y\Big) \Big] ~.
}
One can then follow the same steps as above, to define an $N=2$ supersymmetry algebra in the block theory.  The corresponding rotation will differ by a reflection in the $H_3$ direction in the charge lattice, but otherwise is the same:  
\eqn[Rsoln]{
\cR_0 = 
\left(\begin{matrix}
\half & -\half\sqrt{\frac{n-2}{n}} & \sqrt{\frac{n+1}{2n}} \\[.3cm]
\half\sqrt{\frac{n-2}{n}} & -\frac{n+2}{2n} & -\sqrt{\frac{(n-2)(n+1)}{2n^2}} \\[.3cm] 
\sqrt{\frac{n+1}{2n}} & \sqrt{\frac{(n-2)(n+1)}{2n^2}} & -\frac1n
\end{matrix}\right)  ~.
}
The image of~\eqref{susyNS} under this rotation will be a pair of supersymmetry currents in the type 0 projected block theory.  We see that the chiral GSO projected block theories defined by mutual locality with respect to $\cS^\pm$, or with respect to $\PsiST^\pm$, are essentially identical.


\newsec{Twist operators for the symmetric product }
\label{app:twistops}

In this Appendix, we construct twist operators of the symmetric product.  The twisted sectors are labeled by conjugacy classes of the symmetric group, which are products of cycles where $w$ copies of the block theory are cyclically permuted. 
Our analysis follows~\rcite{Larsen:1999uk}.
Take $w$ copies of linear dilaton theory (with slope $Q_\ell$) plus an extra $\uone$ scalar $Y$ and label the copies by $a=1,\dots,w$.  The cyclic twist $\phi_a(e^{2\pi i}z)=\phi_{a+1}(z)$ with $a$ defined mod $w$ can be diagonalized by discrete Fourier transform
\eqn[phil]{
\phitil_\kappa = \frac1{\sqrt{w}} \sum_a \phi_a\, \exp[2\pi i\kappa a/w]
}
with $\kappa=0,\dots,w-1$; and similarly for $Y$.  The factor of $1/\sqrt{w}$ makes the $\phitil_\kappa$ canonically normalized.  We then have that the $\bZ_w$ twist operator is the product 
\eqn[total bos tw]{
\sigma^{\rm bos} = \prod_{\kappa=1}^{w-1} \sigma^{\rm bos}_{\kappa/w}
}
of the $\kappa/w$ twist operators $\sigma_{\kappa/w}$ that implement $\phitil_\kappa(e^{2\pi i}z)=e^{2\pi i\kappa/w}\phitil_\kappa(z)$ for each of the $\phitil_\kappa$, $\kappa\tight=1,\dots,n\tight-1$; similarly for $Y$.  Note that the linear dilaton is the sum of the linear dilatons of all the $\phi_a$, so lives in the $\kappa=0$ sector; the other sectors have no linear dilaton and are conventional twisted free bosons.  The dilaton slope in the $\phitil_{\kappa=0}$ field is then 
\eqn[qtil]{
\Qtil_\ell = \sqrt{w} \,Q_\ell ~.
}
The total dimension of the twist operator for $\phi$ and $Y$ in the $\kappa$-twisted sectors for $\kappa\ne 0$ is the sum
\eqn[hnonzero]{
2 \sum_{\kappa=1}^{w-1} \frac14\frac{\kappa}{w}\Big(1-\frac{\kappa}{w}\Big) = \frac1{12}\Big(w-\frac1w\Big) ~.
}
In the center-of-mass $\kappa=0$ sector, we append to $\sigma^{\rm bos}$ an exponential operator
\eqn[sigexp]{
\exp\Big[ -\medtilde Q_\ell \,\jtil\, \phi_0 + i\medtilde Q_\ell\, \tilde p_Y Y_0 \Big]
}
where $\phi_0,Y_0$ are the $\kappa=0$ sector fields~\eqref{phil}.  The conformal dimension of this operator is
\eqn[hcomL]{
-\half \Qtil_\ell^2 \jtil (\jtil-1) + \half \Qtil_\ell^2 \tilde p_Y^2
= -\half \Qtil_\ell^2 (\jtil-\hf)^2 + \frac{\Qtil_\ell^2}8 + \half \Qtil_\ell^2 \tilde p_Y^2  ~,
}
leading to a formula for the dimension of twist operators
\eqn[hwbosL]{
h_w^{\rm bos} = -\half \Qtil_\ell^2 (\jtil-\hf)^2 + \frac{\Qtil_\ell^2}8 + \half \Qtil_\ell^2 \tilde p_Y^2 + \frac1{12}\Big(w-\frac1w\Big) ~.
}
Note that if we rescale the momenta as $(\jtil-\hf)=(j-\hf)/w$ and $\tilde p_Y=p_Y/w$,%
\footnote{In this appendix, we will suppress the subscript ``ST'' on the quantum number $j$ to avoid clutter.}
then we can rewrite the conformal dimension as
\eqna[hwbosLa]{
h_w^{\rm bos} &= -\frac{1}{2w} Q_\ell^2 (j-\hf)^2 + \frac{\Qtil_\ell^2}8 + \half \Qtil_\ell^2\, \tilde p_Y^2 + \frac1{12}\Big(w-\frac1w\Big) 
\cr
&=  \frac{Q_\ell^2}{2w}\Big( -j(j-1) + p_Y^2 \Big) + \frac{2+3Q_\ell^2}{24}\Big(w-\frac1w\Big)  ~,
}
which has the form $\frac1w h_1+\frac{c}{24}(w-\frac1w)$ expected on general grounds~\rcite{Klemm:1990df}, with $h_1$ the dimension of an operator from the untwisted sector with $w=1$.

We will also want to twist the fermions $\chi^i$ in the block theory.  
Again going to the discrete Fourier basis as in~\eqref{phil}, one has $\kappa/w$ twists which are generalized spin fields.  The product of such spin fields 
\eqn[total ferm tw]{
\sigma^{\rm ferm}_\pm = \prod_{\kappa=1}^{w-1} \sigma^{\rm ferm}_{\pm\kappa/w}
}
yields an operator of dimension 
\eqna[hfermL]{
h_w^{\rm ferm} &= \sum_{\kappa=1}^{w-1} \frac12\Big(\frac{\kappa}{w}\Big)^2 
= \frac{(w-1)(2w-1)}{12 w} 
\\
&= \frac1{8w} {(w-1)^2} + \frac1{24}\Big(w-\frac1w\Big) ~.
}
As expected from bosonization, the last term on the second line is the dimension of the cyclic twist of a free boson, while the first term in the second line gives the contribution from exponential in the $\kappa=0$ sector that carries the overall fermion charge $\hf(w-1)$. 
The minimal dimension twist operator omits this charge exponential, but as we are interested in BPS twist operators carrying the appropriate R-charge, we will be using operators of the above dimension.%
\footnote{Note that the bosonization that leads to this second expression for the dimension $h_w^{\rm ferm}$ is different from the one that realizes the various spin fields in the first expression in terms of free boson exponentials.  The latter constructs the discrete Fourier eigenmodes $\chi^{~}_\kappa$ of the fermions analogous to $\phil$, and then bosonizes them in terms of bosons $\medtilde H_\kappa$; the former bosonizes the block fermions $\chi^{~}_a$ in terms of bosons $H_a$, which then have a discrete Fourier transform $H_\kappa$.  There is no simple relation between $H_\kappa$ and $\medtilde H_\kappa$.  We will work with the fields $H_\kappa$. }

To summarize, the operator that builds the ground state twist operator of a free $N=2$ superfield resolves via a discrete Fourier transform~\eqref{phil} into the product of twist operators for the twist eigenmodes.  For bosons these twist operators are the standard $\bZ_w$ twist operators that create the $\kappa/w$ twist ground state; the fermions can be bosonized and the fermion twist operator written in terms of the twist operators for the bosons.  

One can then decorate this operator with oscillator excitations as well as exponentials of the center-of-mass momentum.  
The center-of-mass momentum profile has one of two leading forms for a given conformal dimension,  
\eqn[nonnormphi]{
\exp\Big[ -\half\tilde Q_\ell \phi_0 \pm \frac{Q_\ell}{\sqrt w}\Big(j -\half\Big)\phi_0 \Big] ~.
}
The upper sign corresponds to non-normalizable operators, see~\eqref{twistphidep}; the lower sign corresponds to normalizable profiles.

One finds supersymmetric twist operators (under the usual $N=2$ supersymmetry~\eqref{susyNS}) for $p_Y=\pm[j+\hf(w-1)]$ and $\eta_3=\pm\hf(w-1)$
\eqn[htotL]{
h_w = h_w^{\rm bos} + h_w^{\rm ferm} = \frac14\Big((w-1)(1+Q_\ell^2)+2jQ_\ell^2\Big) ~.
}
Indeed these operators are (anti)chiral as one can check from the R-charge 
\eqn[eeeetwo]{
R_\ST =  Q_\ell^2 \,p_Y+\eta_3 = \pm 2h_w ~.
}
The splitting of the $\phi$ dimension into a contribution to the twist ground state energy $\frac18 Q_\ell^2(w-1/w)$ plus the remaining exponential in $\phi$ obscures the chiral nature of this twist operator.  Reverting to the original parametrization in terms of $\jtil,\tilde p_Y$, one sees that the chiral twist operator has the form 
\eqn[chiraltwist]{
\Sigma_{\rm free}^\pm = \exp\Big[ -\frac{Q_\ell}{\sqrt w}\Big(j+\hf(w-1)\Big)\big(\phi\mp iY\big)_{0}  \Big] \, \sigma^{\rm bos}\, \sigma^{\rm ferm}_\pm 
}
which takes the form of the BPS ground state of $\bZ_w$ twisted scalars and fermions, times a chiral exponential of the center-of-mass.
Note that we have assumed a normalizable profile, in order to describe candidate wall operators.

One can perform a similar analysis for the Landau-Ginsburg model, using an $N\tight=2$ Feigin-Fuchs style representation~\rcite{DiVecchia:1985ief,Kato:1986td,Yu:1986xc,Mussardo:1988av,Ito:1989cm}, 
with a scalar $\phi'$ having a linear dilaton with imaginary slope of magnitude $Q_\LG\tight=\sqrt{\frac2n}$, and a compact timelike boson $Y'$.
The explicit map relating the linear dilaton and Feigin-Fuchs theories is
\eqn[LDFFmap]{
Q_\ell\to -iQ_{\LG} 
~~,~~~~ j\to j'+1 
~~,~~~~ \phi\to \phi'
~~,~~~~ Y\to iY' ~.
}
This map results in a bosonic twist field dimension 
\eqna[hwbosFF]{
h_w^{\rm bos} &= \half \Qtil_\LG^2 (\jtil'+\hf)^2 - \frac18{\Qtil_\LG^2} - \half \Qtil_\LG^2 (\tilde p'_Y)^2+ \frac1{12}\Big(w-\frac1w\Big) 
\\[.2cm]
&= \frac{h_1}{w} + \frac{(2-3Q_\LG^2)}{24}\Big(w-\frac1w\Big) 
}
where again
\eqn[hzeroFF]{
h_1 = \frac{Q_\LG^2}{2}\Big( j'(j'+1) - (p'_Y)^2\Big) 
}
and $(\jtil'+\hf)=(j'+\hf)/w$. 
The fermion charge $\eta_\su=\hf(w-1)$ is the same as above.
The R-charge under the usual $N=2$ supersymmetry~\eqref{susyNS} is thus 
\eqn[eeeethree]{
R_\ST = Q_\LG^2 p'_Y + \hf(w-1) ~.
}

Now let's put together the $N=2$ linear dilaton and Landau-Ginsburg results.  Each cyclic twist sector has a twist ground state of dimension 
\eqn[gdtwistdim]{
h_{\rm min} = \frac{c}{24}\Big( w-\frac1w\Big) = \frac{k}{4}\Big( w-\frac1w\Big)
}
that combines the minimum dimension twist operators $\sigma^{\rm bos}$~\eqref{total bos tw} for the bosons and the uncharged twist ground state operator for the fermions (which can be obtained through bosonization~-- see below for the $\bZ_2$ case);  one then decorates this twist operator with exponentials of the form~\eqref{nonnormphi} (with $j=\hf, Q_\ell=\sqrt{\frac{2}{n(n+1)}}$ for the ground state of $\phi$; and $j'=0, Q_\LG=\sqrt{\frac 2n}$ for the Landau-Ginsburg ground state).  One can then build excited states using center-of-mass exponentials, and (un)twisted oscillator raising operators.  The result is equation~\eqref{windingdim}. 

To obtain BPS operators, we set 
\eqna[twistone]{
j = \hf \nu n-j'
~,~~~
&p_Y = -\big(j+\hf(w-1)\big)
\,~,~~~\,
\eta_3=\hf(w-1)
\\[.2cm]
&p_{Y'} =  -j' + \hf(w-1)
~,~~~
\eta_\su=\hf(w-1)
}
(with $\nu\in\bN$) to focus on potential chiral fields, and so that the operator survives the $\bZ_n$ orbifold projection that turns $\bS^1\times LG_n$ into $\sqsphere$. %
We then find a total twist dimension
\eqn[htot]{
h_w = \frac{(w-1)n+2j'+\nu}{2(n+1)}
}
with $R_w=2h_w$ for a chiral operator.
The parameter $\nu$  describes different linear dilaton dressings of the same Landau-Ginsburg primary.%
\footnote{For instance, if one considers the chiral spectrum of the SCFT~\eqref{blockwall} one finds chiral primaries for each $\nu=1,...,n$ corresponding to the various twisted sectors of the $\bZ_{n+1}$ orbifold of the $\frac{\sltwo_{n(n+1)}}\uone$ cigar coset model.}  
The $AdS_3$ worldsheet theory has only chiral operators with $\nu=1$, as we see from~\eqref{YwRchg}.

For the chiral operator $\cY^{(-w)}_{j'}$ of~\ref{sec:BPS} and its FZZ dual representative  $\medtilde\cY_{j'}^{(-w-1)}$, we have
\eqna[Y jwsjst]{
j_\WS = \half+\frac{2j'+1}{2(n+1)} ~~&\longleftrightarrow~~ j_\ST = j'+1
\\
\jtil_\WS = \frac{n-j'}{n+1} \qquad\quad ~~&\longleftrightarrow~~ \jtil_\ST = \frac{n}{2} - j'  ~.
}
where the FZZ dual spin is $\jtil=\frac k2\tight+1\tight-j$, and we have used~\eqref{jSTjWS}.  
The corresponding spacetime operators have scaling dimension $h_\ST = j_\WS\tight-\hf\tight+\frac k2 |w|$ (see equation~\eqref{YhSTwneq0}); their wavefunctions scale as $j_\ST$ for a component of winding $w$ and $\jtil_\ST$ for a component of winding $w\tight-1$ (recall $w<0$ in our conventions).

The spacetime-chiral worldsheet operators $\medtilde\cY_{j'}^{(-w-1)}$ correspond to the twist $w$ 
chiral operators constructed above.  The quantum numbers and radial wavefunctions on the two sides match in the following way:
(1) The two have the same angular harmonic on $\bS^3_\flat$;
(2) The cyclic twist sector $w$ matches the winding $-w$ of this particular representative of the worldsheet operator;
(3) For $\nu=1$ the conformal dimension~\eqref{htot} matches the spacetime conformal dimension~\eqref{YwRchg} for of the worldsheet operator; 
(4) The two have wavefunctions with the same falloff in $\phi$, if we identify the spin $j$ in~\eqref{twistone} of the symmetric orbifold chiral twist operator with the value $\jtil_\ST$ of~\eqref{Y jwsjst} for the FZZ dual representative $\medtilde\cY^{(-w)}_{j'}$. 

We have described twist operators using the language of the standard $N=2$ supersymmetry~\eqref{susyNS}, however we are ultimately interested in the rotated theory with the fermions $\cS^\pm$ or $\PsiST^\pm$.  In principle, one could try to apply the field rotation $\cR$ of appendix~\ref{app:blocksusy} to the $N=2$ free fields, and build the twist superfields, however the $\bZ_w$ twist operators are somewhat simpler to describe in the field basis of the standard $N=2$ supersymmetry.  The map of Appendix~\ref{app:blocksusy} is a simple field redefinition, and thus cannot affect any physical quantity.  We can thus calculate the operators and their correlation functions in the standard basis, and interpret the results in the rotated basis by applying the rotation $\cR$ to the result.

Note that there is an additional field rotation that must be considered.  In the $N=2$ Feigin-Fuchs representation of the Landau-Ginsburg model used above, the standard $N=2$ R-current is a linear combination of $Y'$ and the bosonized complex fermion (since it arises from the free field R-current in~\eqref{susyNS} via the substitution~\eqref{LDFFmap}); however, in the text we worked with the field $Z$ that directly bosonizes the R-current of the Landau-Ginsburg theory, see also~\eqref{susyNS}.  There is indeed a field basis $(Z,X')$ related to $(H_\su,Y')$ of the $N=2$ Feigin-Fuchs representation by a boost transformation (because $X'$ and $Y'$ have negative signature).  In this boosted basis, $Z$ bosonizes the R-current~\eqref{su2bos} of the super-parafermion theory, while the orthogonal boson $X'$ together with $\phi'$ form the Feigin-Fuchs representation of the bosonic $\frac{\sutwo_{n-2}}\uone$ parafermion theory~\rcite{Gerasimov:1989mz,Griffin:1988tf,Distler:1989xv,Bilal:1989py}.
Indeed, the bosonic parafermions $\psi_{\rm pf}^\pm$ appearing in the supercurrents~\eqref{susyNS}, \eqref{susyR} are written as
~\rcite{Nemeschansky:1989wg}
\eqna[pfbos]{
\psi^\pm_{\rm pf} = \frac{1}{\sqrt{2}} \bigg(\partial X' \mp
\sqrt{\frac{n}{n-2}}\,\partial\phi' \bigg) \,
\exp\Big[\pm i \sqrt{\frac{2}{n-2}}\, X' \Big] ~.
}
The boost transformation of the fields is given by 
\eqna[FFrot]{
H_\su = \frac{\sqrt{n}\,Z - \sqrt{2}\,X'}{\sqrt{n-2}}
~~&,~~~~
Y' = \frac{\sqrt{n}\,X' - \sqrt{2}\,Z}{\sqrt{n-2}} ~,
\\[.2cm]
Z = \frac{\sqrt{n}\,H_\su + \sqrt{2}\,Y'}{\sqrt{n-2}}
~~&,~~~~
X' = \frac{\sqrt{n}\,Y' + \sqrt{2}\,H_\su}{\sqrt{n-2}} ~,
}
which we write $\cR_{\LG}\cdot(H_\su,Y') = (Z,X')$.  
Then the combined transformation
\eqn[Rtot]{
\cR_\tot = \cR\cdot\cR_{\LG}
}
defines the field redefinition from the field space basis $(H_3,H_\su,Y',Y)$ to the field space basis directly related to the worldsheet formalism in sections~\ref{sec:typeIIshort}--\ref{sec:wall}.



\subsec{The marginal \texorpdfstring{$\bZ_2$}{} twist operator}
\label{app:Z2twist}

Let us be a bit more explicit for the $\bZ_2$ twisted sector, which is the key ingredient in the construction of the wall in section~\ref{sec:wall}.  We can bosonize the fermions in two copies of the block theory in terms of $H_3^{(i)},H_\su^{(i)}$, $i=1,2$; then we can write everything we need if we can construct the symmetric orbifold of two free bosons.  

Consider then the symmetric orbifold $\cM=\big(\bS^1_\sfX\big)^2/\bZ_2$ of two bosons $\sfX_{(1)},\sfX_{(2)}$, each of radius $R_\sfX$, and define (anti)symmetric linear combinations
\eqn[XSA]{
\sfX_S=\frac1{\sqrt2}\big(\sfX_{(1)}+\sfX_{(2)}\big)
~~,~~~~
\sfX_A=\frac1{\sqrt2}\big(\sfX_{(1)}-\sfX_{(2)}\big)~.
}
The diagonal current $\uone_S$ defined by
\eqn[JX]{ 
J_\sfX=i\partial \sfX_{(1)} + i\partial \sfX_{(2)} = i\sqrt2\,\partial \sfX_S
}
is conserved and present in the orbifold theory, and so we can use the coset construction to write any operator in the symmetric product $\cM$ as an operator in the $\uone_S$ current algebra theory, times an operator in the quotient theory $\cM/\uone_S$.  The quotient theory is basically that of the $\bZ_2$ twisted boson $\sfX_A$.

The spectrum of states is determined from the partition function~\rcite{Klemm:1990df} 
\eqna[ZsymX]{
\cZ_\cM &= \sum_{h,\bar h} d(h,\bar h)\,q^{h-\frac c{24}}\, \bar q^{\bar h-\frac c{24}}
\\[.1cm]
&=
\half\Big(\cZ_\sfX(\tau,\bar\tau) + \cZ_\sfX(2\tau,2\bar\tau)\Big) + \half\Big(\cZ_\sfX(\hf\tau,\hf\bar\tau) +
\cZ_\sfX\big(\hf(\tau\tight+1),\hf(\bar\tau\tight+1)\big)\Big) 
}
where $q=e^{2\pi i\tau}$, $d(h,\bar h)$ is the degeneracy of states with conformal weight $(h,\bar h)$, and $\cZ_\sfX$ is the partition function of the $U(1)_\sfX$ theory
\eqn[ZX]{
\cZ_\sfX = \frac{1}{\eta(q)\bar\eta(\bar q)}
\sum_{m,n\in\bZ} \, q^{\half p^2} \, \bar q^{\half \bar p^2}
~~,~~~~ \eta(q) = q^{-\frac1{24}}\prod_{n=1}^\infty\frac{1}{1-q^n} ~~,~~~~
p,\bar p = \frac n{R_\sfX} \pm \half mR_\sfX  ~.
}
In~\eqref{ZsymX}, the first term on the second line is the contribution of the untwisted sector of the symmetric orbifold, while the second term is the contribution of the twisted sector, which exhibits the expected form of twisted scaling dimensions $h=\hf h_1 + \frac 1{16}$ of a $\bZ_2$ symmetric orbifold with block theory spectrum $\{h_1\}$ and $c=1$.

We see that the twisted sector ground state has dimension $h=\frac1{16}$ (see the discussion around equation~\eqref{hfermL}) and can be thought of as the dimension of the $\bZ_2$ twist operator $\sigma_{\sfX_A}$ for $\sfX_A$, however the ground state is non-degenerate because the fixed locus of $\cM$ has only a single connected component.%
\footnote{This non-degeneracy of the twist ground state is an example of the fact that the theory is not the tensor product of a twisted scalar on $\bS^1/\bZ_2$ (which has two degenerate twisted sector ground states) and the center-of-mass $\uone$.}
If we set $R_\sfX=1$ in~\eqref{ZX}, the radius for the bosons $H_3,H_\su$, the lowest charged states in the twisted sector are associated to the spin fields of dimension $h_1=\frac18$ in the untwisted theory; these twisted operators have dimension $h=\hf h_1+\frac1{16}=\frac 18$, and correspond to operators 
\eqn[fermZ2tw]{
\sigma_\pm^{\rm ferm} = e^{\pm \frac{i}{2\sqrt2} \sfX_S}\,\sigma_{\sfX_A}
}
where the exponential of dimension $h=\frac1{16}$ carries the $U(1)$ charge. 
The operators $\sigma_\pm^{\rm ferm}$ have the dimension~\eqref{hfermL}, with $w=2$.  More generally, we can consider the above construction for other radii and realize the $\bZ_2$ twist operators for the bosons $Y,Y',X',Z$ \etc; and we can decompactify, add a linear dilaton, and give the twist operator a center-of-mass momentum in order to realize the twist operators for $\phi,\phi'$.

Now consider the chiral $\bZ_2$ twist operators of the block theory $\bR_\phi\times\bS^3_\flat$ with $w=2,j'=0,\nu=1$.  It turns out that the charge vector 
\eqna[Z2 twist chg]{
p_\phi = \half n
~~,~~~~
&p_Y = -\frac{1}{\sqrt{2k}}
~~,~~~
\eta_3 = \half
~~,~~~~
\eta_\su = \half
\\[.2cm]
p_{\phi'} = 0
~~~~,~~~~
&p_{Y'} = \frac{1}{\sqrt{2n}} 
~~~,~~~
p_{X'} = 0
~~,~~~~
p_Z = \half a 
}
is invariant under the rotation~\eqref{Rsoln}, and thus is the same under the exotic realization of supersymmetry in the block theory and in the theory with block fermions $\chi^i$ and standard supersymmetry.

Finally, we can assemble the various ingredients to write the overall dimension $(\hf,\hf)$ chiral $\bZ_2$ twist operators as
\eqna[Z2chiraltwist]{
\Sigma^\pm &= \Sigma^\pm_{\rm free}\Sigma^\pm_{\LG}
\\[.2cm]
\Sigma^\pm_{\rm free} &= \exp\Big[-\frac{1}{2\sqrt k} \big( \phi_S \mp i Y_S \big) \pm \frac i{2\sqrt2} H_{3,S}  \Big] \times \big(\sigma^{~}_{\phi_A} \, \sigma^{~}_{Y_A} \, \sigma^{~}_{H_{3,A}}\big) 
\\[.2cm]
\Sigma^\pm_{\LG} &= \exp\Big[-\frac{1}{2\sqrt n} \big( \phi'_S \mp i Y'_S \big) \pm \frac i{2\sqrt2} H_{\su,S}\Big] \times \big(\sigma^{~}_{\phi'_A} \, \sigma^{~}_{Y'_A} \, \sigma^{~}_{H_{\su,A}} \big) ~,
}
where we have suppressed the right-moving contribution, which has the same form, and the~$\pm$ signs in the exponent are all correlated.  Through the binary choice of overall sign in the exponent for each chirality, one realizes all the $(c,c)$, $(a,a)$, $(c,a)$ and $(a,c)$ dimension $(\hf,\hf)$ operators.

We then wish to act on these dimension $(\hf,\hf)$ operators with the supercurrent raising operators $\cG^-_{-\half}$ and $\bar\cG^-_{-\half}$.  To see how this works, consider for instance complex bosons $\Upsilon^\pm=\phi\pm iY$ and their superpartners $\chi^\pm$ under the standard $N=2$ free field supercurrent~\eqref{susyNS}.
We have two copies $\Upsilon^\pm_{(1)},\Upsilon^\pm_{(2)}$ \etc. for the two blocks being sewn together, and (anti)symmetric linear combinations
\eqna[Ysa]{
\Upsilon^\pm_S &= \frac1{\sqrt2}\big( \Upsilon^\pm_{(1)} + \Upsilon^\pm_{(2)} \big)
~~\;,~~~~
\Upsilon^\pm_A = \frac1{\sqrt2}\big( \Upsilon^\pm_{(1)} - \Upsilon^\pm_{(2)} \big) 
\\[.2cm]
\chi^\pm_S &= \frac1{\sqrt2}\big( \chi^\pm_{(1)} + \chi^\pm_{(2)} \big)
~~\;,~~~~
\chi^\pm_A = \frac1{\sqrt2}\big( \chi^\pm_{(1)} - \chi^\pm_{(2)} \big) 
}
that diagonalize the twist. Inverting this transformation and plugging into equation~\eqref{susyNS}, the total supercurrent for these fields is then
\eqna[GtotUpsilon]{&\hskip 2cm
\cG^\pm = \cG^\pm_{(1)} + \cG^\pm_{(2)} = \cG^\pm_{S} + \cG^\pm_{A}
\\[.2cm]
&\cG^{\pm}_{S} =\pm\chi^\pm_S\,\partial\Upsilon^\mp_S \pm \sqrt2\, Q_\ell\,\partial \chi^\pm_S
~~,~~~~
\cG^{\pm}_{A} =\pm\chi^\pm_A\,\partial\Upsilon^\mp_A   ~.
} 
The supercurrent contribution $\cG^\mp_A$ acts on $\Sigma^\pm$ by the raising operator $(\Upsilon^\pm_A)_{-\half}$ while the zero mode of $\chi^\mp_A$ acts as a gamma matrix to flip the sign of the $H_3$ charge of the ground state.
In addition, the action of $\cG^\mp_S$ on $\Sigma^\pm$ measures its momentum along $\Upsilon^\mp$ and acts on the ground state twist operator by the fermion raising mode $\big(\chi^\mp_S\big)_{-\half}$, and so contributes if that momentum is nonzero; this is the term responsible for the violation of holomorphy for $\chi^\pm$ in section~\ref{sec:genops}.  One has a similar story for the Landau-Ginsburg contribution in the Feigin-Fuchs representation.


\newsec{The BPS spectrum of the deformed symmetric product
\label{app:halfBPS}}


The single-string 1/2-BPS spectrum exhibited in section~\ref{sec:BPS} for the type IIB (2,2) theory suggests a multi-string BPS spectrum in which, say, all the strings are in $(c,c)$ states associated to the chiral operators $\cY_{j'}^{(w)}$, or all in $(c,a)$ states associated to the twisted chiral operators ${\medhat\cY}^{(w)}_{j'}$.  In this appendix, we describe a conjectural picture of the 1/2-BPS spectrum, based on an extrapolation of the multi-string BPS spectrum deduced from the worldsheet to the full theory.  We will concentrate on the $(c,c)$ spectrum.

Before discussing multi-string BPS states, we will first discuss how the wavefunction of single-string 1/2-BPS states at large $\phi$ fits into the symmetric product structure. The normalizable chiral operators correspond to the worldsheet operators $\cY_{j'}^{(-w)}$ in \eqref{IIB22 Ywneq0 susypt}. It is natural for the leading asymptotic of the operator with $\sltwo$ winding $w_{\sl}=-w$ to lie in the $\mathbb{Z}_{w+1}$ twisted sector of the spacetime CFT. More precisely, we consider the FZZ duals $\medtilde\cY_{j'}^{(-w-1)}$ of the worldsheet operators $\cY_{j'}^{(-w)}$, since these have $\sltwo$ windings $w_{\sl}\leq -1$, and as we saw in Appendix~\ref{app:twistops}, the wavefunctions of these FZZ duals agree with those of the cyclic BPS twist operators of the spacetime CFT. Let's assume this structure for the moment and explore its consequences.

There is, however, a subtlety we ought to mention.  At large $\phi$, the block CFT has $N=2$ supersymmetry. Since the R-charges of the operators $\cY_{j'}^{(w=0)}$ lie in the range $0<R_{ST}<k$, $N=2$ spectral flow in the block applied to their antichiral conjugates shifts the R-charge by $\frac13 c_{\rm block}=2k$, and tells us that there should exist additional normalizable operators whose asymptotics at large $\phi$ lie in the block theory, with R-charges in the range $k<R_{ST}<2k$. This would seem to suggest that there should be extra chiral operators, which are not given by $\cY_{j'}^{(-w)}$.

Of course, there are chiral operators with R-charges in the range $k<R_{ST}<2k$, namely $\cY_{j'}^{(w=-1)}$, whose FZZ duals are in winding two.  We propose that these are the additional operators we are looking for, and not some entirely new class of BPS operators (a thorough search of the worldsheet spectrum leads us to believe we have exhibited the complete chiral spectrum in section~\ref{sec:BPS}).  

In order to understand what is going on, it is useful to recall our discussion of the wall operator in section~\ref{sec:wall}. There we saw that the normalizable operator $\cY_{j'=0}^{(-1)}$ has a FZZ dual with $\sltwo$ winding $w_{\sl}=-2$, and therefore the wall has a contribution in both the untwisted and $\mathbb{Z}_2$ twisted sectors. 

A similar story is at work here. The FZZ dual of $\cY_{j'}^{(0)}$ has $\sltwo$ winding $w_{\sl}=-1$, and therefore we identify it as being in the untwisted sector (\ie\ the block theory) and having R-charge in the range $0<R_{ST}<k$. On the other hand, the worldsheet operator $\cY_{j'}^{(-1)}$ also has winding $w_\sl=-1$ (while its FZZ dual has winding $w_\sl=-2$).  Furthermore, the R-charges of $\cY_{j'}^{(-1)}$ lie precisely in the range $k<R_{ST}<2k$ and are related by $N=2$ spectral flow in the block to those of (the charge conjugate of) $\cY_{j'}^{(0)}$.

In the end, it seems there is no doubling of the spectrum~-- there are two sets of worldsheet operators with unit winding, each covering half the range of spacetime R-charges in the block theory.  But one set is FZZ dual to winding zero, while the other is FZZ dual to winding two.  The BPS states on the worldsheet do not have a unique winding number, because FZZ duality tells us that a given normalizable state has support in at least two winding sectors.  Similarly, although our understanding of the spacetime CFT is somewhat more primitive at the moment, the fact that the wall is built from a $\bZ_2$ twist operator suggests that normalizable states in the spacetime theory are not supported entirely in a single twist sector.  The suggestion is that we identify worldsheet winding with spacetime twist, at least when both are nonzero, and that the set $\cY^{(w)}_{j'}$ comprises the full set of single-string chiral operators.

Because of the absence of an $\sltwo$ invariant state, BPS multi-string states (where each string in the state is in the $(c,c)$ ring) have energies of order $p$.  For generic states having this energy, the perturbative picture breaks down; however, for BPS multi-string states, one expects non-renormalization theorems to protect the state, so that at least the state counting is accurate even if other aspects of the states (such as their wavefunctions) receive modifications.

This result should hold even when the multi-particle state has an energy that scales with the central charge.  The 1/2-BPS spectrum is then the Fock space of such strings which we can characterize via a tensor product state 
\eqn[halfBPSnoncrit2]{
\prod_{w,j'} \big| N_{w,j'} \big\rangle  ~.
}
We adopt a convention for describing the state where the quantum number $w$ is the order of a cyclic twist in the spacetime CFT, which matches the winding $-w$ of the of the operator $\medtilde\cY^{(-w)}$
The total central charge of the spacetime theory is $6pk$, where the total string winding $p$ is partitioned into the number $N_{w,j'}$ of strings with winding $w$ and R-charge $\frac{2j'+1+(w-1)n}{n+1}$,%
\footnote{The winding $w$ in \YwRchg\ is offset by one because it is the FZZ dual representative of the worldsheet operator $\cY^{(w)}_{j'}$ that matches the spacetime CFT.}
with
\eqn[pval]{
p = \sum_{w,j'} w N_{w,j'} ~.
}
A symmetric product CFT has just such a structure, however to verify this proposal will require a better understanding of how the wall deformation leads to normalizable states (for instance how it allows only $\nu=1$ in~\eqref{twistone}).  One can identify a set of chiral operators in the twisted sectors of the symmetric orbifold which match the above spectrum, by comparing the single-string spectrum of equation~\eqref{YwRchg} with the corresponding R-charges of cyclic twist operators of the symmetric orbifold, equation~\eqref{htot}.

Assuming this structure holds, the spacetime 1/2-BPS state with minimum R-charge consists of $p$ copies of the $j'=0,w=1$ single-string state ($w=-1$ for the FZZ dual worldsheet operator that matches the corresponding operator in the block theory), having total R-charge 
\eqn[Rmin]{R_{\ST,\rm min}=\frac{p}{n+1}~.
}
The number of 1/2-BPS states grows with R-charge as one partitions the available choices of $j'$ and $w$ out of a winding budget totalling $p$, reaching a maximum at $\hf R_{\ST,0}$, where $R_{\ST,0}$ is the maximal R-charge allowed by the unitarity of $\NN=2$ SCFT
\eqn[Rzero]{
R_{\ST,0} = \frac c3 = \frac{2pn}{n+1}
~.
}
Among the states around R-charge $\hf R_{\ST,0}$ is the maximally twisted ground state consisting of a single string of winding $w=p$.%
\footnote{Somewhat confusingly also called the long string sector of the spacetime CFT, a terminology we avoid here.}

In the range $R_\ST\in(\hf R_{\ST,0},R_{\ST,0})$, one has a reflected copy of the spectrum in the range $R_\ST\in(0,\hf R_{\ST,0})$.  The chiral spectrum is always symmetric about its midpoint, because a unit of spectral flow in the spacetime $N=2$ R-symmetry shifts the R-charge by $R_\ST\to R_\ST+c/3$, and maps the antichiral spectrum to the chiral spectrum (this fact was used in the discussion of the chiral spectrum in the block theory above).  Thus the first half of the antichiral spectrum of the full spacetime theory maps to the top half of the chiral spectrum.  In particular, the maximal R-charge attained by an actual state in the theory is 
\eqn[Rmax]{
R_{\ST,\rm max} = R_{\ST,0} - \frac{p}{n+1} =\frac{p(2n-1)}{n+1} ~.
}
Note that this state has the same quantum numbers as the multi-string state created by $p$ copies of the maximally charged unit winding operator $\cY^{(-1)}_{j'=\frac n2-1}$. As discussed above, these operators are related by the approximate spectral flow in the block theory at large $\phi$ to the minimal charge antichiral operator conjugate to $\medtilde\cY^{(-1)}_{j'=0}$.  Taking $p$ copies of this relation then relates the minimal and maximal R-charges in the full theory. 

In the critical dimension, each $\hf$-BPS state sources a different geometry, from global $AdS_3$ for the CFT vacuum to the maximally twisted state which sits at the BTZ threshold~\rcite{Lunin:2001fv,Kanitscheider:2007wq}. Stringy aspects of these geometries are now well-understood~\rcite{Martinec:2020gkv}, and have been extended to particular $k<1$ examples in~\rcite{Brennan:2020bju}.  It would be interesting to understand the general case.



\end{appendices}


\vskip 2cm
\bibliography{fivebranes}

\end{document}